\documentclass{./elsarticle}
\biboptions{sort&compress}

\usepackage{amssymb}
\usepackage{doi}
\usepackage{hyperref}
\usepackage{amsmath}
\usepackage{subcaption}
\usepackage{siunitx}
\usepackage{setspace}
\usepackage[left]{lineno}
\usepackage{booktabs}
\usepackage{appendix}
\usepackage{setspace}
\usepackage{graphicx} %
\usepackage{subcaption}
\usepackage{stackengine}
\usepackage{xcolor}   
\definecolor{mypurple}{RGB}{128, 0, 128}

\author[1]{Wei Wen}
\fnmark[1]
\ead{wenwei@nuaa.edu.cn}
\address[1]{College of Energy and Power Engineering, Nanjing University of Aeronautics
and Astronautics, Nanjing 210016, China}
\author[1]{Wenkai Qi}
\fnmark[2]
\ead{qwkai@nuaa.edu.cn}
\author[1]{Weidong Wen}
\fnmark[3]

\nonumnote{E-mail addresses: wenwei@nuaa.edu.cn (Wei Wen).}

\begin{document}
\begin{frontmatter}



\title{Hessian-Enhanced Alternating Frequency/Time method for Computing Resonance Backbone Curves}




\onehalfspacing
\begin{abstract}
Computing resonance and anti-resonance backbone curves in complex nonlinear mechanical systems is of high engineering relevance but remains computationally challenging, especially for large finite-element (FE) models. Existing manifold-based approaches often rely on polynomial parameterizations, limiting their effectiveness for general smooth, non-polynomial nonlinearities. To overcome these limitations, we develop a direct optimization framework that employs a Lagrange multiplier formulation to determine the resonance backbone curve on the response surface constrained by the harmonic balance governing equations. Crucially, solving this formulation efficiently requires second-order sensitivity information. Therefore, the primary innovation of this work is the derivation of a analytical Hessian Tensor for generic $C^2$-continuous nonlinear elements. This is achieved by combining an extended Alternating Frequency/Time (AFT) method for computing second-order derivatives with local-coordinate tensor transformations. By integrating this analytical Hessian into the solver, the proposed framework ensures robust convergence and significantly reduces runtime, making it practical for large-scale models where numerical differentiation is computationally prohibitive. The method is validated on three benchmarks of increasing complexity: a two-degree-of-freedom (2-DOF) system with cubic nonlinearity, a beam with cubic stiffness or hyperbolic tangent (tanh) friction nonlinearities, and an industrial-scale finite element model of a compressor bladed disk (blisk) with a friction ring damper. Results demonstrate that the proposed framework accurately and efficiently computes both resonance and anti-resonance backbone curves, providing a robust frequency-domain tool for structures with non-polynomial nonlinearities.
\end{abstract}


\begin{keyword}
Resonance backbone curves \sep Nonlinear vibration \sep Harmonic Balance Method \sep Extended Alternating Frequency/Time (AFT) \sep Analytical Hessian Tensor \sep $C^2$-continuous nonlinearity
\end{keyword}

\end{frontmatter}
\onehalfspacing
\section{Introduction}
\label{sec:intro}

Nonlinearity is ubiquitous in modern high-performance mechanical systems, arising notably from geometric effects and frictional contact interfaces. Unlike linear systems, these structures lack the superposition property and exhibit complex dynamic behaviors, including bifurcations \cite{guckenheimer1983nonlinear,seydel2009practical}, internal resonances \cite{kerschen2005method,hill2016influence}, and quasi-periodic motions \cite{li2022nonlinear,wu2025efficient}. Due to this complexity, general closed-form solutions are rarely attainable \cite{kerschen2006past}. Perturbation techniques, such as the method of multiple scales \cite{nayfeh2005resolving} and the method of averaging \cite{sanders2007averaging,mahmoud1993generalized}, serve as fundamental tools for elucidating nonlinear mechanisms and conducting qualitative analysis of low-dimensional, weakly nonlinear systems. However, these classical perturbation techniques struggle with the high dimensionality and strong nonlinearities characteristic of industrial engineering models. 
Consequently, to achieve accurate quantitative predictions for such complex systems, numerical methods become indispensable. The most straightforward approach is direct time integration (e.g., Newmark-$\beta$) \cite{newmark1959method}. While robust, this method treats the response as an initial value problem, necessitating computationally expensive time-marching to simulate the decay of transients before reaching the steady state. To bypass this inefficient transient phase, the shooting method \cite{goodman1956numerical} reformulates the periodic response as a two-point boundary value problem, accelerating convergence by iteratively correcting initial states to enforce periodicity directly. To eliminate the burden of time integration entirely, frequency-domain approaches have gained prominence in engineering practice. In particular, the Harmonic Balance Method (HBM) \cite{krack2019harmonic} avoids time integration altogether by transforming the nonlinear differential equations into algebraic systems via Fourier expansion, thereby offering superior efficiency for high-dimensional models. 

Despite the efficiency of HBM, characterizing the system's global behavior typically requires sweeping the excitation frequency and amplitude over a wide range. Because the response of a nonlinear system is energy-dependent \cite{rosenbergNormalModesNonlinear1962}, using these conventional methods necessitates exhaustive, repetitive computations to map the response landscape under varying excitation levels. This process becomes computationally prohibitive for large-scale finite element models. 
In practical engineering, robust design and vibration control often focus critically on the response characteristics at resonance—typically the most dangerous operating scenario—rather than the entire frequency spectrum. Therefore, developing methods to directly compute the resonance backbone curves (the locus of resonance peaks) is of paramount importance for dynamic performance evaluation and effective vibration mitigation.

The resonance backbone curve is intrinsically linked to the concept of Nonlinear Normal Modes (NNMs). Fundamentally, both representations characterize the amplitude-dependent frequency of periodic oscillations. Specifically, for weakly damped systems, Cenedese and Haller \cite{cenedeseHowConservativeBackbone2020} rigorously demonstrated that the resonance backbone of the forced response asymptotically tracks the NNM backbone of the underlying conservative system. The definition of NNMs has evolved significantly, originating from Rosenberg's \cite{rosenbergNormalModesNonlinear1962} concept of ``vibrations in unison", which restricts all degrees of freedom to reach equilibrium simultaneously. To overcome this limitation, Shaw and Pierre \cite{shaw1991non,shaw1994invariant} redefined NNMs as invariant manifolds in phase space, a geometric generalization that accommodates the phase differences and damping effects inherent in non-conservative systems \cite{jiang2004large,renson2014effective}. More recently, Haller and Ponsioen \cite{haller2016nonlinear} introduced Spectral Submanifolds (SSMs) to address the existence and uniqueness issues of invariant manifold. Distinct from these manifold-based formulations, Laxalde and Thouverez \cite{laxalde2009complex} defined NNMs as pseudo-periodic motions using a complex modal framework, and Krack \cite{krack2015nonlinear} defined NNMs based on periodic orbits extended to non-conservative systems via negative modal damping. 

Building on the fundamental correspondence between forced resonance and NNMs, researchers typically compute resonance backbones via two primary NNM-based pathways. The first strategy involves constructing reduced-order models (ROMs), where techniques like SSMs \cite{jainHowComputeInvariant2022,ponsioen2020model} or complex nonlinear modes \cite{laxalde2009complex,joannin2017nonlinear} project the dynamics onto a low-dimensional manifold, upon which the forced response is computed efficiently \cite{breunung2018explicit,cenedese2022data,li2023model,malyshev2025nonlinear,farvandi2025nonlinear}. Alternatively, the second pathway relies on energy or phase lag criteria. Approaches such as the energy balance method \cite{hill2015interpreting,hill2016analytical,sun2021extended} or force-appropriation \cite{peeters2011modal,renson2018force,pacini2025understanding} predict the intersection of the forced response curve with the conservative NNM backbone directly, bypassing full response surface generation.
However, applying these NNM-based pathways to complex engineering structures presents distinct challenges. Manifold-based methods are predominantly tailored for geometric nonlinearities; their application to non-smooth systems (e.g., contact and friction) is inherently limited by the analyticity requirements of polynomial expansions, often necessitating data-driven reduced-order models to bypass these theoretical bottlenecks. Conversely, methods based on pseudo-periodic motions typically enforce a priori phase-lag conditions. This assumption restricts their prediction accuracy in scenarios governed by strong nonlinear damping \cite{cenedeseHowConservativeBackbone2020}, where the actual phase evolution deviates from the prescribed modal form.

Beyond NNM-based constructions, resonance backbones can be determined through optimization formulations \cite{kernevez1987optimization}. Geometrically, under single-harmonic forcing, resonance points correspond to stationary points of a response functional (e.g., maximum amplitude or energy) constrained by the governing equations. To systematically trace the locus of these stationary points across varying energy levels, parameter continuation techniques \cite{li2018staged,dankowicz2013recipes} are required. Following this perspective, recent studies have begun to combine optimization with model reduction. Li et al. \cite{liFastComputationCharacterization2024a} integrated adjoint-based parameter continuation with SSM-based models to extract backbones. 

While promising, most existing optimization-based methods fail to fully exploit higher-order information to accelerate convergence, primarily because deriving the exact analytical second-order gradient expressions (the Hessian) remains mathematically formidable for complex nonlinear systems. Nevertheless, for robust convergence on high-fidelity models, the precise analytical Hessian is indispensable ---a capability that remains largely absent in standard harmonic balance frameworks.

To address this challenge, we employ the HBM, which remains the dominant numerical framework for periodic responses. A pivotal advancement in this domain was the Alternating Frequency/Time (AFT) method, originally proposed by Cameron and Griffin \cite{cameron1989alternating}. By utilizing the Fast Fourier Transform (FFT) to alternate between domains, the AFT framework evaluates nonlinear forces directly in the time domain, thereby circumventing the computationally prohibitive convolution of Fourier coefficients required by pure frequency-domain approaches \cite{nacivetDynamicLagrangianFrequency2003,lee2025nonlinear}. This architecture has proven effective for modeling non-smooth discontinuities common in engineering. This capability makes it indispensable for analyzing the complex dynamics of bladed disks with dry friction interfaces \cite{krack2012robust,siewert2005vibrational,petrovadvanced,petrov2004method}. 

Crucially, the AFT framework facilitates the derivation of the analytical Jacobian matrix, particularly for friction contact elements \cite{petrovAnalyticalFormulationFriction2002,afzalAnalyticalCalculationJacobian2016b}, which accelerates the convergence of Newton-Raphson iterations. This capability is routinely integrated with Component Mode Synthesis (CMS) \cite{craig1968coupling} and interface reduction techniques—such as the Dual Craig-Bampton method \cite{rixen2004dual}—to handle large-scale finite element models with localized nonlinearities. However, standard formulations stop at first-order derivatives.

Building upon this foundation, this paper extends the conventional AFT method to compute the frequency-domain analytical Hessian tensor. By integrating this second-order information, we establish a direct optimization framework where the HBM algebraic equations serve as the governing constraints. Crucially, this approach significantly enhances numerical robustness, rendering the analysis of complex models computationally feasible.

To demonstrate the generality of the proposed framework, we apply it to three distinct test cases: a two-degree-of-freedom oscillator with cubic stiffness \cite{krack2019harmonic}; a cantilever beam with cubic stiffness or hyperbolic tangent (tanh) friction nonlinearities, which serves as a widely utilized benchmark for nonlinear solvers \cite{thouverez2003presentation,arslan2011parametric}; and a finite element model of a compressor blisk equipped with ring dampers, where the friction interface is approximated via a tanh regularization function. Given the computational cost of high-fidelity finite element models, a preliminary model order reduction is essential. Drawing upon our previous work \cite{wen2025complex}, we employ a reduction strategy based on a complex-valued reduction basis, which has been shown to be effective for reducing the degrees of freedom in nonlinear bladed disk systems.

The remainder of this paper is organized as follows. Section \ref{sec:Optimization-Based method} presents the theoretical framework for computing backbone curves and the associated response characteristics, establishing the equations of motion for forced vibration and the constrained optimization formulation based on Lagrange multipliers. 
Section \ref{sec:Hessian Tensor} details the numerical implementation, focusing on the derivation of the analytical Hessian tensor required by the optimization solver. Key steps for efficient assembly are summarized, including the evaluation of second-order element derivatives via an extended AFT workflow and the subsequent element-to-global coordinate transformations. 
Section \ref{sec:Numerical Results} validates the proposed approach through three case studies of increasing complexity: a two-DOF system with cubic stiffness, a cantilever beam with cubic stiffness or friction nonlinearities, and a large-scale blisk finite element model with localized friction nonlinearities approximated by the tanh regularization function. These examples illustrate the method's applicability to both displacement- and velocity-dependent nonlinearities across models ranging from low-order oscillators to high-fidelity FE systems. 
Finally, Section \ref{Conclusion} concludes the paper by summarizing the main findings, discussing computational efficiency and scalability, and outlining limitations and potential extensions.    

\section{Optimization-Based Computation of Forced Backbone Curves via AFT} \label{sec:Optimization-Based method}
This section presents the theoretical framework for computing resonance backbone curves. The problem is formulated as a constrained optimization task where the frequency-domain algebraic equations, derived via the HBM, serve as the governing constraints. Section  \ref{ssec:model and equation} outlines the discretization process, transforming time-domain differential equations into algebraic systems through Fourier-Galerkin projection. Section \ref{Constrained Optimization} details the optimization formulation. An amplitude response functional is formulated, and the Lagrangian is constructed by appending the HBM governing equations using Lagrange multipliers. The first-order optimality conditions are then derived by computing the gradients. Section \ref{ssec:Numerical Continuation}  describes the numerical solution strategy, specifically how parameter continuation techniques are employed to solve the resulting system of equations and trace the resonance backbone curve across the response surface. 
\subsection{Governing Equations of Forced Vibration} \label{ssec:model and equation}
The general form of the time-domain governing equations of motion for a discretized nonlinear system is given by:
\begin{equation}
\boldsymbol{M} \ddot{\boldsymbol{q}}(t)+\boldsymbol{C} \dot{\boldsymbol{q}}(t)+\boldsymbol{K} \boldsymbol{q}(t)+\boldsymbol{f}_\mathrm{n l}(\dot{\boldsymbol{q}}, \boldsymbol{q})=\boldsymbol{f}_\mathrm{e x}(t),
\end{equation}
where $\boldsymbol{M}$, $\boldsymbol{C}$, and $\boldsymbol{K}$ are the mass, damping, and  stiffness matrices, respectively. $\boldsymbol{f}_\mathrm{n l}$, $\boldsymbol{f}_\mathrm{e x}$, and $\boldsymbol{q}$ represent the time-domain nonlinear force vector, external excitation vector, and structural displacement vector, respectively. We seek periodic responses and approximate the relevant time-dependent quantities by truncated Fourier series. Specifically, we expand:
\begin{equation}
\begin{aligned}
\boldsymbol{q}(t) & =\boldsymbol{Q}^{0}+\sum_{n=1}^{N_{\mathrm{h}}} (\boldsymbol{Q}^{\mathrm{c}, n} \cos (n \omega t)+\boldsymbol{Q}^{\mathrm{s}, n} \sin (n \omega t)), \\
\boldsymbol{f}_\mathrm{n l}(t) & =\boldsymbol{F}_{n l}^{0}+\sum_{n=1}^{N_{\mathrm{h}}} (\boldsymbol{F}_\mathrm{n l}^{\mathrm{c}, n} \cos (n \omega t)+\boldsymbol{F}_\mathrm{n l}^{\mathrm{s}, n} \sin (n \omega t)), \\
\boldsymbol{f}_\mathrm{e x}(t) & =\boldsymbol{F}_{n l}^{0}+\sum_{n=1}^{N_{\mathrm{h}}} (\boldsymbol{F}_\mathrm{e x}^{\mathrm{c}, n} \cos (n \omega t)+\boldsymbol{F}_\mathrm{e x}^{\mathrm{s}, n} \sin (n \omega t)).
\end{aligned}
\end{equation}

Here, ${N_{\mathrm{h}}}$ denotes the number of retained harmonics. The superscripts $\mathrm{c}$ and $\mathrm{s}$ represent the cosine and sine components, respectively, and the superscript $n$ indicates the $n$-th harmonic. By applying the Fourier-Galerkin projection, the time-domain differential equations are transformed into a system of frequency-domain algebraic equations in terms of the Fourier coefficients:
\begin{equation}
\boldsymbol{D}(\omega) \boldsymbol{Q}+\boldsymbol{F}_\mathrm{nl}(\boldsymbol{Q}, \omega)-\boldsymbol{F}_\mathrm{ex}=\boldsymbol{0},
\end{equation}
where $\boldsymbol{Q}$, $\boldsymbol{F}_\mathrm{nl}$, and $\boldsymbol{F}_\mathrm{ex}$ denote the vectors of Fourier coefficients for displacement, nonlinear forces, and external excitation, respectively, and $\boldsymbol{D}(\omega)$ is the block-diagonal dynamic stiffness matrix. The detailed matrix structures and vector arrangements are provided in \ref{Appendix A}.

The solutions of Equation (3) constitute the forced vibration responses computed using the HBM. Subsequently, we treat this solution set as the dynamic constraint manifold to determine the locus of response extrema---namely, the resonance backbone curve.
\subsection{Constrained Optimization Formulation} \label{Constrained Optimization}
In this framework, the determination of resonance backbones is formulated as a constrained optimization problem. The primary objective is to maximize a scalar measure of the response amplitude (the norm of the Fourier coefficients) subject to the governing equations derived in Equation (3). By constructing the Lagrangian, the problem is transformed into finding the stationarity conditions that define the locus of peak response points with respect to the excitation frequency. Under standard regularity assumptions, this locus rigorously corresponds to the resonance backbone curve. Drawing upon the methodology in \cite{li2020optimization}, we integrate this formulation with parameter continuation to allow for the robust tracking of these extremal curves and their associated mode shapes.

As outlined, the computation of resonance backbone curves and corresponding modes is cast as a response-extremization problem. Hence, we start our derivation with the general theory of constrained optimization via Lagrange multipliers:
\begin{equation}
\Phi(\boldsymbol{u}, \boldsymbol{\mu})=0.
\end{equation}

Let $\boldsymbol{u}$ and $\boldsymbol{\mu}$ denote the state and parameter vectors, respectively. In the context of computing resonance backbones, the parameter vector $\boldsymbol{\mu}$ includes/comprises both the excitation frequency and the forcing amplitude. Consequently, Equation (3) implicitly defines a response surface relating the system response to these control parameters. The specific formulation is given by:
\begin{equation}
\begin{aligned}
 & \boldsymbol{R}(\boldsymbol{Q}, \omega, \alpha)=0, \\
\boldsymbol{R}(\boldsymbol{Q}, \omega, \alpha)= & \boldsymbol{D}(\omega) \cdot \boldsymbol{Q}+\boldsymbol{F}_{n l}(\boldsymbol{Q}, \omega)-\alpha \boldsymbol{F}_{e x}. 
\end{aligned}
\end{equation}

Here, $\omega$ is the excitation frequency, $\alpha$ denotes the forcing level, and $\boldsymbol{F}_\mathrm{ex} $ is the normalized excitation vector. When using the continuation method to trace the response surface, parameter bounds are defined as follows:
\begin{equation}
\begin{aligned}
& \alpha \in\left[\alpha_{\mathrm{min} }, \alpha_{\mathrm{max} }\right], \\
& \omega \in\left[\omega_{\mathrm{min} }, \omega_{\mathrm{max} }\right].
\end{aligned}
\end{equation}

To extract the extremum curve, we first define the response metric. In this work, the response is quantified as the Euclidean norm of the Fourier coefficients for the monitored $k$-th degree of freedom.
\begin{equation}
E=\frac{\sqrt{2}}{2} \sqrt{{{Q}_k^{0}}^{2}+\sum_{n=1}^{N_{\mathrm{h}}} {{Q}_k^{\mathrm{c}, n}}^{2}+{{Q}_k^{\mathrm{s}, n}}^{2}}.
\end{equation}

Geometrically, the ridge of the response surface coincides with the resonance backbone. Consequently, determining the backbone and corresponding modes is formulated as a constrained optimization problem seeking surface extrema. To solve this via the method of Lagrange multipliers, we introduce the Lagrangian function as:
\begin{equation}
L=E+\eta_E\left(E-E_0\right)+\eta_\alpha\left(\alpha-\alpha_0\right)+\eta_\omega\left(\omega-\omega_0\right)+\boldsymbol{\lambda}^{\top} \boldsymbol{R},
\end{equation}
where $E_0$, $\alpha_0$, and $\omega_0$ are the introduced auxiliary variables, and $\eta_E$, $\eta_\alpha$, $\eta_\omega$, and $\boldsymbol{\lambda}$ represent the Lagrange multipliers. By setting the partial derivatives of the Lagrangian function with respect to the variables $\mathbf{Q}$, $\alpha$, $\omega$, and the corresponding Lagrange multipliers to zero, we obtain the following equations:
\begin{equation}
\left\{\begin{array}{l}
E-E_0=0, \alpha-\alpha_0=0, \omega-\omega_0=0, \boldsymbol{R}=\mathbf{0}, \\
\frac{\partial E}{\partial \boldsymbol{Q}} \eta_E+\left(\frac{\partial \boldsymbol{R}}{\partial \boldsymbol{Q}}\right)^{\top} \boldsymbol{\lambda}=\mathbf{0}, \\
\frac{\partial E}{\partial \omega} \eta_E+\eta_\omega+\left(\frac{\partial \boldsymbol{R}}{\partial \omega}\right)^{\top} \boldsymbol{\lambda}=0, \\
\frac{\partial E}{\partial \alpha} \eta_E+\eta_\alpha+\left(\frac{\partial \boldsymbol{R}}{\partial \alpha}\right)^{\top} \boldsymbol{\lambda}=0.
\end{array}\right.
\end{equation}
\subsection{Numerical Continuation Strategy} \label{ssec:Numerical Continuation}
The solution of the resulting augmented system is obtained via parameter continuation techniques. The specific numerical continuation strategy is detailed as follows \cite{liFastComputationCharacterization2024a} :
\begin{itemize}
\item[1)] Computation of the initial forced response: With the excitation parameter fixed at $\alpha_\mathrm{min}$, the forced response curve is computed using $\omega_0$ as the continuation parameter within the range $[\omega_\mathrm{min}, \omega_\mathrm{max}]$. In this stage, all Lagrange multipliers are zero, and the adjoint equations are satisfied trivially. The COCO package \cite{dankowicz2013recipes} is employed to automatically detect the fold point of the forced response curve, which corresponds to the maximum amplitude response.
\item[2)] Initialization of the optimization branch: Subsequently, a branch switch is performed at the detected fold point. The auxiliary variable $\eta_E$ is selected as the continuation parameter over the interval $[0, 1]$. The continuation proceeds until $\eta_E = 1$. During this step, only the Lagrange multipliers evolve linearly, while the state variables, excitation frequency, and excitation amplitude remain constant.
\item[3)] Tracing the resonance backbone: Finally, the forced vibration resonance backbone curve is computed by using $\alpha$ as the continuation parameter within the range $[\alpha_{min}, \alpha_{max}]$.
\end{itemize}

The Newton-Raphson method is employed to solve Equation (9). This requires evaluating its Jacobian matrix with respect to the state variables and parameters. Since the equation explicitly involves the first-order partial derivatives of the governing residual $\boldsymbol{R}$ (with respect to state variables and excitation frequency), the construction of the Jacobian necessitates further differentiation, yielding the following second-order partial derivatives:
\begin{equation}
\frac{\partial^2 E}{\partial \boldsymbol{Q}^2}, \frac{\partial^2 \boldsymbol{R}}{\partial \boldsymbol{Q}^2}, \frac{\partial^2 \boldsymbol{R}}{\partial \omega \partial \boldsymbol{Q}}, \frac{\partial^2 \boldsymbol{R}}{\partial \boldsymbol{Q} \partial \omega},\frac{\partial^2 \boldsymbol{R}}{\partial {\omega}^2}.
\end{equation}

Specifically, these terms are derived by assembling the second-order derivatives of the nonlinear element forces. To facilitate this, the classical AFT method must be extended beyond its standard capability of evaluating nonlinear forces and stiffness matrices \cite{cameron1989alternating}. The proposed extension enables the efficient evaluation of element Hessian tensors, which are then assembled into the global Jacobian.
\section{Computation of the Hessian Tensor via the Extended AFT Method} \label{sec:Hessian Tensor}
To facilitate the efficient computation of the analytical Hessian, we adopt an element-based approach. Whether the nonlinearity is global or local, it can be decomposed into discrete nonlinear elements. Consequently, we first derive the second-order derivatives for a generic single-degree-of-freedom element using the extended AFT method. For large-scale systems, these elemental contributions are subsequently assembled into the global Hessian tensor via standard coordinate transformations.

This section details the computation of the Hessian matrix for nonlinear elements via the extended AFT method. Section \ref{sec:Classical AFT} briefly reviews the fundamental principles of the classical AFT framework and presents the component formulas for the stiffness matrix of the nonlinear element. Section \ref{sec:Extended AFT and Hessian} extends this classical framework to derive the components of the elemental Hessian Tensor (second-order derivatives). Finally, Section   \ref{sec:Coordinate Transformation} utilizes tensor coordinate transformations to map the locally computed stiffness matrices and Hessian Tensor to the global coordinate system. These assembled matrices are essential for constructing the Jacobian of the governing system Equation (9), thereby accelerating the computation of resonance backbone curves and their associated mode shapes.

\subsection{The Classical AFT Method} \label{sec:Classical AFT}
The classical AFT method is an efficient technique for evaluating nonlinear forces and assembling element stiffness contributions. Its key idea is to circumvent deriving explicit frequency-domain expressions by computing the nonlinear terms in the time domain first. These time-domain quantities are then mapped to the frequency domain—together with the associated stiffness contributions—using a Discrete Fourier Transform (DFT). This enables the rapid construction of the Jacobian for the frequency-domain response Equation (3). In the following, the expressions for the frequency-domain nonlinear terms and the nonlinear element stiffness matrices are presented via the classical AFT method.

In the time domain, the general expression for the nonlinear element force is:
\begin{equation}
f_\mathrm{nl}=f_\mathrm{nl}(x, \dot{x}).
\end{equation}

By uniformly sampling the system variables over a single period, the displacement samples can be expressed as:
\begin{equation}
\begin{array}{ll}
&x[l]  =x(\Delta T \cdot l) \quad l=0, \ldots, N-1  \\
&\Delta T  =\frac{T}{N}, \quad T=\frac{2 \pi}{\omega},
\end{array}
\end{equation}
where $N$ is the number of sampling points, $l$ denotes the index of the sampling sequence, $\Delta T$ is the sampling interval, and $T$ denotes the vibration period. The nonlinear force samples can be regarded as functions of the displacement and velocity samples:
\begin{equation}
f_\mathrm{nl}[l]=f_\mathrm{nl}(x[1], x[2], \cdots, x[l], \cdots, x[N], \dot{x}[1], \dot{x}[2], \cdots, \dot{x}[l], \cdots, \dot{x}[N]).
\end{equation}
For non-time-delayed nonlinear systems, the expression for the nonlinear force can be directly written as follows:
\begin{equation}
f_\mathrm{nl}[l]=f_\mathrm{nl}(x[l], \dot{x}[l]) \quad l=0,1, \cdots, N-1.
\end{equation}

The DFT is applied to the time-domain samples of the displacement and nonlinear force:
\begin{equation}
\begin{aligned}
& X[n]=\sum_{l=0}^{N-1} x[l] e^{\left(\frac{-i 2 \pi}{N} n l\right)} \quad n=0,1, \cdots, N-1 .\\
& {F_\mathrm{n l}}[n]=\sum_{l=0}^{N-1} f_\mathrm{n l}[l] e^{\left(\frac{-i 2 \pi}{N} n l\right)} \quad n=0,1, \cdots, N-1.
\end{aligned}
\end{equation}

The AFT method uses the DFT of periodic time samples to obtain frequency-domain quantities. We state the correspondence between DFT coefficients and Fourier series coefficients. As this result is fundamental in signal processing, a detailed derivation is provided in \ref{Appendix B}.
\begin{equation}
\begin{aligned}
& X^0= \Re\left(\frac{X[0]}{N}\right),  \\
& X^{\mathrm{c}, n} = 2\Re{\left(\frac{X[n]}{N}\right)}, \\
& X^{\mathrm{s}, n} = 2\Im{\left(\frac{X[n]}{N}\right)}.
\end{aligned}
\end{equation}
Here $X[n]$ denotes the $n$-th frequency-domain sample of the DFT sequence for the SDOF displacement $x$. Similarly, for the nonlinear term ${f}_\mathrm{n l}$:
\begin{equation}
\begin{aligned}
& {F_\mathrm{n l}}^0= \Re{\left(\frac{{F_\mathrm{n l}}[0]}{N}\right)},  \\
& {F_\mathrm{n l}}^{\mathrm{c}, n} = 2\Re{\left(\frac{{F_\mathrm{n l}}[n]}{N}\right)}, \\
& {F_\mathrm{n l}}^{\mathrm{s}, n} = 2\Im{\left(\frac{{F_\mathrm{n l}}[n]}{N}\right)}.
\end{aligned}
\end{equation}

We expand the $x$ and ${f}_\mathrm{n l}$ into Fourier series:
\begin{equation}
\begin{gathered}
x(t)=X^0+\sum_{k=1}^{\infty}\left[X^{\mathrm{c}, k} \cos (k \omega t)+X^{\mathrm{s}, k} \sin (k \omega t)\right], \\
f_\mathrm{n l}(t)={F_\mathrm{n l}}^0+\sum_{k=1}^{\infty}\left[{F_\mathrm{n l}}^{\mathrm{c}, k} \cos (k \omega t)+{F_\mathrm{n l}}^{\mathrm{s}, k} \sin (k \omega t)\right].
\end{gathered}
\end{equation}

Similarly, the Fourier expansion of $\dot{x}$ and $ \ddot{x}$ are given by:
\begin{equation}
\begin{aligned}
& \dot{x}(t)=\sum_{k=1}^{\infty} k \omega\left[-X^{\mathrm{c}, k} \sin (k \omega t)+X^{\mathrm{s}, k} \cos (k \omega t)\right], \\
& \ddot{x}(t)=\sum_{k=1}^{\infty}(k \omega)^2\left[-X^{\mathrm{c}, k} \cos (k \omega t)-X^{\mathrm{s}, k} \sin (k \omega t)\right].
\end{aligned}
\end{equation}
The displacement and nonlinear force coefficients are expressed as column vectors:
\begin{equation}
\begin{aligned}
& \boldsymbol{X}=\left[X^0 ; X^{\mathrm{c}, 1} ; X^{\mathrm{s}, 1} ; \cdots ; X^{\mathrm{c}, N_\mathrm{h}} ; X^{\mathrm{s}, N_\mathrm{h}}\right], \\
& \boldsymbol{F_\mathrm{n l}}=\left[F_\mathrm{n l}{ }^0 ; F_\mathrm{n l}{ }^{\mathrm{c}, 1} ; F_\mathrm{n l}{ }^{\mathrm{s}, 1} ; \cdots ; F_{n l}{ }^{\mathrm{c}, N_\mathrm{h}} ; F_\mathrm{nl}{ }^{\mathrm{s}, N_\mathrm{h}}\right].
\end{aligned}
\end{equation}

Then, in the frequency domain, the nonlinear element force vector and stiffness matrix can be expressed as follows:
\begin{equation}
\begin{gathered}
\boldsymbol{F_\mathrm{n l}}(\boldsymbol{X}, \omega), \\
{\left[\frac{\partial \boldsymbol{F_\mathrm{n l}}(\boldsymbol{X}, \omega)}{\partial \boldsymbol{X}}, \frac{\partial \boldsymbol{F_\mathrm{n l}}(\boldsymbol{X}, \omega)}{\partial \omega}\right]}.
\end{gathered}
\end{equation}

We obtain the derivatives $\frac{\partial \boldsymbol{F_\mathrm{n l}}}{\partial \boldsymbol{X}}$ using the time-domain partial derivatives of the nonlinear force with respect to displacement and velocity. Applying the chain rule, the frequency-domain element stiffness matrix is given by:
\begin{equation}
\frac{\partial \boldsymbol{F_\mathrm{n l}}}{\partial \boldsymbol{X}}=\frac{\partial \boldsymbol{F_\mathrm{n l}}}{\partial f_\mathrm{n l}} \frac{\partial f_\mathrm{n l}}{\partial x} \frac{\partial x}{\partial \boldsymbol{X}}+\frac{\partial \boldsymbol{F_\mathrm{n l}}}{\partial f_\mathrm{n l}} \frac{\partial f_\mathrm{n l}}{\partial \dot{x}} \frac{\partial \dot{x}}{\partial \boldsymbol{X}}.
\end{equation}

By substituting the established mapping between DFT sequences and Fourier series coefficients into the chain rule formulation, we derive the explicit expressions for the components of the element stiffness matrix:
\begin{equation}
\begin{array}{l}
\frac{\partial F_\mathrm{n l}^{\mathrm{c}, k}}{\partial X^{\mathrm{c}, n}}=\frac{2}{N} \sum_{l=0}^{N-1} \sum_{m=0}^{N-1}\left\{\begin{array}{l}
\frac{\partial f_\mathrm{n l}(l)}{\partial x(m)} \cos \left(\frac{2 \pi k l}{N}\right) \cos \left(\frac{2 \pi n m}{N}\right) \\
-(n \omega) \frac{\partial f_\mathrm{n l}(l)}{\partial \dot{x}(m)} \cos \left(\frac{2 \pi k l}{N}\right) \sin \left(\frac{2 \pi n m}{N}\right)
\end{array}\right\} \\
\frac{\partial F_\mathrm{n l}^{\mathrm{c}, k}}{\partial X^{\mathrm{s}, n}}=\frac{2}{N} \sum_{l=0}^{N-1} \sum_{m=0}^{N-1}\left\{\begin{array}{l}
\frac{\partial f_\mathrm{n l}(l)}{\partial x(m)} \cos \left(\frac{2 \pi k l}{N}\right) \sin \left(\frac{2 \pi n m}{N}\right) \\
+(n \omega) \frac{\partial f_\mathrm{n l}(l)}{\partial \dot{x}(m)} \cos \left(\frac{2 \pi k l}{N}\right) \cos \left(\frac{2 \pi n m}{N}\right)
\end{array}\right\} \quad k, n \neq 0.\\
\frac{\partial F_\mathrm{n l}^{\mathrm{s}, k}}{\partial X^{\mathrm{c}, n}}=\frac{2}{N} \sum_{l=0}^{N-1} \sum_{m=0}^{N-1}\left\{\begin{array}{l}
\frac{\partial f_\mathrm{n l}(l)}{\partial x(m)} \sin \left(\frac{2 \pi k l}{N}\right) \cos \left(\frac{2 \pi n m}{N}\right) \\
-(n \omega) \frac{\partial f_\mathrm{n l}(l)}{\partial \dot{x}(m)} \sin \left(\frac{2 \pi k l}{N}\right) \sin \left(\frac{2 \pi n m}{N}\right)
\end{array}\right\} \\
\frac{\partial F_\mathrm{n l}^{\mathrm{s}, k}}{\partial X^{\mathrm{s}, n}}=\frac{2}{N} \sum_{l=0}^{N-1} \sum_{m=0}^{N-1}\left\{\begin{array}{l}
\frac{\partial f_{n l}(l)}{\partial x(m)} \sin \left(\frac{2 \pi k l}{N}\right) \sin \left(\frac{2 \pi n m}{N}\right) \\
+(n \omega) \frac{\partial f_{n l}(l)}{\partial \dot{x}(m)} \sin \left(\frac{2 \pi k l}{N}\right) \cos \left(\frac{2 \pi n m}{N}\right)
\end{array}\right\}
\end{array}
\end{equation}

In the case that $k$ or $n$ is zero, the following holds:
\begin{equation}
\begin{array}{l}
\frac{\partial F_\mathrm{n l}^0}{\partial X^0}=\frac{1}{N} \sum_{l=0}^{N-1} \sum_{m=0}^{N-1} \frac{\partial f_\mathrm{n l}(l)}{\partial x(m)} \\
\frac{\partial F_\mathrm{n l}^0}{\partial X^{\mathrm{c}, n}}=\frac{1}{N} \sum_{l=0}^{N-1} \sum_{m=0}^{N-1}\left\{\frac{\partial f_\mathrm{n l}(l)}{\partial x(m)} \cos \left(\frac{2 \pi n m}{N}\right)-(n \omega) \frac{\partial f_\mathrm{n l}(l)}{\partial \dot{x}(m)} \sin \left(\frac{2 \pi n m}{N}\right)\right\} \\
\frac{\partial F_\mathrm{n l}^0}{\partial X^{\mathrm{s}, n}}=\frac{1}{N} \sum_{l=0}^{N-1} \sum_{m=0}^{N-1}\left\{\frac{\partial f_\mathrm{n l}(l)}{\partial x(m)} \sin \left(\frac{2 \pi n m}{N}\right)+(n \omega) \frac{\partial f_\mathrm{n l}(l)}{\partial \dot{x}(m)} \cos \left(\frac{2 \pi n m}{N}\right)\right\} \\
\frac{\partial F_\mathrm{n l}^{\mathrm{c}, k}}{\partial X^0}=\frac{2}{N} \sum_{l=0}^{N-1} \sum_{m=0}^{N-1} \frac{\partial f_\mathrm{n l}(l)}{\partial x(m)} \cos \left(\frac{2 \pi k l}{N}\right) \\
\frac{\partial F_\mathrm{n l}^{\mathrm{s}, k}}{\partial X^0}=\frac{2}{N} \sum_{l=0}^{N-1} \sum_{m=0}^{N-1} \frac{\partial f_\mathrm{n l}(l)}{\partial x(m)} \sin \left(\frac{2 \pi k l}{N}\right)
\end{array} .
\end{equation}

For non-time-delayed systems, the $l$-th sample of the nonlinear force depends exclusively on the $l$-th sample of the displacement or velocity. Consequently, the aforementioned equation can be simplified to the following form:
\begin{equation}
\begin{array}{l}
\frac{\partial F_\mathrm{n l}^{\mathrm{c}, k}}{\partial X^{\mathrm{c}, n}}=\frac{2}{N} \sum_{l=0}^{N-1}\left\{\begin{array}{l}
\frac{\partial f_\mathrm{n l}}{\partial x}(l) \cos \left(\frac{2 \pi k l}{N}\right) \cos \left(\frac{2 \pi n l}{N}\right) \\
-(n \omega) \frac{\partial f_\mathrm{n l}}{\partial \dot{x}}(l) \cos \left(\frac{2 \pi k l}{N}\right) \sin \left(\frac{2 \pi n l}{N}\right)
\end{array}\right\} \\
\frac{\partial F_\mathrm{n l}^{\mathrm{c}, k}}{\partial X^{\mathrm{s}, n}}=\frac{2}{N} \sum_{l=0}^{N-1}\left\{\begin{array}{l}
\frac{\partial f_\mathrm{n l}}{\partial x}(l) \cos \left(\frac{2 \pi k l}{N}\right) \sin \left(\frac{2 \pi n l}{N}\right) \\
+(n \omega) \frac{\partial f_\mathrm{n l}}{\partial \dot{x}}(l) \cos \left(\frac{2 \pi k l}{N}\right) \cos \left(\frac{2 \pi n l}{N}\right)
\end{array}\right\} \quad k, n \neq 0 \\
\frac{\partial F_\mathrm{n l}^{\mathrm{s}, k}}{\partial X^{\mathrm{c}, n}}=\frac{2}{N} \sum_{l=0}^{N-1}\left\{\begin{array}{l}
\frac{\partial f_\mathrm{n l}}{\partial x}(l) \sin \left(\frac{2 \pi k l}{N}\right) \cos \left(\frac{2 \pi n l}{N}\right) \\
-(n \omega) \frac{\partial f_\mathrm{n l}}{\partial \dot{x}}(l) \sin \left(\frac{2 \pi k l}{N}\right) \sin \left(\frac{2 \pi n l}{N}\right)
\end{array}\right\} \\
\frac{\partial F_\mathrm{n l}^{\mathrm{s}, k}}{\partial X^{\mathrm{s}, n}}=\frac{2}{N} \sum_{l=0}^{N-1}\left\{\begin{array}{l}
\frac{\partial f_\mathrm{n l}}{\partial x}(l) \sin \left(\frac{2 \pi k l}{N}\right) \sin \left(\frac{2 \pi n l}{N}\right) \\
+(n \omega) \frac{\partial f_\mathrm{n l}}{\partial \dot{x}}(l) \sin \left(\frac{2 \pi k l}{N}\right) \cos \left(\frac{2 \pi n l}{N}\right)
\end{array}\right\}
\end{array}
\end{equation}
and
\begin{equation}
\begin{array}{l}
\frac{\partial F_\mathrm{n l}^0}{\partial X^0}=\frac{1}{N} \sum_{l=0}^{N-1} \frac{\partial f_\mathrm{n l}}{\partial x}(l) \\
\frac{\partial F_\mathrm{n l}^0}{\partial X^{\mathrm{c}, n}}=\frac{1}{N} \sum_{l=0}^{N-1}\left\{\frac{\partial f_\mathrm{nl}}{\partial x}(l) \cos \left(\frac{2 \pi n l}{N}\right)-(n \omega) \frac{\partial f_\mathrm{nl}}{\partial \dot{x}}(l) \sin \left(\frac{2 \pi n l}{N}\right)\right\} \\
\frac{\partial F_\mathrm{n l}^0}{\partial X^{\mathrm{s}, n}}=\frac{1}{N} \sum_{l=0}^{N-1}\left\{\frac{\partial f_\mathrm{n l}}{\partial x}(l) \sin \left(\frac{2 \pi n l}{N}\right)+(n \omega) \frac{\partial f_\mathrm{n l}}{\partial \dot{x}}(l) \cos \left(\frac{2 \pi n l}{N}\right)\right\}. \\
\frac{\partial F_\mathrm{n l}^{\mathrm{c}, k}}{\partial X^0}=\frac{2}{N} \sum_{l=0}^{N-1} \frac{\partial f_\mathrm{n l}}{\partial x}(l) \cos \left(\frac{2 \pi k l}{N}\right) \\
\frac{\partial F_\mathrm{n l}^{\mathrm{s}, k}}{\partial X^0}=\frac{2}{N} \sum_{l=0}^{N-1} \frac{\partial f_\mathrm{n l}}{\partial x}(l) \sin \left(\frac{2 \pi k l}{N}\right)
\end{array}
\end{equation}

For velocity-dependent nonlinear forces, the partial derivative of the Fourier coefficients with respect to the excitation frequency is non-zero:
\begin{equation}
\begin{array}{l}
\frac{\partial F_\mathrm{n l}^0}{\partial \omega}=\frac{1}{N} \sum_{l=0}^{N-1} \sum_{n=1}^{N_h}\left\{\begin{array}{l}
-n X^{\mathrm{c}, n} \frac{\partial f_\mathrm{n l}}{\partial \dot{x}}(l) \sin \left(\frac{2 \pi n l}{N}\right) \\
+n X^{\mathrm{s}, n} \frac{\partial f_\mathrm{n l}}{\partial \dot{x}}(l) \cos \left(\frac{2 \pi n l}{N}\right)
\end{array}\right\} \\
\frac{\partial F_\mathrm{n l}^{\mathrm{c}, k}}{\partial \omega}=\frac{2}{N} \sum_{l=0}^{N-1} \sum_{n=1}^{N_h}\left\{\begin{array}{l}
-n X^{\mathrm{c}, n} \frac{\partial f_\mathrm{n l}}{\partial \dot{x}}(l) \cos \left(\frac{2 \pi k l}{N}\right) \sin \left(\frac{2 \pi n l}{N}\right) \\
+n X^{\mathrm{s}, n} \frac{\partial f_\mathrm{n l}}{\partial \dot{x}}(l) \cos \left(\frac{2 \pi k l}{N}\right) \cos \left(\frac{2 \pi n l}{N}\right)
\end{array}\right\} \quad k \neq 0 .\\
\frac{\partial F_\mathrm{n l}^{\mathrm{s}, k}}{\partial \omega}=\frac{2}{N} \sum_{l=0}^{N-1} \sum_{n=1}^{N_h}\left\{\begin{array}{l}
-n X^{\mathrm{c}, n} \frac{\partial f_\mathrm{n l}}{\partial \dot{x}}(l) \sin \left(\frac{2 \pi k l}{N}\right) \sin \left(\frac{2 \pi n l}{N}\right) \\
+n X^{\mathrm{s}, n} \frac{\partial f_\mathrm{n l}}{\partial \dot{x}}(l) \sin \left(\frac{2 \pi k l}{N}\right) \cos \left(\frac{2 \pi n l}{N}\right)
\end{array}\right\}
\end{array}
\end{equation}

The foregoing equations constitute the expressions for the nonlinear element stiffness matrix within the framework of the classical AFT method. We now proceed to extend the AFT method to evaluate the Hessian tensor of the nonlinear element.
\subsection{Extended AFT Method for Hessian Tensor Evaluation} \label{sec:Extended AFT and Hessian}
Assembling the Jacobian matrix for the governing equations of the resonance backbone curve given in Equation (9) necessitates the evaluation of the second-order partial derivatives of the nonlinear force Fourier coefficients with respect to the displacement coefficients. These global second-order terms are assembled from the elemental Hessian terms. 

The derivation of these high-order derivatives parallels that of the element stiffness matrix, exploiting the linearity of the Fourier transform. For the sake of brevity, the following discussion focuses on non-time-delayed systems; the extension to time-delayed cases is straightforward.

By applying the chain rule and omitting terms that vanish, the element second-order partial derivatives can be obtained via the following equation:
\begin{equation}
\begin{aligned}
\frac{\partial^2 \boldsymbol{F_\mathrm{n l}}}{\partial \boldsymbol{X}^2}= & \frac{\partial F_\mathrm{n l}}{\partial f_\mathrm{n l}} \frac{\partial^2 f_\mathrm{n l}}{\partial x^2}\left(\frac{\partial x}{\partial \boldsymbol{X}}\right)^2+\frac{\partial \boldsymbol{F_\mathrm{n l}}}{\partial f_\mathrm{n l}} \frac{\partial^2 f_\mathrm{n l}}{\partial \dot{x}^2}\left(\frac{\partial \dot{x}}{\partial \boldsymbol{X}}\right)^2 \\
& +\frac{\partial \boldsymbol{F_\mathrm{n l}}}{\partial f_\mathrm{n l}} \frac{\partial^2 f_\mathrm{n l}}{\partial x \partial \dot{x}} \frac{\partial x}{\partial \boldsymbol{X}} \frac{\partial \dot{x}}{\partial \boldsymbol{X}}+\frac{\partial \boldsymbol{F_\mathrm{n l}}}{\partial f_\mathrm{n l}} \frac{\partial^2 f_\mathrm{n l}}{\partial \dot{x} \partial x} \frac{\partial \dot{x}}{\partial \boldsymbol{X}} \frac{\partial x}{\partial \boldsymbol{X}}.
\end{aligned}
\end{equation}

According to Clairaut's theorem, mixed partial derivatives are equal within a region assuming $C^2$ continuity. Consequently, provided that the second-order partial derivatives of the nonlinear force function with respect to displacement and velocity are continuous, the last two terms on the right-hand side of the above equation are equal.

Analogously, the evaluation of second-order partial derivative terms relies on time-domain samples. Building upon the analytical expression in Equation (28), the components of these second-order derivatives are derived by further differentiating the previously obtained element stiffness matrix components (maintaining the assumption of a non-time-delayed system):
\begin{equation}
\frac{\partial^2 {F_\mathrm{n l}}^{\mathrm{c}, k}}{\partial X^{\mathrm{c}, n} \partial X^{\mathrm{c}, r}}=\frac{2}{N} \sum_{l=0}^{N-1}\left\{\begin{array}{l}
\frac{\partial^2 f_\mathrm{n l}}{\partial x^2}(l) \cos \left(\frac{2 \pi k l}{N}\right) \cos \left(\frac{2 \pi n l}{N}\right) \cos \left(\frac{2 \pi r l}{N}\right) \\
-(r \omega) \frac{\partial^2 f_\mathrm{n l}}{\partial x \partial \dot{x}}(l) \cos \left(\frac{2 \pi k l}{N}\right) \cos \left(\frac{2 \pi n l}{N}\right) \sin \left(\frac{2 \pi r l}{N}\right) \\
+(n \omega)(r \omega) \frac{\partial^2 f_\mathrm{n l}}{\partial \dot{x}^2}(l) \cos \left(\frac{2 \pi k l}{N}\right) \sin \left(\frac{2 \pi n l}{N}\right) \sin \left(\frac{2 \pi r l}{N}\right) \\
-(n \omega) \frac{\partial^2 f_\mathrm{n l}}{\partial \dot{x} \partial x}(l) \cos \left(\frac{2 \pi k l}{N}\right) \sin \left(\frac{2 \pi n l}{N}\right) \cos \left(\frac{2 \pi r l}{N}\right)
\end{array}\right\},
\end{equation}
\begin{equation}
\frac{\partial^2 {F_\mathrm{n l}}^{\mathrm{c}, k}}{\partial X^{\mathrm{c}, n} \partial X^{\mathrm{s}, r}}=\frac{2}{N} \sum_{l=0}^{N-1}\left\{\begin{array}{l}
\frac{\partial^2 f_\mathrm{n l}}{\partial x^2}(l) \cos \left(\frac{2 \pi k l}{N}\right) \cos \left(\frac{2 \pi n l}{N}\right) \sin \left(\frac{2 \pi r l}{N}\right) \\
+(r \omega) \frac{\partial^2 f_\mathrm{n l}}{\partial x \partial \dot{x}}(l) \cos \left(\frac{2 \pi k l}{N}\right) \cos \left(\frac{2 \pi n l}{N}\right) \cos \left(\frac{2 \pi r l}{N}\right) \\
-(n \omega)(r \omega) \frac{\partial^2 f_\mathrm{n l}}{\partial \dot{x}^2}(l) \cos \left(\frac{2 \pi k l}{N}\right) \sin \left(\frac{2 \pi n l}{N}\right) \cos \left(\frac{2 \pi r l}{N}\right) \\
-(n \omega) \frac{\partial^2 f_\mathrm{n l}}{\partial \dot{x} \partial x}(l) \cos \left(\frac{2 \pi k l}{N}\right) \sin \left(\frac{2 \pi n l}{N}\right) \sin \left(\frac{2 \pi r l}{N}\right)
\end{array}\right\},
\end{equation}
\begin{equation}
\frac{\partial^2 {F_\mathrm{n l}}^{\mathrm{c}, k}}{\partial X^{\mathrm{s}, n} \partial X^{\mathrm{c}, r}}=\frac{2}{N} \sum_{l=0}^{N-1}\left\{\begin{array}{l}
\frac{\partial^2 f_\mathrm{n l}}{\partial x^2}(l) \cos \left(\frac{2 \pi k l}{N}\right) \sin \left(\frac{2 \pi n l}{N}\right) \cos \left(\frac{2 \pi r l}{N}\right) \\
-(r \omega) \frac{\partial^2 f_\mathrm{n l}}{\partial x \partial \dot{x}}(l) \cos \left(\frac{2 \pi k l}{N}\right) \sin \left(\frac{2 \pi n l}{N}\right) \sin \left(\frac{2 \pi r l}{N}\right) \\
-(n \omega)(r \omega) \frac{\partial^2 f_\mathrm{n l}}{\partial \dot{x}^2}(l) \cos \left(\frac{2 \pi k l}{N}\right) \cos \left(\frac{2 \pi n l}{N}\right) \sin \left(\frac{2 \pi r l}{N}\right) \\
+(n \omega) \frac{\partial^2 f_\mathrm{n l}}{\partial \dot{x} \partial x}(l) \cos \left(\frac{2 \pi k l}{N}\right) \cos \left(\frac{2 \pi n l}{N}\right) \cos \left(\frac{2 \pi r l}{N}\right)
\end{array}\right\},
\end{equation}
\begin{equation}
\frac{\partial^2 {F_\mathrm{n l}}^{\mathrm{c}, k}}{\partial X^{\mathrm{s}, n} \partial X^{\mathrm{s}, r}}=\frac{2}{N} \sum_{l=0}^{N-1}\left\{\begin{array}{l}
\frac{\partial^2 f_\mathrm{n l}}{\partial x^2}(l) \cos \left(\frac{2 \pi k l}{N}\right) \sin \left(\frac{2 \pi n l}{N}\right) \sin \left(\frac{2 \pi r l}{N}\right) \\
+(r \omega) \frac{\partial^2 f_\mathrm{n l}}{\partial x \partial \dot{x}}(l) \cos \left(\frac{2 \pi k l}{N}\right) \sin \left(\frac{2 \pi n l}{N}\right) \cos \left(\frac{2 \pi r l}{N}\right) \\
+(n \omega)(r \omega) \frac{\partial^2 f_\mathrm{n l}}{\partial \dot{x}^2}(l) \cos \left(\frac{2 \pi k l}{N}\right) \cos \left(\frac{2 \pi n l}{N}\right) \cos \left(\frac{2 \pi r l}{N}\right) \\
+(n \omega) \frac{\partial^2 f_\mathrm{n l}}{\partial \dot{x} \partial x}(l) \cos \left(\frac{2 \pi k l}{N}\right) \cos \left(\frac{2 \pi n l}{N}\right) \sin \left(\frac{2 \pi r l}{N}\right)
\end{array}\right\},
\end{equation}

\begin{equation}
\frac{\partial^2 {F_\mathrm{n l}}^{\mathrm{s}, k}}{\partial X^{\mathrm{c}, n} \partial X^{\mathrm{c}, r}}=\frac{2}{N} \sum_{l=0}^{N-1}\left\{\begin{array}{l}
\frac{\partial^2 f_\mathrm{n l}}{\partial x^2}(l) \sin \left(\frac{2 \pi k l}{N}\right) \cos \left(\frac{2 \pi n l}{N}\right) \cos \left(\frac{2 \pi r l}{N}\right) \\
-(r \omega) \frac{\partial^2 f_\mathrm{n l}}{\partial x \partial \dot{x}}(l) \sin \left(\frac{2 \pi k l}{N}\right) \cos \left(\frac{2 \pi n l}{N}\right) \sin \left(\frac{2 \pi r l}{N}\right) \\
+(n \omega)(r \omega) \frac{\partial^2 f_\mathrm{n l}}{\partial \dot{x}^2}(l) \sin \left(\frac{2 \pi k l}{N}\right) \sin \left(\frac{2 \pi n l}{N}\right) \sin \left(\frac{2 \pi r l}{N}\right) \\
-(n \omega) \frac{\partial^2 f_\mathrm{n l}}{\partial \dot{x} \partial x}(l) \sin \left(\frac{2 \pi k l}{N}\right) \sin \left(\frac{2 \pi n l}{N}\right) \cos \left(\frac{2 \pi r l}{N}\right)
\end{array}\right\},
\end{equation}
\begin{equation}
\frac{\partial^2 {F_\mathrm{n l}}^{\mathrm{s}, k}}{\partial X^{\mathrm{c}, n} \partial X^{\mathrm{s}, r}}=\frac{2}{N} \sum_{l=0}^{N-1}\left\{\begin{array}{l}
\frac{\partial^2 f_\mathrm{n l}}{\partial x^2}(l) \sin \left(\frac{2 \pi k l}{N}\right) \cos \left(\frac{2 \pi n l}{N}\right) \sin \left(\frac{2 \pi r l}{N}\right) \\
+(r \omega) \frac{\partial^2 f_\mathrm{n l}}{\partial x \partial \dot{x}}(l) \sin \left(\frac{2 \pi k l}{N}\right) \cos \left(\frac{2 \pi n l}{N}\right) \cos \left(\frac{2 \pi r l}{N}\right) \\
-(n \omega)(r \omega) \frac{\partial^2 f_\mathrm{n l}}{\partial \dot{x}^2}(l) \sin \left(\frac{2 \pi k l}{N}\right) \sin \left(\frac{2 \pi n l}{N}\right) \cos \left(\frac{2 \pi r l}{N}\right) \\
-(n \omega) \frac{\partial^2 f_\mathrm{n l}}{\partial \dot{x} \partial x}(l) \sin \left(\frac{2 \pi k l}{N}\right) \sin \left(\frac{2 \pi n l}{N}\right) \sin \left(\frac{2 \pi r l}{N}\right)
\end{array}\right\},
\end{equation}
\begin{equation}
\frac{\partial^2 {F_\mathrm{n l}}^{\mathrm{s}, k}}{\partial X^{\mathrm{s}, n} \partial X^{\mathrm{c}, r}}=\frac{2}{N} \sum_{l=0}^{N-1}\left\{\begin{array}{l}
\frac{\partial^2 f_\mathrm{n l}}{\partial x^2}(l) \sin \left(\frac{2 \pi k l}{N}\right) \sin \left(\frac{2 \pi n l}{N}\right) \cos \left(\frac{2 \pi r l}{N}\right) \\
-(r \omega) \frac{\partial^2 f_\mathrm{n l}}{\partial x \partial \dot{x}}(l) \sin \left(\frac{2 \pi k l}{N}\right) \sin \left(\frac{2 \pi n l}{N}\right) \sin \left(\frac{2 \pi r l}{N}\right) \\
-(n \omega)(r \omega) \frac{\partial^2 f_\mathrm{n l}}{\partial \dot{x}^2}(l) \sin \left(\frac{2 \pi k l}{N}\right) \cos \left(\frac{2 \pi n l}{N}\right) \sin \left(\frac{2 \pi r l}{N}\right) \\
+(n \omega) \frac{\partial^2 f_\mathrm{n l}}{\partial \dot{x} \partial x}(l) \sin \left(\frac{2 \pi k l}{N}\right) \cos \left(\frac{2 \pi n l}{N}\right) \cos \left(\frac{2 \pi r l}{N}\right)
\end{array}\right\},
\end{equation}
\begin{equation}
\frac{\partial^2 {F_\mathrm{n l}}^{\mathrm{s}, k}}{\partial X^{\mathrm{s}, n} \partial X^{\mathrm{s}, r}}=\frac{2}{N} \sum_{l=0}^{N-1}\left\{\begin{array}{l}
\frac{\partial^2 f_\mathrm{n l}}{\partial x^2}(l) \sin \left(\frac{2 \pi k l}{N}\right) \sin \left(\frac{2 \pi n l}{N}\right) \sin \left(\frac{2 \pi r l}{N}\right) \\
+(r \omega) \frac{\partial^2 f_\mathrm{n l}}{\partial x \partial \dot{x}}(l) \sin \left(\frac{2 \pi k l}{N}\right) \sin \left(\frac{2 \pi n l}{N}\right) \cos \left(\frac{2 \pi r l}{N}\right) \\
+(n \omega)(r \omega) \frac{\partial^2 f_\mathrm{n l}}{\partial \dot{x}^2}(l) \sin \left(\frac{2 \pi k l}{N}\right) \cos \left(\frac{2 \pi n l}{N}\right) \cos \left(\frac{2 \pi r l}{N}\right) \\
+(n \omega) \frac{\partial^2 f_\mathrm{n l}}{\partial \dot{x} \partial x}(l) \sin \left(\frac{2 \pi k l}{N}\right) \cos \left(\frac{2 \pi n l}{N}\right) \sin \left(\frac{2 \pi r l}{N}\right)
\end{array}\right\},
\end{equation}
\begin{equation}
\begin{aligned}
\frac{\partial^2 F_\mathrm{n l}^0}{\partial X^0 \partial X^{\mathrm{c}, r}} & =\frac{1}{N} \sum_{l=0}^{N-1}\left\{\frac{\partial^2 f_\mathrm{n l}}{\partial x^2}(l) \cos \left(\frac{2 \pi r l}{N}\right)-(r \omega) \frac{\partial^2 f_\mathrm{n l}}{\partial x \partial \dot{x}}(l) \sin \left(\frac{2 \pi r l}{N}\right)\right\}, \\
\frac{\partial^2 F_\mathrm{n l}^0}{\partial X^0 \partial X^{\mathrm{s}, r}} & =\frac{1}{N} \sum_{l=0}^{N-1}\left\{\frac{\partial^2 f_\mathrm{n l}}{\partial x^2}(l) \sin \left(\frac{2 \pi r l}{N}\right)+(r \omega) \frac{\partial^2 f_\mathrm{n l}}{\partial x \partial \dot{x}}(l) \cos \left(\frac{2 \pi r l}{N}\right)\right\},
\end{aligned}
\end{equation}
\begin{equation}
\frac{\partial^2 F_\mathrm{n l}^0}{\partial X^{\mathrm{c}, n} \partial X^{\mathrm{c}, r}}=\frac{1}{N} \sum_{l=0}^{N-1}\left\{\begin{array}{l}
\frac{\partial^2 f_\mathrm{n l}}{\partial x^2}(l) \cos \left(\frac{2 \pi n l}{N}\right) \cos \left(\frac{2 \pi r l}{N}\right) \\
-(r \omega) \frac{\partial^2 f_\mathrm{n l}}{\partial x \partial \dot{x}}(l) \cos \left(\frac{2 \pi n l}{N}\right) \sin \left(\frac{2 \pi r l}{N}\right) \\
-(n \omega) \frac{\partial^2 f_\mathrm{n l}}{\partial \dot{x} \partial x}(l) \sin \left(\frac{2 \pi n l}{N}\right) \cos \left(\frac{2 \pi r l}{N}\right) \\
+(n \omega)(r \omega) \frac{\partial^2 f_\mathrm{n l}}{\partial \dot{x}^2}(l) \sin \left(\frac{2 \pi n l}{N}\right) \sin \left(\frac{2 \pi r l}{N}\right)
\end{array}\right\},
\end{equation}
\begin{equation}
\frac{\partial^2 F_\mathrm{n l}^0}{\partial X^{\mathrm{c}, n} \partial X^{\mathrm{s}, r}}=\frac{1}{N} \sum_{l=0}^{N-1}\left\{\begin{array}{l}
\frac{\partial^2 f_\mathrm{n l}}{\partial x^2}(l) \cos \left(\frac{2 \pi n l}{N}\right) \sin \left(\frac{2 \pi r l}{N}\right) \\
+(r \omega) \frac{\partial^2 f_\mathrm{n l}}{\partial x \partial \dot{x}}(l) \cos \left(\frac{2 \pi n l}{N}\right) \cos \left(\frac{2 \pi r l}{N}\right) \\
+(n \omega) \frac{\partial^2 f_\mathrm{n l}}{\partial \dot{x} \partial x}(l) \sin \left(\frac{2 \pi n l}{N}\right) \sin \left(\frac{2 \pi r l}{N}\right) \\
-(n \omega)(r \omega) \frac{\partial^2 f_\mathrm{n l}}{\partial \dot{x}^2}(l) \sin \left(\frac{2 \pi n l}{N}\right) \cos \left(\frac{2 \pi r l}{N}\right)
\end{array}\right\},
\end{equation}
\begin{equation}
\frac{\partial^2 F_\mathrm{n l}^0}{\partial X^{\mathrm{s}, n} \partial X^{\mathrm{c}, r}}=\frac{1}{N} \sum_{l=0}^{N-1}\left\{\begin{array}{l}
\frac{\partial^2 f_\mathrm{n l}}{\partial x^2}(l) \sin \left(\frac{2 \pi n l}{N}\right) \cos \left(\frac{2 \pi r l}{N}\right) \\
-(r \omega) \frac{\partial^2 f_\mathrm{n l}}{\partial x \partial \dot{x}}(l) \sin \left(\frac{2 \pi n l}{N}\right) \sin \left(\frac{2 \pi r l}{N}\right) \\
+(n \omega) \frac{\partial^2 f_\mathrm{n l}}{\partial \dot{x} \partial x}(l) \cos \left(\frac{2 \pi n l}{N}\right) \cos \left(\frac{2 \pi r l}{N}\right) \\
-(n \omega)(r \omega) \frac{\partial^2 f_\mathrm{n l}}{\partial \dot{x}^2}(l) \cos \left(\frac{2 \pi n l}{N}\right) \sin \left(\frac{2 \pi r l}{N}\right)
\end{array}\right\},
\end{equation}
\begin{equation}
\frac{\partial^2 F_\mathrm{n l}^0}{\partial X^{\mathrm{s}, n} \partial X^{\mathrm{s}, r}}=\frac{1}{N} \sum_{l=0}^{N-1}\left\{\begin{array}{l}
\frac{\partial^2 f_\mathrm{n l}}{\partial x^2}(l) \sin \left(\frac{2 \pi n l}{N}\right) \sin \left(\frac{2 \pi r l}{N}\right) \\
+(r \omega) \frac{\partial^2 f_\mathrm{n l}}{\partial x \partial \dot{x}}(l) \sin \left(\frac{2 \pi n l}{N}\right) \cos \left(\frac{2 \pi r l}{N}\right) \\
+(n \omega) \frac{\partial^2 f_\mathrm{n l}}{\partial \dot{x} \partial x}(l) \cos \left(\frac{2 \pi n l}{N}\right) \sin \left(\frac{2 \pi r l}{N}\right) \\
+(n \omega)(r \omega) \frac{\partial^2 f_\mathrm{n l}}{\partial \dot{x}^2}(l) \cos \left(\frac{2 \pi n l}{N}\right) \cos \left(\frac{2 \pi r l}{N}\right)
\end{array}\right\},
\end{equation}
\begin{equation}
\frac{\partial^2 F_\mathrm{n l}^{\mathrm{c}, k}}{\partial X^0 \partial X^{\mathrm{c}, r}}=\frac{2}{N} \sum_{l=0}^{N-1}\left\{\begin{array}{l}
\frac{\partial^2 f_\mathrm{n l}}{\partial x^2}(l) \cos \left(\frac{2 \pi k l}{N}\right) \cos \left(\frac{2 \pi r l}{N}\right) \\
-(r \omega) \frac{\partial^2 f_\mathrm{n l}}{\partial x^2}(l) \cos \left(\frac{2 \pi k l}{N}\right) \sin \left(\frac{2 \pi r l}{N}\right)
\end{array}\right\},
\end{equation}
\begin{equation}
\frac{\partial^2 F_\mathrm{n l}^{\mathrm{c}, k}}{\partial X^0 \partial X^{\mathrm{s}, r}}=\frac{2}{N} \sum_{l=0}^{N-1}\left\{\begin{array}{l}
\frac{\partial^2 f_\mathrm{n l}}{\partial x^2}(l) \cos \left(\frac{2 \pi k l}{N}\right) \sin \left(\frac{2 \pi r l}{N}\right) \\
+(r \omega) \frac{\partial^2 f_\mathrm{n l}}{\partial x^2}(l) \cos \left(\frac{2 \pi k l}{N}\right) \cos \left(\frac{2 \pi r l}{N}\right)
\end{array}\right\},
\end{equation}
\begin{equation}
\frac{\partial^2 F_\mathrm{n l}^{\mathrm{s}, k}}{\partial X^0 \partial X^{\mathrm{c}, r}}=\frac{2}{N} \sum_{l=0}^{N-1}\left\{\begin{array}{l}
\frac{\partial^2 f_\mathrm{n l}}{\partial x^2}(l) \sin \left(\frac{2 \pi k l}{N}\right) \cos \left(\frac{2 \pi r l}{N}\right) \\
-(r \omega) \frac{\partial^2 f_\mathrm{n l}}{\partial x^2}(l) \sin \left(\frac{2 \pi k l}{N}\right) \sin \left(\frac{2 \pi r l}{N}\right)
\end{array}\right\},
\end{equation}
\begin{equation}
\frac{\partial^2 F_\mathrm{n l}^{\mathrm{s}, k}}{\partial X^0 \partial X^{\mathrm{s}, r}}=\frac{2}{N} \sum_{l=0}^{N-1}\left\{\begin{array}{l}
\frac{\partial^2 f_\mathrm{n l}}{\partial x^2}(l) \sin \left(\frac{2 \pi k l}{N}\right) \sin \left(\frac{2 \pi r l}{N}\right) \\
+(r \omega) \frac{\partial^2 f_\mathrm{n l}}{\partial x^2}(l) \sin \left(\frac{2 \pi k l}{N}\right) \cos \left(\frac{2 \pi r l}{N}\right)
\end{array}\right\}.
\end{equation}

In cases where the nonlinear force exhibits velocity dependence (e.g., tanh friction), the governing terms become explicitly dependent on the excitation frequency. Consequently, it is necessary to evaluate the second-order mixed partial derivatives involving $\omega$. These terms are derived via the chain rule as follows:
\begin{equation}
\begin{aligned}
& \frac{\partial \boldsymbol{F_\mathrm{n l}}}{\partial \omega}=\frac{\partial \boldsymbol{F_\mathrm{n l}}}{\partial f_\mathrm{n l}} \frac{\partial f_\mathrm{n l}}{\partial \dot{x}} \frac{\partial \dot{x}}{\partial \omega}, \\
& \frac{\partial^2 \boldsymbol{F_\mathrm{n l}}}{\partial \omega \partial \boldsymbol{X}}=\frac{\partial \boldsymbol{F_\mathrm{n l}}}{\partial f_\mathrm{n l}} \frac{\partial f_\mathrm{n l}}{\partial \dot{x}} \frac{\partial^2 \dot{x}}{\partial \omega \partial \boldsymbol{X}}+\frac{\partial \boldsymbol{F_\mathrm{n l}}}{\partial f_\mathrm{n l}} \frac{\partial^2 f_\mathrm{n l}}{\partial \dot{x} \partial x} \frac{\partial x}{\partial \boldsymbol{X}} \frac{\partial \dot{x}}{\partial \omega}+\frac{\partial \boldsymbol{F_\mathrm{n l}}}{\partial f_\mathrm{n l}} \frac{\partial^2 f_\mathrm{n l}}{\partial \dot{x}^2} \frac{\partial \dot{x}}{\partial \boldsymbol{X}} \frac{\partial \dot{x}}{\partial \omega}, \\
& \frac{\partial^2 \boldsymbol{F_\mathrm{n l}}}{\partial \boldsymbol{X} \partial \omega}=\frac{\partial \boldsymbol{F_\mathrm{n l}}}{\partial f_{n l}} \frac{\partial f_{n l}}{\partial \dot{x}} \frac{\partial^2 \dot{x}}{\partial \boldsymbol{X} \partial \omega}+\frac{\partial \boldsymbol{F_\mathrm{n l}}}{\partial f_\mathrm{n l}} \frac{\partial^2 f_\mathrm{n l}}{\partial \dot{x}^2} \frac{\partial \dot{x}}{\partial \omega} \frac{\partial \dot{x}}{\partial \boldsymbol{X}}+\frac{\partial \boldsymbol{F_\mathrm{n l}}}{\partial f_\mathrm{n l}} \frac{\partial^2 f_\mathrm{n l}}{\partial x \partial \dot{x}^2} \frac{\partial \dot{x}}{\partial \omega} \frac{\partial x}{\partial \boldsymbol{X}}, \\
& \frac{\partial^2 \boldsymbol{F_\mathrm{n l}}}{\partial \omega \partial \omega}=\frac{\partial \boldsymbol{F_\mathrm{n l}}}{\partial f_\mathrm{n l}} \frac{\partial^2 f_\mathrm{n l}}{\partial \dot{x}^2} \frac{\partial \dot{x}}{\partial \omega} \frac{\partial \dot{x}}{\partial \omega}.
\end{aligned}
\end{equation}

For the sake of brevity, the detailed derivation of the second-order derivative terms involving the excitation frequency is provided in \ref{Appendix C}.

The Jacobian entries in Equation (10) are assembled by superimposing the stiffness and Hessian contributions of the nonlinear elements onto the linear dynamic stiffness matrix. When a nonlinear element involves multiple degrees of freedom—typically representing localized interfaces—its elemental stiffness matrices and Hessian tensors must first be transformed to the global coordinate system prior to assembly. 
\subsection{Nonlinear Element Coordinate Transformation} \label{sec:Coordinate Transformation}
\begin{figure}
  \centering
  \includegraphics[width=0.7\textwidth]{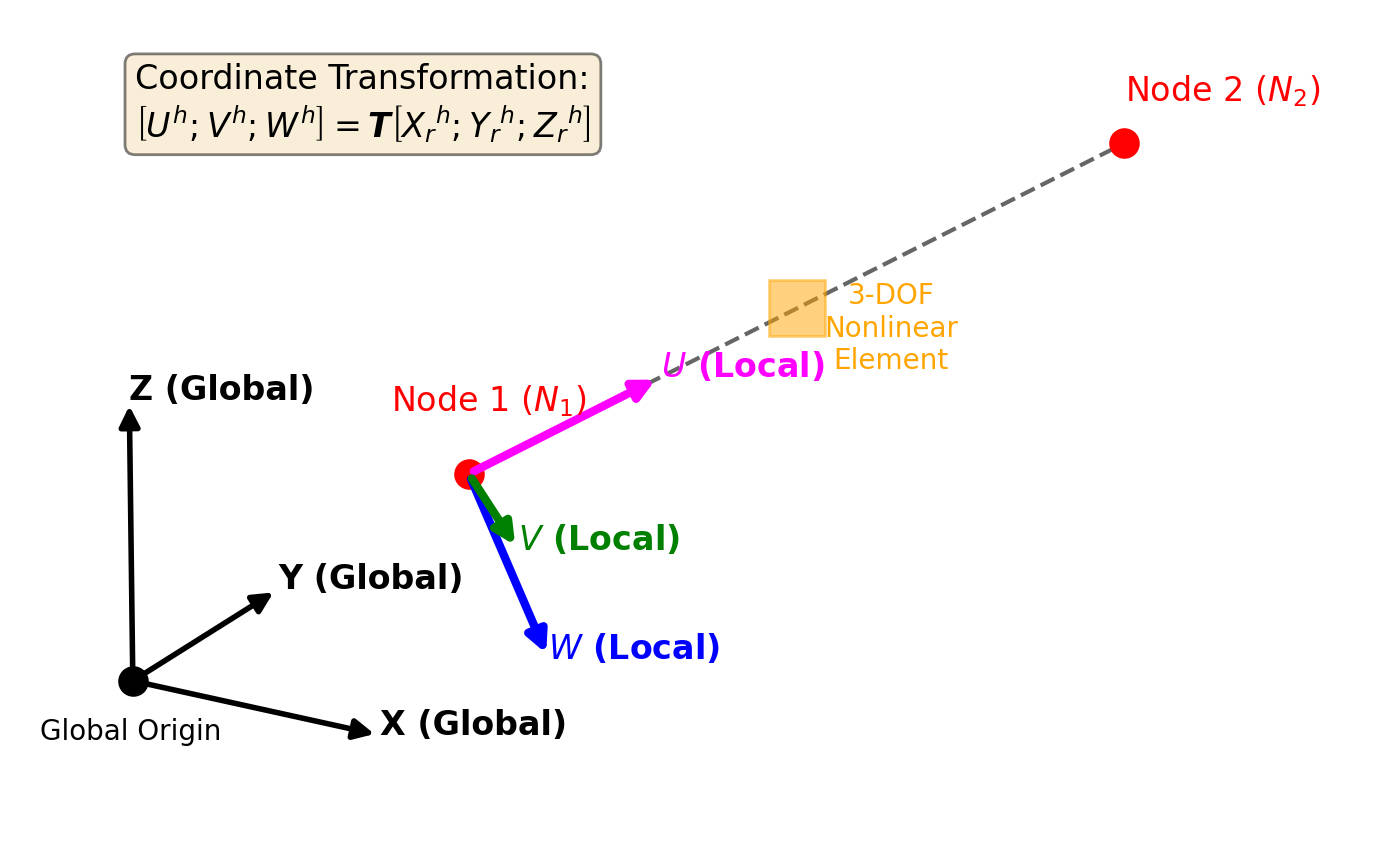}
  \caption{Schematic representation of the 3-DOF nonlinear element and the definition of local and global coordinate systems.}
  \label{fig:3-DOF nonlinear element}
\end{figure}

This subsection details the coordinate transformation framework for a representative three-degree-of-freedom (3-DOF) nonlinear element. This element characterizes the nonlinear interaction forces between two nodes, formulated in terms of their relative displacements and velocities along three orthogonal axes. The geometric configuration and the associated coordinate systems are illustrated in Figure~\ref{fig:3-DOF nonlinear element}.

The frequency-domain relative displacement vector of the element is given by:
\begin{equation}
\boldsymbol{Q}_\mathrm{r}=\left[\boldsymbol{Q}_\mathrm{r}{ }^0 ; \boldsymbol{Q}_\mathrm{r}{ }^{\mathrm{c}, 1} ; \boldsymbol{Q}_\mathrm{r}{ }^{\mathrm{s}, 1} ; \cdots ; \boldsymbol{Q}_\mathrm{r}{ }^{\mathrm{c}, N_\mathrm{h}} ; \boldsymbol{Q}_\mathrm{r}{ }^{\mathrm{s}, N_\mathrm{h}}\right],
\end{equation}
where
\begin{equation}
\begin{gathered}
\boldsymbol{Q}_\mathrm{r}^0=\left[X_\mathrm{r}^0 ; Y_\mathrm{r}^0 ; Z_\mathrm{r}^0\right] ,\\
\boldsymbol{Q}_\mathrm{r}^{\mathrm{c}, k}=\left[X_\mathrm{r}^{\mathrm{c}, k} ; Y_\mathrm{r}^{\mathrm{c}, k} ; Z_\mathrm{r}^{\mathrm{c}, k}\right], \\
\boldsymbol{Q}_\mathrm{r}^{\mathrm{s}, k}=\left[X_\mathrm{r}^{\mathrm{s}, k} ; Y_\mathrm{r}^{\mathrm{s}, k} ; Z_\mathrm{r}^{\mathrm{s}, k}\right], \\
\boldsymbol{X}_\mathrm{r}=\boldsymbol{X}_1-\boldsymbol{X}_2, \\
\boldsymbol{Y}_\mathrm{r}=\boldsymbol{Y}_1-\boldsymbol{Y}_2 ,\\
\boldsymbol{Z}_\mathrm{r}=\boldsymbol{Z}_1-\boldsymbol{Z}_2.
\end{gathered}
\end{equation}

Here $\boldsymbol{X}_1, \boldsymbol{Y}_1, \boldsymbol{Z}_1, \boldsymbol{X}_2, \boldsymbol{Y}_2, \boldsymbol{Z}_2$ denote the frequency-domain components at the two nodes of the element in the global coordinate system. Let $\boldsymbol{U}, \boldsymbol{V}, \boldsymbol{W}$ denote the displacement vectors in the local coordinate system, and let $\boldsymbol{T}$ be the coordinate transformation matrix from the local to the global coordinate system. The relationship between the displacement vectors is given by:
\begin{equation}
\left[ {{U^h};{V^h};{W^h}} \right] = \boldsymbol{T}\left[ {{X_\mathrm{r}}^h;{Y_\mathrm{r}}^h;{Z_\mathrm{r}}^h} \right],
\end{equation}
\begin{equation}
\left[ {{F_\mathrm{u}}^h;{F_\mathrm{v}}^h;{F_\mathrm{w}}^h} \right] = {\boldsymbol{T}^{T}}\left[ {{F_\mathrm{x}}^h;{F_\mathrm{y}}^h;{F_\mathrm{z}}^h} \right].
\end{equation}
The superscript $h$ represents the harmonic order, while the superscript $T$ denotes the transpose. This derivation relies on the orthogonality of the rotation matrix. For cylindrical or spherical local coordinates, where the transformation matrix is invertible, the equation holds by substituting the inverse for the transpose. Similarly, the transformation law of the element stiffness matrix is expressed as:
\begin{equation}
\boldsymbol{K}_\mathrm{global} = ({\boldsymbol{I}_{2{N_\mathrm{h}} + 1}} \otimes {\boldsymbol{T}^T}){\boldsymbol{K}_\mathrm{local}}({\boldsymbol{I}_{2{N_\mathrm{h}} + 1}} \otimes \boldsymbol{T}),
\end{equation}
where $\otimes$ denotes the Kronecker product. The row and column ordering of the global stiffness matrix $\boldsymbol{K}_{\mathrm{global}}$ corresponds to the vector structure of $\boldsymbol{Q}_{\mathrm{r}}$. The expression for the stiffness matrix $\boldsymbol{K}_{\mathrm{local}}$ in the local coordinate system is given by:
\begin{equation}
\left[\begin{array}{ccccccc}
\frac{\partial F_\mathrm{u}^0}{\partial U^0} & \frac{\partial F_\mathrm{u}^0}{\partial V^0} & \frac{\partial F_\mathrm{u}^0}{\partial W^0} & \cdots & \frac{\partial F_\mathrm{u}^0}{\partial U^{\mathrm{s}, N_\mathrm{h}}} & \frac{\partial F_\mathrm{u}^0}{\partial V^{\mathrm{s}, N_\mathrm{h}}} & \frac{\partial F_\mathrm{u}^0}{\partial W^{\mathrm{s}, N_h}} \\
\frac{\partial F_\mathrm{v}^0}{\partial U^0} & \frac{\partial F_\mathrm{v}^0}{\partial V^0} & \frac{\partial F_\mathrm{v}^0}{\partial W^0} & \cdots & \frac{\partial F_\mathrm{v}^0}{\partial U^{\mathrm{s}, N_\mathrm{h}}} & \frac{\partial F_\mathrm{v}^0}{\partial V^{\mathrm{s}, N_\mathrm{h}}} & \frac{\partial F_\mathrm{v}^0}{\partial W^{\mathrm{s}, N_\mathrm{h}}} \\
\frac{\partial F_\mathrm{w}^0}{\partial U^0} & \frac{\partial F_\mathrm{w}^0}{\partial V^0} & \frac{\partial F_\mathrm{w}^0}{\partial W^0} & \cdots & \frac{\partial F_\mathrm{w}^0}{\partial U^{\mathrm{s}, N_\mathrm{h}}} & \frac{\partial F_\mathrm{v}^0}{\partial V^{\mathrm{s}, N_\mathrm{h}}} & \frac{\partial F_\mathrm{w}^0}{\partial W^{\mathrm{s}, N_\mathrm{h}}} \\
\cdots & \cdots & \cdots & \cdots & \cdots & \cdots & \cdots \\
\frac{\partial F_\mathrm{u}^{\mathrm{s}, N_\mathrm{h}}}{\partial U^0} & \frac{\partial F_\mathrm{u}^{\mathrm{s}, N_\mathrm{h}}}{\partial V^0} & \frac{\partial F_\mathrm{u}^{\mathrm{s}, N_\mathrm{h}}}{\partial W^0} & \cdots & \frac{\partial F_\mathrm{u}^{\mathrm{s}, N_\mathrm{h}}}{\partial U^{\mathrm{s}, N_\mathrm{h}}} & \frac{\partial F_\mathrm{u}^{\mathrm{s}, N_\mathrm{h}}}{\partial V^{\mathrm{s}, N_\mathrm{h}}} & \frac{\partial F_\mathrm{u}^{\mathrm{s}, N_\mathrm{h}}}{\partial W^{\mathrm{s}, N_\mathrm{h}}} \\
\frac{\partial F_\mathrm{v}^{\mathrm{s}, N_\mathrm{h}}}{\partial U^0} & \frac{\partial F_\mathrm{v}^{\mathrm{s}, N_\mathrm{h}}}{\partial V^0} & \frac{\partial F_\mathrm{v}^{\mathrm{s}, N_\mathrm{h}}}{\partial W^0} & \cdots & \frac{\partial F_\mathrm{v}^{\mathrm{s}, N_\mathrm{h}}}{\partial U^{\mathrm{s}, N_\mathrm{h}}} & \frac{\partial F_\mathrm{v}^{\mathrm{s}, N_\mathrm{h}}}{\partial V^{\mathrm{s}, N_\mathrm{h}}} & \frac{\partial F_\mathrm{v}^{\mathrm{s}, N_\mathrm{h}}}{\partial W^{\mathrm{s}, N_\mathrm{h}}} \\
\frac{\partial F_\mathrm{w}{ }^{\mathrm{s}, N_\mathrm{h}}}{\partial U^0} & \frac{\partial F_\mathrm{w}{ }^{\mathrm{s}, N_\mathrm{h}}}{\partial V^0} & \frac{\partial F_\mathrm{w}{ }^{\mathrm{s}, N_\mathrm{h}}}{\partial W^0} & \cdots & \frac{\partial F_\mathrm{w}{ }^{\mathrm{s}, N_\mathrm{h}}}{\partial U^{\mathrm{s}, N_\mathrm{h}}} & \frac{\partial F_\mathrm{w}{ }^{\mathrm{s}, N_\mathrm{h}}}{\partial V^{\mathrm{s}, N_\mathrm{h}}} & \frac{\partial F_\mathrm{w}{ }^{\mathrm{s}, N_\mathrm{h}}}{\partial W^{\mathrm{s}, N_\mathrm{h}}}
\end{array}\right].
\end{equation}

Every term in the above stiffness matrix can be evaluated via the AFT method derived in the preceding subsection.

Regarding the Hessian of the nonlinear element, if the transformation between the local and global coordinate systems is linear (e.g., rotation or translation between Cartesian systems), tensor transformation rules can be applied. Note that the element Hessian tensor here is effectively a type-(1, 2) tensor with a dimension of $3 \cdot (2N_h+1)$. And it can be viewed as a collection of $3 \cdot (2N_h+1)$ matrices, each of dimension $3 \cdot (2N_h+1)$. Consequently, its coordinate transformation follows the tensor transformation law:
\begin{equation}
H_{m n}^p=\sum_{k, i, j} L_k^{T p}(L)_m^i(L)_n^j H_{i j}^{\prime k}.
\end{equation}

Here, $\boldsymbol{H}'$ and $\boldsymbol{H}$ denote the third-order tensors representing the second derivatives of the nonlinear force in the local and global coordinate systems, respectively. The indices $p, m, n$ denote the tensor indices in the global system, while $k, j, i$ denote those in the local system. Specifically, in the term $\frac{\partial^2 {F_\mathrm{x}}^0}{\partial X^{\mathrm{c},m} \partial Y^{\mathrm{s},n}}$, the index $p$ corresponds to the nonlinear force component (here, the 0-th harmonic in the X-direction); $m$ corresponds to the variable of differentiation for the first derivative (here, the cosine coefficient of the $m$-th harmonic in the X-direction); and $n$ corresponds to that of the second derivative (here, the sine coefficient of the $n$-th harmonic in the Y-direction).
And $\boldsymbol{L}$ denotes the multi-harmonic coordinate transformation matrix of the nonlinear element:
\begin{equation}
\boldsymbol{L} = {\boldsymbol{I}_{2{N_\mathrm{h}} + 1}} \otimes \boldsymbol{T}.
\end{equation}

To provide an explicit representation of the Hessian, the local tensor $\boldsymbol{H}'$ is decomposed into matrix slices. We define $\boldsymbol{H}_{::}^{\prime {\mathrm{u},\left(\mathrm{c},k\right)}}$ as the matrix representing the second-order derivatives of the harmonic coefficients ${F_\mathrm{u}}^{\mathrm{c}, k}$ with respect to the local displacement vector. It takes the following form:
\begin{equation}
\resizebox{\hsize}{!}{$
\left[\begin{array}{ccccccc}
\frac{\partial^2 F_\mathrm{u}^{\mathrm{c}, k}}{\partial^2 U^0} & \frac{\partial^2 F_\mathrm{u}^{\mathrm{c}, k}}{\partial V^0 \partial U^0} & \frac{\partial^2 F_\mathrm{u}^{\mathrm{c}, k}}{\partial W^0 \partial U^0} & \cdots & \frac{\partial^2 F_\mathrm{u}^{\mathrm{c}, k}}{\partial U^{\mathrm{s}, N_\mathrm{h}} \partial U^0} & \frac{\partial^2 F_\mathrm{u}^{\mathrm{c}, k}}{\partial V^{\mathrm{s}, N_\mathrm{h}} \partial U^0} & \frac{\partial^2 F_\mathrm{u}^{\mathrm{c}, k}}{\partial W^{\mathrm{s}, N_\mathrm{h}} \partial U^0} \\
\frac{\partial^2 F_\mathrm{v}^{\mathrm{c}, k}}{\partial V^0 \partial U^0} & \frac{\partial^2 F_\mathrm{v}^{\mathrm{c}, k}}{\partial^2 V^0} & \frac{\partial^2 F_\mathrm{v}^{\mathrm{c}, k}}{\partial V^0 \partial W^0} & \cdots & \frac{\partial^2 F_\mathrm{v}^{\mathrm{c}, k}}{\partial V^0 \partial U^{\mathrm{s}, N_\mathrm{h}}} & \frac{\partial^2 F_\mathrm{v}^{\mathrm{c}, k}}{\partial V^0 \partial V^{\mathrm{s}, N_\mathrm{h}}} & \frac{\partial^2 F_\mathrm{v}^{\mathrm{c}, k}}{\partial V^0 \partial W^{\mathrm{s}, N_\mathrm{h}}} \\
\frac{\partial^2 F_\mathrm{w}^{\mathrm{c}, k}}{\partial W^0 \partial U^0} & \frac{\partial^2 F_\mathrm{w}^{\mathrm{c}, k}}{\partial W^0 \partial V^0} & \frac{\partial^2 F_\mathrm{w}^{\mathrm{c}, k}}{\partial^2 W^0} & \cdots & \frac{\partial^2 F_\mathrm{w}^{\mathrm{c}, k}}{\partial W^0 \partial U^{\mathrm{s}, N_\mathrm{h}}} & \frac{\partial^2 F_\mathrm{v}^{\mathrm{c}, k}}{\partial W^0 \partial V^{\mathrm{s}, N_\mathrm{h}}} & \frac{\partial^2 F_\mathrm{w}^{\mathrm{c}, k}}{\partial W^0 \partial W^{\mathrm{s}, N_\mathrm{h}}} \\
\cdots & \cdots & \cdots & \cdots & \cdots & \cdots & \cdots \\
\frac{\partial^2 F_\mathrm{u}^{\mathrm{c}, k}}{\partial U^{\mathrm{s}, N_\mathrm{h}} \partial U^0} & \frac{\partial^2 F_\mathrm{u}^{\mathrm{c}, k}}{\partial U^{\mathrm{s}, N_\mathrm{h}} \partial V^0} & \frac{\partial^2 F_\mathrm{u}^{\mathrm{c}, k}}{\partial U^{\mathrm{s}, N_\mathrm{h}} \partial W^0} & \cdots & \frac{\partial^2 F_\mathrm{u}^{\mathrm{c}, k}}{\partial^2 U^{\mathrm{s}, N_\mathrm{h}}} & \frac{\partial^2 F_\mathrm{u}^{\mathrm{c}, k}}{\partial U^{\mathrm{s}, N_\mathrm{h}} \partial V^{\mathrm{s}, N_\mathrm{h}}} & \frac{\partial^2 F_\mathrm{u}^{\mathrm{c}, k}}{\partial U^{\mathrm{s}, N_\mathrm{h}} \partial W^{\mathrm{s}, N_\mathrm{h}}} \\
\frac{\partial^2 F_\mathrm{v}^{\mathrm{c}, k}}{\partial V^{\mathrm{s}, N_\mathrm{h}} \partial U^0} & \frac{\partial^2 F_\mathrm{v}^{\mathrm{c}, k}}{\partial V^{\mathrm{s}, N_\mathrm{h}} \partial V^0} & \frac{\partial^2 F_\mathrm{v}^{\mathrm{c}, k}}{\partial V^{\mathrm{s}, N_\mathrm{h}} \partial W^0} & \cdots & \frac{\partial^2 F_\mathrm{v}^{\mathrm{c}, k}}{\partial V^{\mathrm{s}, N_\mathrm{h}} \partial U^{\mathrm{s}, N_\mathrm{h}}} & \frac{\partial^2 F_\mathrm{v}^{\mathrm{c}, k}}{\partial^2 V^{\mathrm{s}, N_\mathrm{h}}} & \frac{\partial^2 F_\mathrm{v}^{\mathrm{c}, k}}{\partial V^{\mathrm{s}, N_\mathrm{h}} \partial W^{\mathrm{s}, N_\mathrm{h}}} \\
\frac{\partial^2 F_\mathrm{w}{ }^{\mathrm{c}, k}}{\partial W^{\mathrm{s}, N_\mathrm{h}} \partial U^0} & \frac{\partial^2 F_\mathrm{w}^{\mathrm{c}, k}}{\partial W^{\mathrm{s}, N_\mathrm{h}} \partial V^0} & \frac{\partial^2 F_\mathrm{w}^{\mathrm{c}, k}}{\partial W^{\mathrm{s}, N_\mathrm{h}} \partial W^0} & \cdots & \frac{\partial^2 F_\mathrm{w}^{\mathrm{c}, k}}{\partial W^{\mathrm{s}, N_\mathrm{h}} \partial U^{\mathrm{s}, N_\mathrm{h}}} & \frac{\partial^2 F_\mathrm{w}^{\mathrm{c}, k}}{\partial W^{\mathrm{s}, N_\mathrm{h}} \partial V^{\mathrm{s}, N_\mathrm{h}}} & \frac{\partial^2 F_\mathrm{w}^{\mathrm{c}, k}}{\partial^2 W^{\mathrm{s}, N_\mathrm{h}}}
\end{array}\right]
$.}
\end{equation}

Equation~(53) transforms the Hessian from the local to the global coordinate system. Since the resulting Hessian still represents derivatives with respect to relative coordinates, it must subsequently be mapped to the absolute degrees of freedom of the two connected nodes. Finally, these nodal contributions are utilized to construct the Jacobian matrix for Equation~(9), where the elemental stiffness and Hessian tensors are assembled into the global system.

\section{Numerical results} \label{sec:Numerical Results}
This section presents three numerical examples to evaluate the accuracy, computational efficiency, and applicability of the proposed algorithm. Section~\ref{2-DOF System with Cubic} validates the method using a simple 2-DOF oscillator with cubic stiffness nonlinearity. Section~\ref{Beam Element Model with cubic} applies the algorithm to a benchmark cantilever beam structure featuring cubic stiffness or tanh friction nonlinearities. Finally, Section~\ref{Blisk Element Model} demonstrates the method's capability on a large-scale finite element model of an industrial compressor bladed disk equipped with a friction ring damper.
\subsection{2-DOF System with Cubic Spring Nonlinearity} \label{2-DOF System with Cubic}
\begin{figure}[b]
  \centering
  \includegraphics[width=0.9\textwidth]{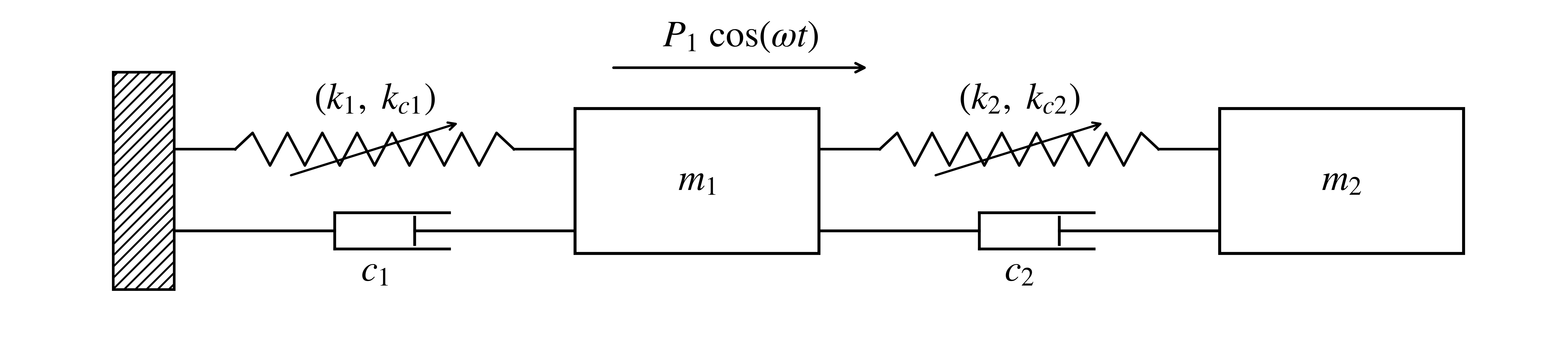}
  \caption{A schematic diagram of a forced, 2-DOF oscillator with cubic nonlinearity.}
  \label{fig:2-DOF oscillator}
\end{figure}

The study begins by examining the dynamic characteristics of a 2-DOF oscillator featuring cubic nonlinearity, a classic benchmark problem. The system configuration is illustrated in Figure~\ref{fig:2-DOF oscillator}, where $k_{\mathrm{c}}$ denotes the cubic stiffness coefficient. The model parameters are consistent with the corresponding example in the open-source software NLvib \cite{krack2019harmonic}, as listed in Table \ref{tab:1}.  

\begin{table}[htbp]
\caption[Table]{The parameters of the 2-DOF system with cubic nonlinearity}\label{tab:1}
    \centering
    \begin{tabular}{l c c c c c c c c } 
        \toprule 
        Parameters & $m_1$ & $m_2$ & $k_1$ & $k_2$ & $k_{\mathrm{c}1}$ & $k_{\mathrm{c}2}$ & $c_1$ & $c_2$\\
        \midrule 
        Valves & $1.05 $ & $1.05$ & $1$ & $0.0453$ & $1$ & $0.0042$ & $0.002$ & $0.013$ \\
        \bottomrule 
    \end{tabular}
\end{table}

In alignment with the reference example in NLvib, the analysis employs 7 harmonics and 256 samples per cycle to compute both the resonance and anti-resonance backbone curves.

\begin{figure}
  \centering
  \includegraphics[width=0.7\textwidth]{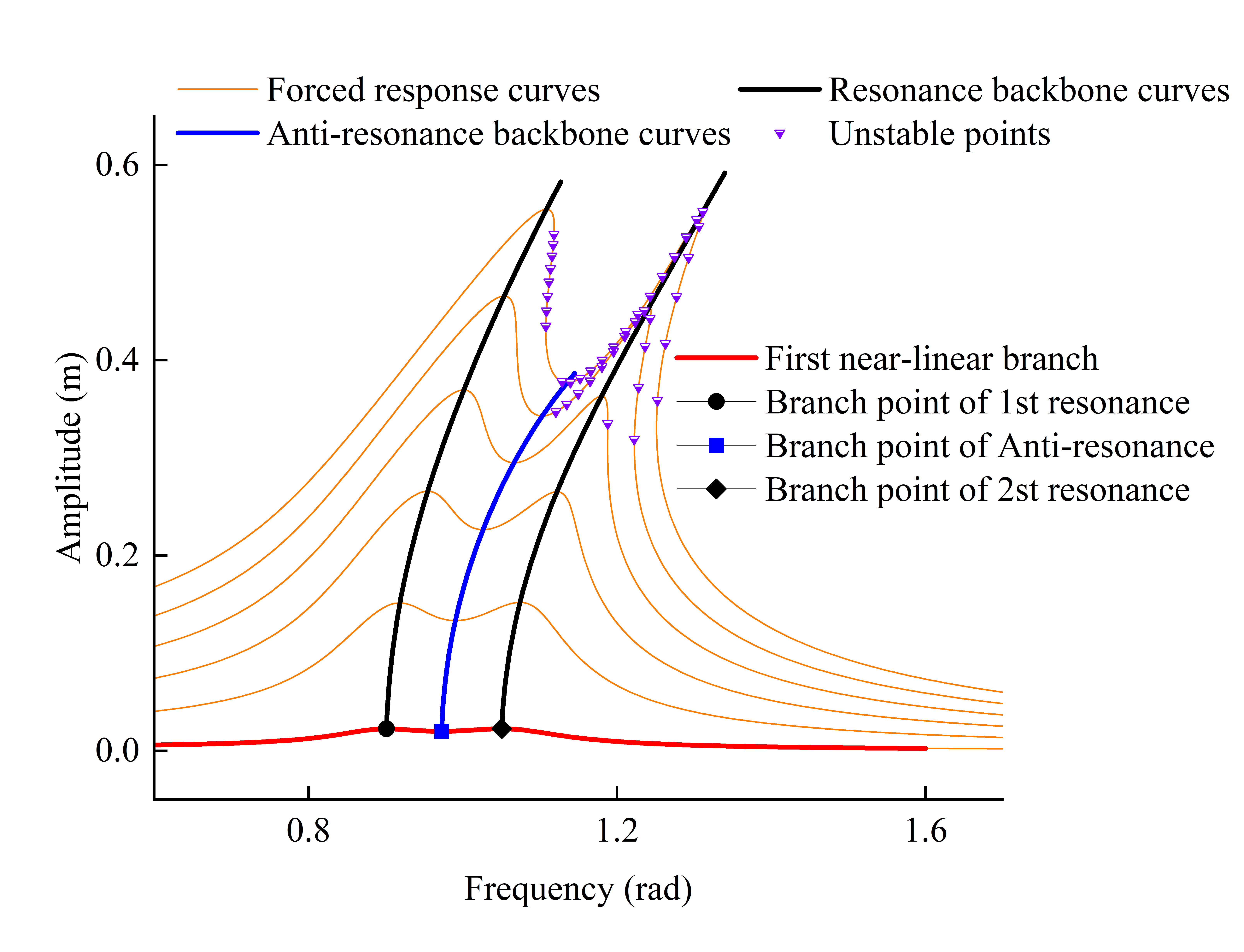}
  \caption{Nonlinear frequency response analysis of the two-degree-of-freedom oscillator. The computed backbone curves (thick lines) are superimposed on the reference forced response curves (thin orange lines) obtained via NLvib. The thick red line denotes the first near-linear branch used for initialization. Markers on the red line indicate branch points for the 1st resonance (\textcolor{black}{$\bullet$}), 2nd resonance ($\blacklozenge$), and anti-resonance (\textcolor{blue}{$\blacksquare$}). The resulting thick black and blue lines represent the resonance and anti-resonance backbone curves, respectively. Purple triangles (\textcolor{mypurple}{$\blacktriangledown$}) denote unstable solutions.}
  \label{fig:2-DOF 2D response}
\end{figure}

Figure~\ref{fig:2-DOF 2D response} presents the results obtained using the proposed method. The thick red line corresponds to the first solution branch calculated during the parameter continuation step (Section~\ref{ssec:Numerical Continuation}) at a low amplitude level. The black and blue markers indicate the resonance and anti-resonance branching points, respectively, identified via the Lagrange multiplier continuation. Subsequently, the thick black and blue lines represent the computed resonance and anti-resonance backbone curves, corresponding to the final continuation phase.

For validation purposes, NLvib was employed to calculate the forced response curves at various excitation levels (depicted as thin orange lines) along with the unstable points (purple markers). It is observed that the backbone curves accurately track all resonance peaks and robustly traverse the unstable regions. 

\begin{figure}
  \centering
  \includegraphics[width=0.8\textwidth]{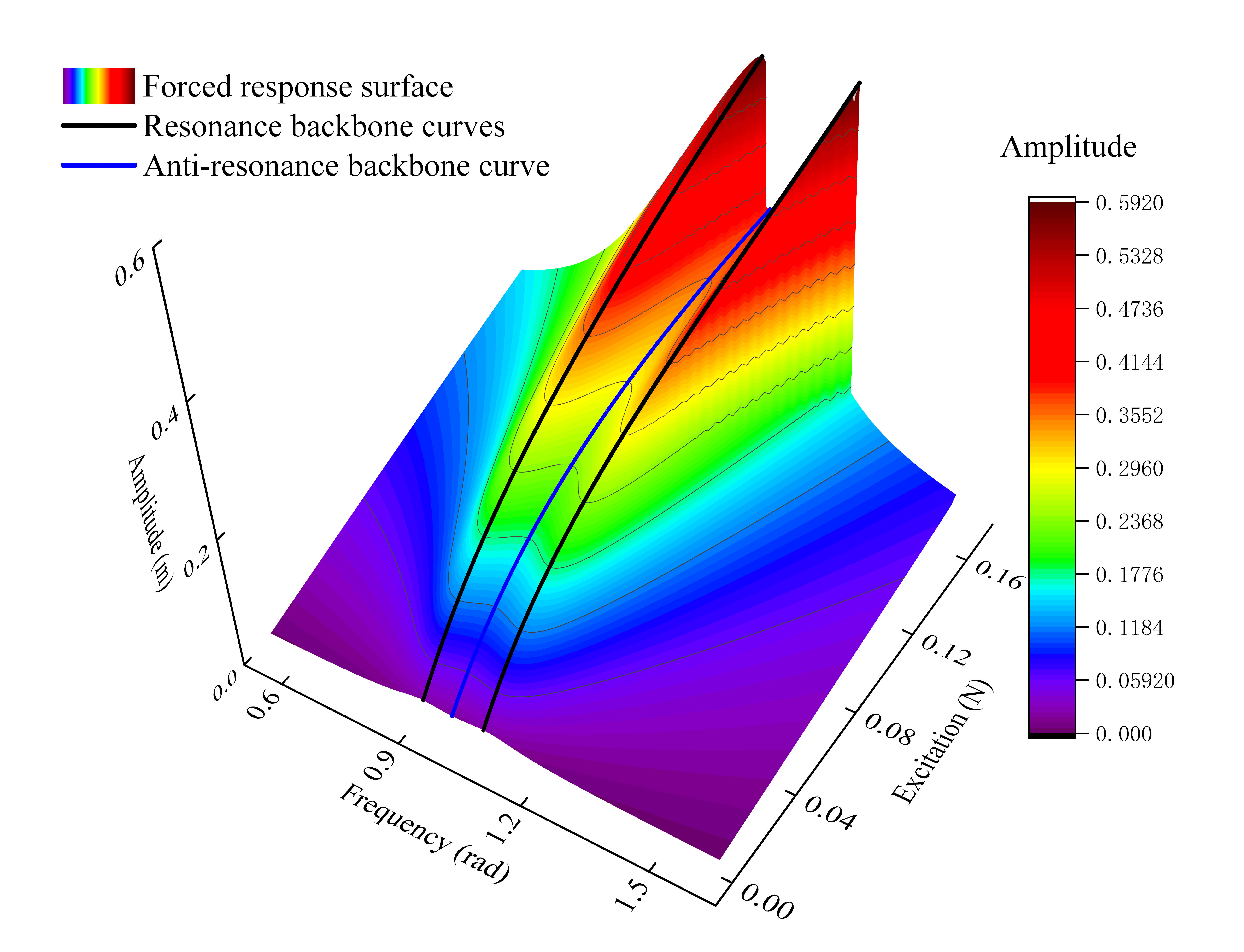}
  \caption{Forced response surface with superimposed backbone curves. The color-mapped surface illustrates the steady-state vibration amplitude versus excitation frequency and force magnitude. The computed resonance backbone curves (black lines) and anti-resonance backbone curve (blue line) are overlaid on the surface. Notably, the backbone curves precisely correspond to the topological ridges (local maxima) and the trench (local minima) of the response surface, demonstrating the global characterization of the nonlinear dynamics.}
  \label{fig:2-DOF FRS}
\end{figure}

Finally, by collecting a comprehensive set of frequency response points, the frequency response surface with respect to the excitation level is constructed in Figure~\ref{fig:2-DOF FRS}. As illustrated, the computed resonance and anti-resonance backbone curves connect all the local extrema, aligning precisely with the ridges and trenches of the response surface.
\begin{figure}[htbp]
    \centering
    \begin{subfigure}[b]{0.48\textwidth}
        \includegraphics[width=\linewidth]{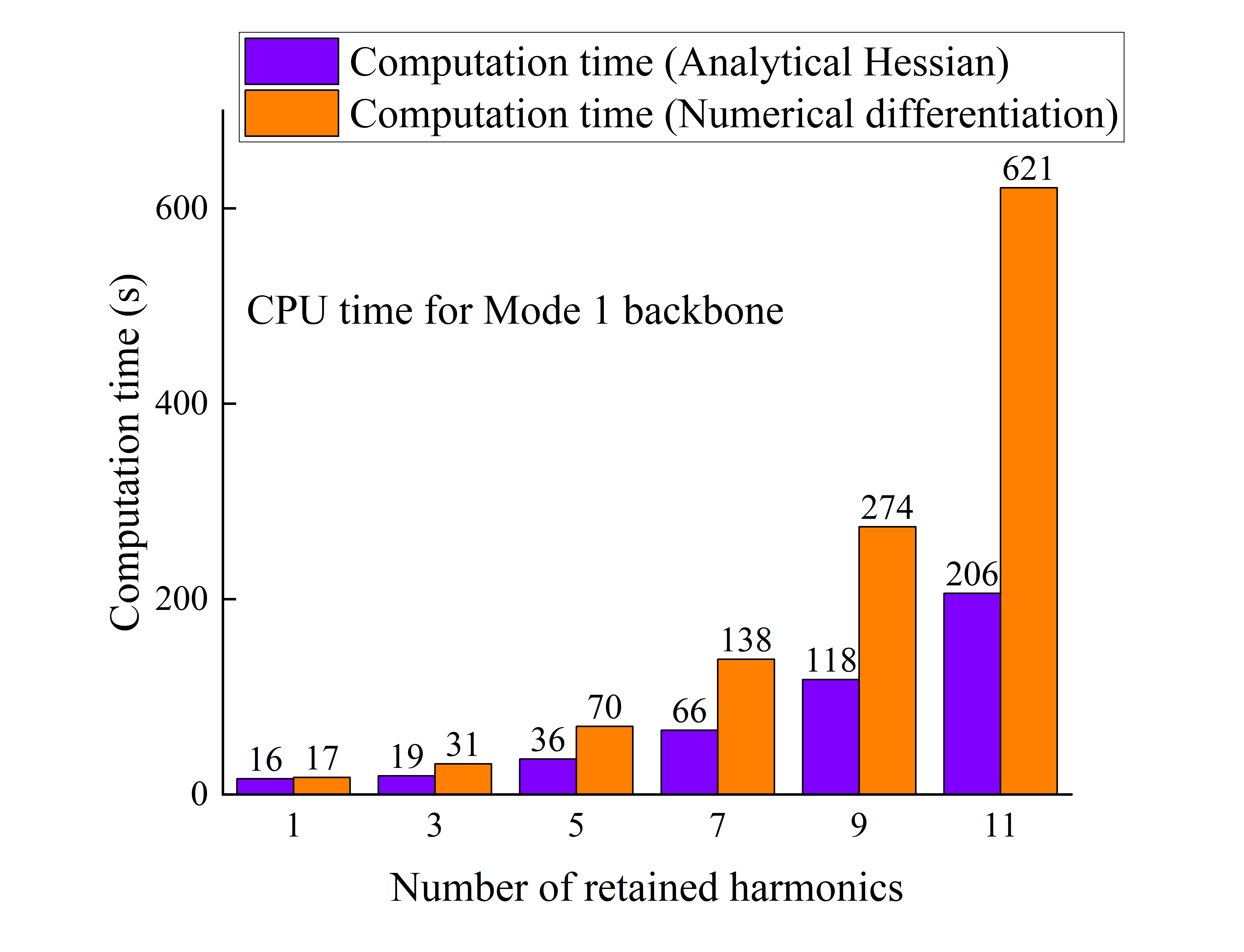}
        \caption{Computational cost for the first mode}        
        \label{fig:sub1}
    \end{subfigure}
    \hfill
    \begin{subfigure}[b]{0.48\textwidth}
        \includegraphics[width=\linewidth]{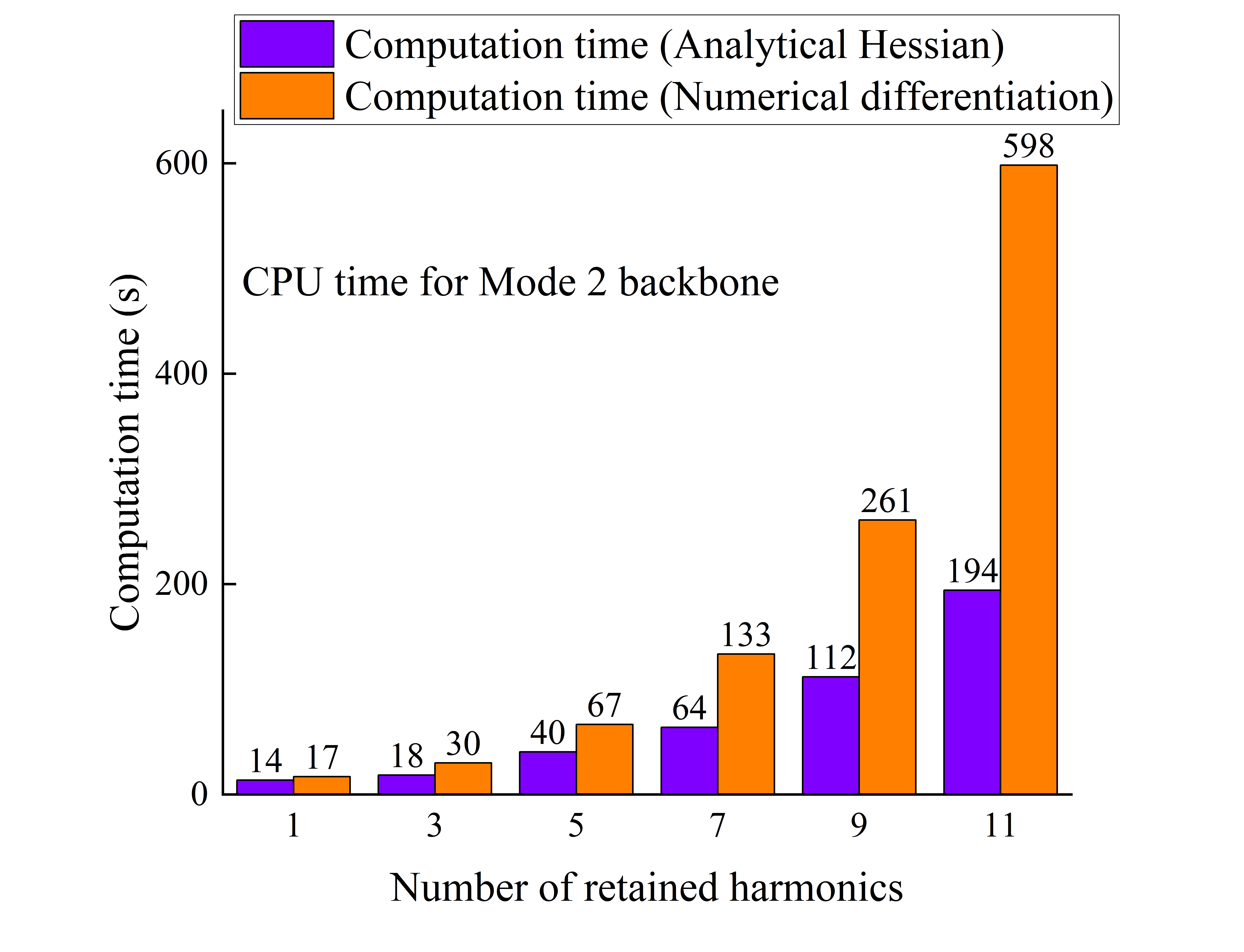}
        \caption{Computational cost for the second mode}         
        \label{fig:sub2}
    \end{subfigure}
    
    \caption{Comparison of computational efficiency between the proposed analytical Hessian and numerical differentiation. The purple bars denote the proposed method using the analytical Hessian, while the orange bars represent the conventional method based on numerical differentiation.}
  \label{fig:2-DOF computing time}    
\end{figure}


A critical advantage of the proposed framework is the explicit derivation of the analytical Hessian. Figure~\ref{fig:2-DOF computing time} quantifies this benefit by comparing the computation time required to trace backbone curves using the analytical Hessian versus standard numerical differentiation with increasing numbers of retained harmonics. The results demonstrate that the proposed method significantly outperforms the numerical counterpart, with efficiency gains becoming increasingly pronounced as the number of retained harmonics grows.

\subsection{Beam Finite Element Model with Cubic Stiffness or Tanh Friction Nonlinearities} \label{Beam Element Model with cubic}
\begin{figure}
  \centering
  \includegraphics[width=1\textwidth]{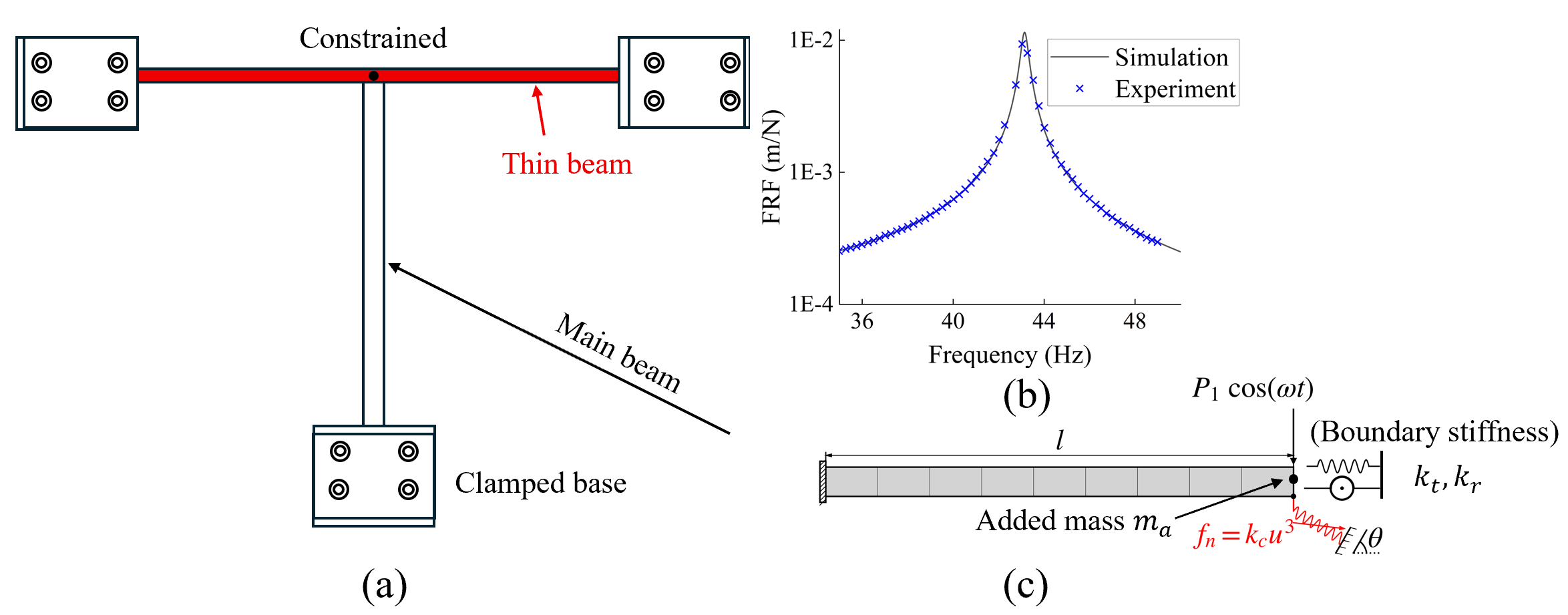}
  \caption{Calibration of the present finite element model against benchmark experimental data \cite{arslan2011parametric}.
(a) Schematic of the test rig (adapted from \cite{arslan2011parametric}), where the free end of the main beam is constrained between two thin beams (highlighted in red) to induce cubic stiffness nonlinearity.
(b) Comparison between the simulated FRF obtained in this study and the experimental data cited from \cite{arslan2011parametric}, which was measured at a low excitation level (near-linear regime). This linear comparison allows for the identification of the boundary stiffness.
(c) The calibrated finite element model. The linear springs ($k_t, k_r$) and a tip added mass ($m_a$) are introduced to represent the boundary stiffness identified through the calibration process in (b). $\theta$ denotes the orientation angle of the cubic spring (set to $\theta=0$ in the calculation without loss of generality).} 
  \label{fig:beam model with cubic stiffness}
\end{figure}

To further validate the effectiveness and applicability of the proposed algorithm to discretized continuous structures, this study employs a cantilever beam model. Two distinct nonlinear scenarios are investigated: a cubic stiffness nonlinearity and a tanh-based friction model. The system configuration, specifically depicting the cubic stiffness case at the free end, is illustrated in Figure~\ref{fig:beam model with cubic stiffness} (the material properties and geometric dimensions are listed in Table~\ref{tab:2}).

\begin{table}[htbp]
\caption[Table]{Material and geometric parameters of the beam.}\label{tab:2}
    \centering
    \begin{tabular}{c c c} 
        \toprule 
        Young' modulus (Gpa) & Density (kg/$m^3$) & Poisson's ratio 
        \\
        \midrule 
        210 & 7850 & 0.3 
        \\
        \midrule         
        Length (m) & Bending thickness (m) & Width (m)  
        \\
        \midrule 
        0.42 & 0.008 & 0.012 \\      
        \bottomrule 
    \end{tabular}
\end{table}

To align the numerical model with the benchmark experimental setup~\cite{arslan2011parametric}, the finite element discretization was calibrated by incorporating additional boundary flexibility and tip inertia. As detailed in Table~\ref{tab:3}, a translational spring ($k_t$) and a rotational spring ($k_r$) are introduced to model the effective boundary constraints, while $m_a$ represents the added tip mass. Furthermore, $\theta$ denotes the orientation angle of the equivalent cubic nonlinear spring (set to $\theta=0$ in the calculation without loss of generality). This calibration procedure ensures that the dynamic behavior in the low-amplitude linear regime remains consistent with experimental observations.
\begin{table}[htbp]
\caption[Table]{Finite element model settings and calibration parameters.}\label{tab:3}
    \centering
    \begin{tabular}{c c c c c c} 
        \toprule 
        Number of elements & $k_t$ (N/m) & $k_r$ (N$\cdot$m/rad) & $m_a$ (kg) &  $k_c$ (N/$m^3$) & $\theta$
        \\
        \midrule 
        9 & 6566 & 33.67 & 0.0756 & $2 \cdot 10^8$ & 0
        \\  
        \bottomrule 
    \end{tabular}
\end{table}

\begin{figure}
  \centering
  \includegraphics[width=0.8\textwidth]{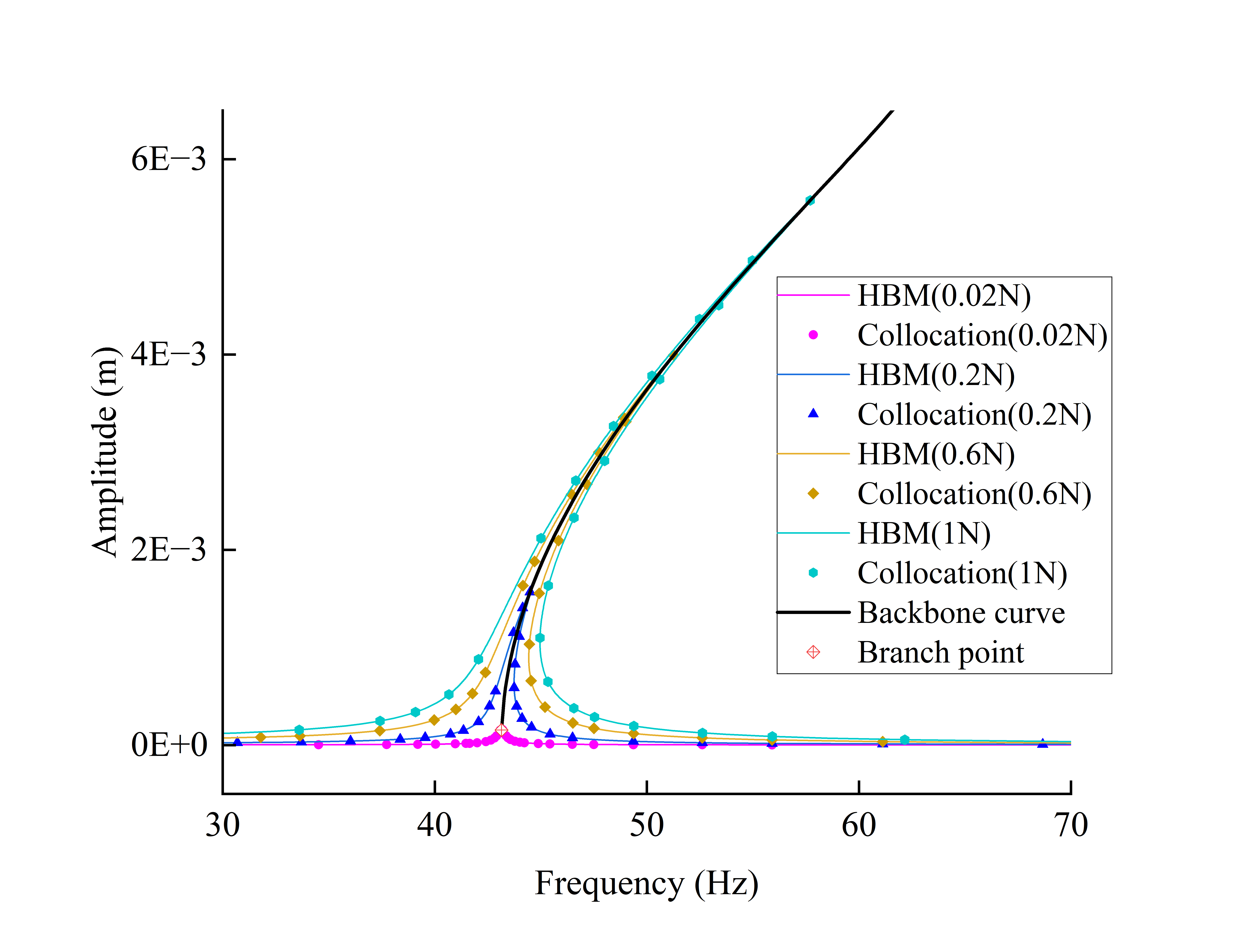}
  \caption{Frequency response curves of the displacement at the monitoring DOF of the beam model under different excitation levels. The black curve represents the resonant backbone curve, which precisely passes through the resonance peaks of the forced responses. The solid lines and markers denote the results calculated by the HBM and the Collocation Method, respectively.} 
  \label{fig:beam model with cubic stiffness FRCs}
\end{figure}

Retaining harmonics up to the third order, the resonant backbone curve corresponding to the first nonlinear mode was computed and validated against the Harmonic Balance Method (implemented in NLvib \cite{krack2019harmonic}) and the Collocation method (implemented in COCO \cite{dankowicz2013recipes}). As shown in Figure~\ref{fig:beam model with cubic stiffness FRCs}, the calculated backbone curve precisely passes through the resonance peaks obtained under various excitation levels.

\begin{figure}
  \centering
  \includegraphics[width=0.8\textwidth]{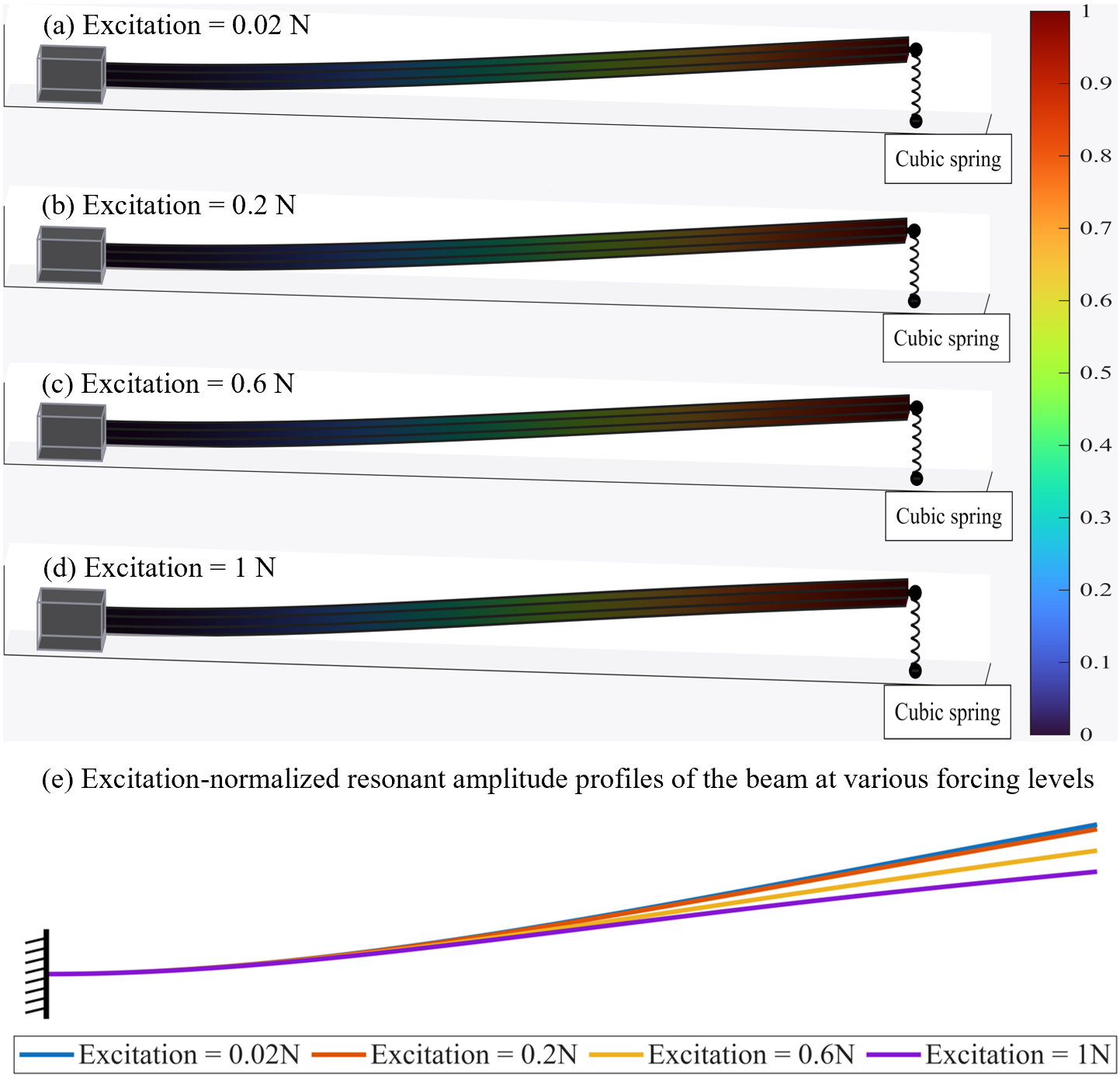}
  \caption{Visualization of the resonant mode shapes under varying excitation levels. (a)-(d) Excitation-normalized displacement contours of the beam at resonance for forcing amplitudes ranging from 0.02 N to 1 N. These snapshots are captured at the instant of maximum displacement (zero velocity at the monitored DOF). (e) Comparison of the normalized beam deflection profiles under different excitation levels. Notably, the normalized tip displacement decreases as the excitation level increases, indicating an increase in the equivalent stiffness of the cubic spring—a hallmark characteristic of structural hardening behavior.} 
 \label{fig:beam model with cubic stiffness resonant mode shapes}
\end{figure}

Additionally, the resonant mode shapes corresponding to the resonance peaks at four distinct excitation levels are illustrated in Figure~\ref{fig:beam model with cubic stiffness resonant mode shapes}. These vibration profiles correspond precisely to the computed backbone curve, clearly demonstrating the hardening characteristic as the nonlinear stiffness increases with the excitation amplitude.

\begin{figure}
  \centering
    \includegraphics[width=\textwidth]{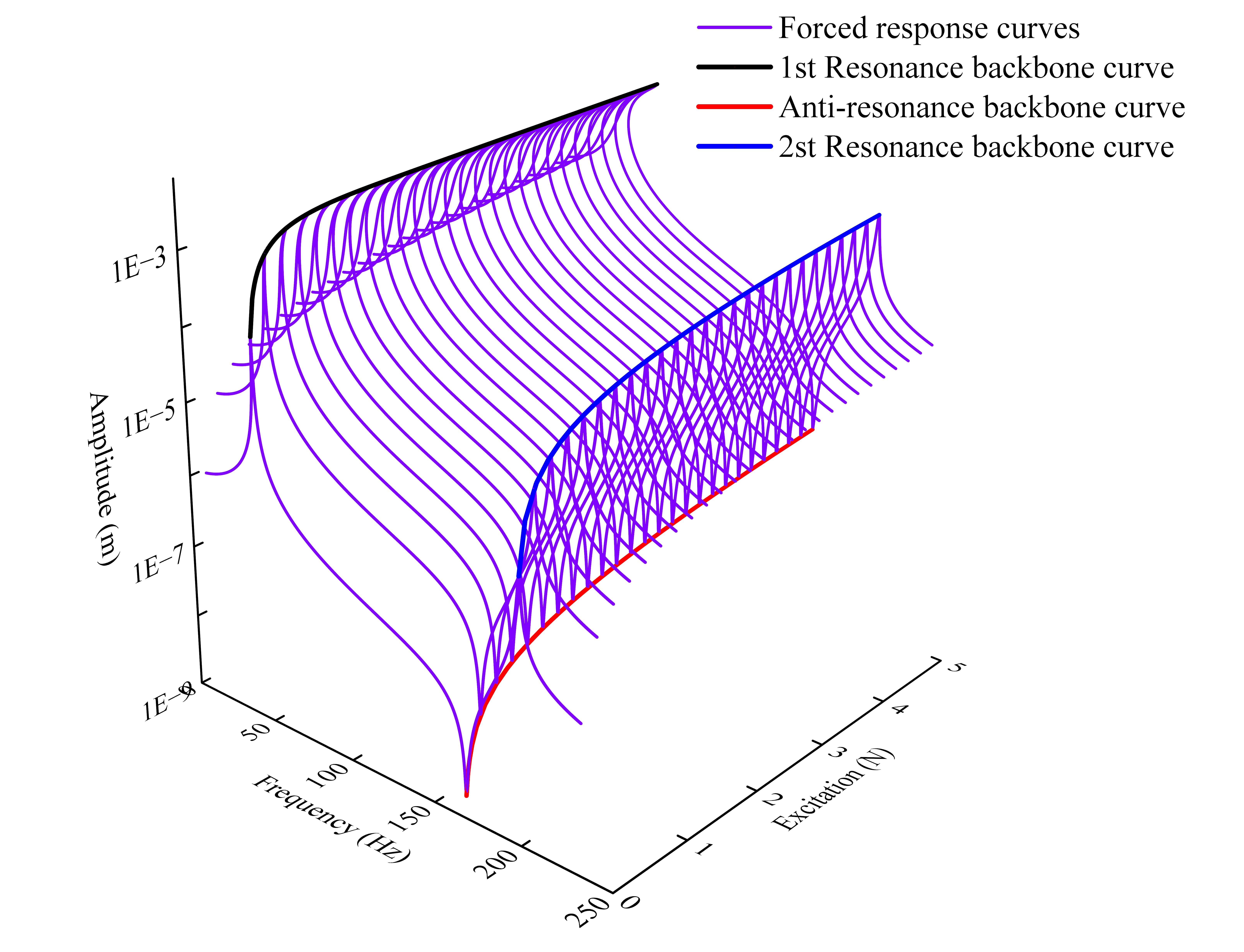}
  \caption{3D comparison of the forced frequency response curves (FRCs) with the computed resonance and anti-resonance backbone curves. The purple lines represent the FRCs obtained by sweeping the excitation frequency at various force levels. The black and blue lines denote the 1st and 2nd resonance backbones, respectively, while the red line indicates the anti-resonance backbone. The perfect alignment of the response extrema with the backbone curves demonstrates the robust accuracy of the proposed method across a wide range of excitation amplitudes.} 
  \label{fig:beam model with cubic stiffness FRCs 3D}
\end{figure}

Subsequently, the forced vibration response was computed over an extended frequency range using refined excitation increments, facilitating a three-dimensional comparison with the backbone curves. As illustrated in Figure~\ref{fig:beam model with cubic stiffness FRCs 3D}, all calculated resonance and anti-resonance points coincide precisely with the backbone curves, indicating that the computational accuracy remains robust across the entire investigated domain.

\begin{figure}
  \centering
  \includegraphics[width=0.8\textwidth]{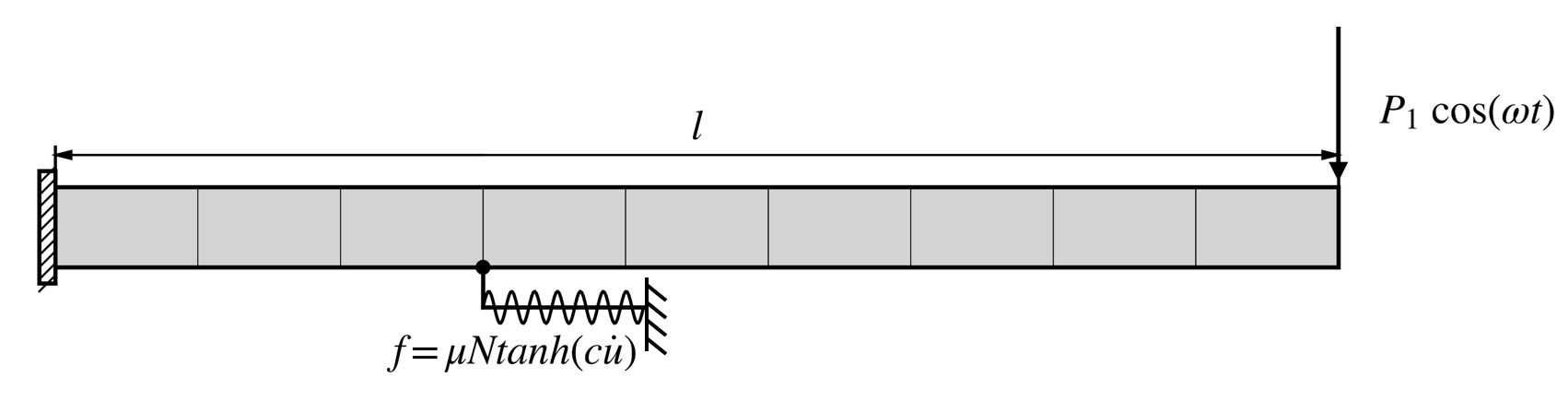}
  \caption{Beam finite element model featuring a hyperbolic tangent (tanh) friction nonlinearity. The underlying calibrated structure corresponds exactly to that in Figure~\ref{fig:beam model with cubic stiffness}, serving to validate the method's applicability to friction elements. Note that for visual clarity, the graphical markers for the boundary stiffness and added mass are omitted, though these calibrated parameters remain active in the numerical model.} 
  \label{fig:beam model with tanh nonlinearity}
\end{figure}

To further demonstrate the versatility of the proposed framework in handling non-polynomial nonlinearities, the cubic stiffness element is substituted with a localized tanh-based friction model~\cite{krack2019harmonic,mostaghel1997representations,petrov2003generic}, positioned at one-third of the beam span from the fixed boundary (see Figure~\ref{fig:beam model with tanh nonlinearity}). The governing expression for the nonlinear friction force is provided in Equation~(56). In this case study, the limiting friction force is set to $3$~N, while the smoothing parameter $c$—which governs the steepness of the approximation to the ideal signum function—is fixed at $6 \times 10^{-4}$.
\begin{equation}
f_{n l}=\mu N \tanh \left( \frac{\dot{x}}{\epsilon}\right).
\end{equation}

\begin{figure}[htbp]
    \centering
    \begin{subfigure}[b]{0.7\textwidth} 
        \centering
        \includegraphics[width=\textwidth]{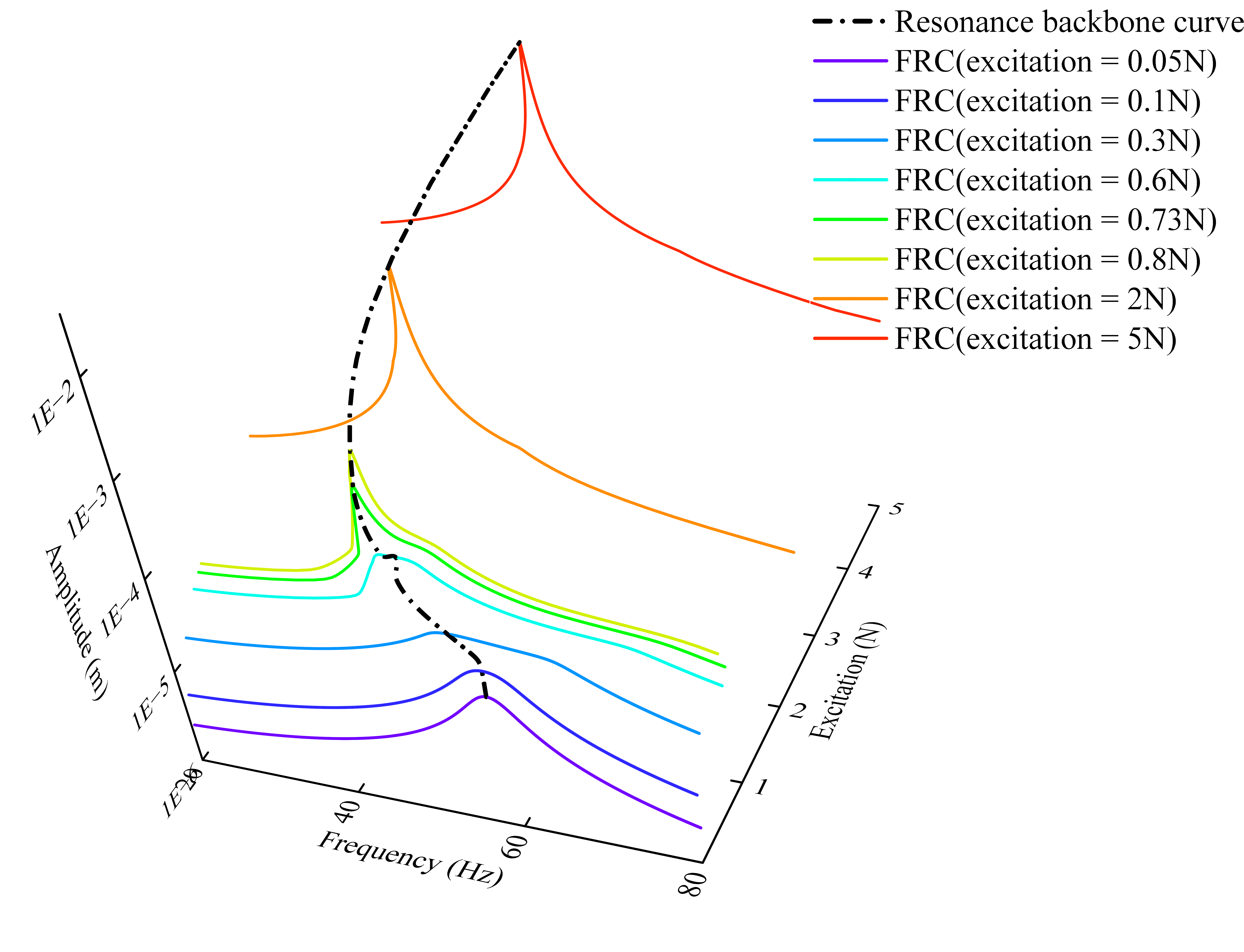}
        \caption{Resonance backbone curve and FRCs}        
    \end{subfigure}
    
    \vspace{0.5cm} 
    
    \begin{subfigure}[b]{0.48\textwidth}
        \centering    
        \includegraphics[width=\textwidth]{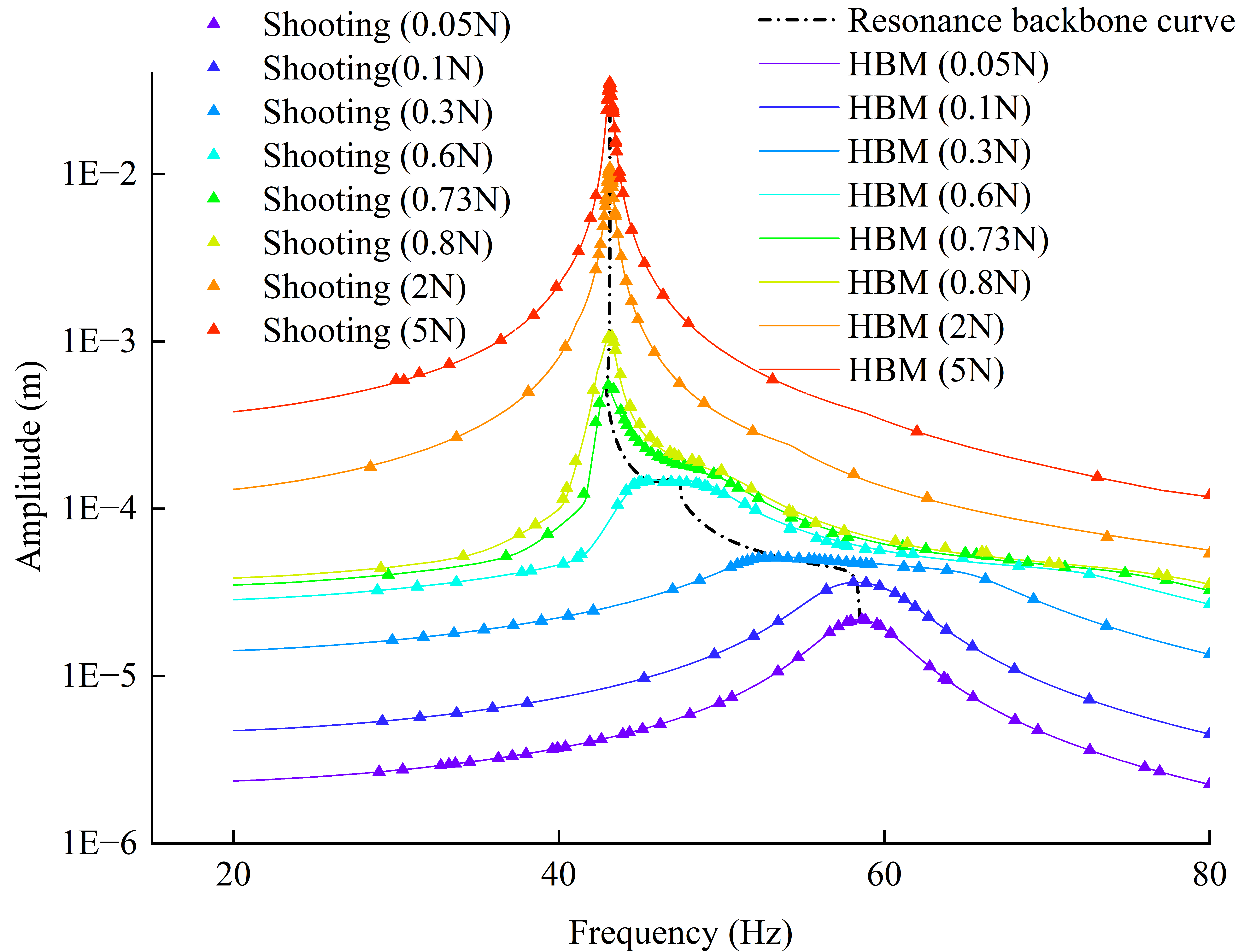}
        \caption{Validation against HBM and Shooting methods}        
    \end{subfigure}
    \hfill  
    \begin{subfigure}[b]{0.48\textwidth}
        \centering      
        \includegraphics[width=\textwidth]{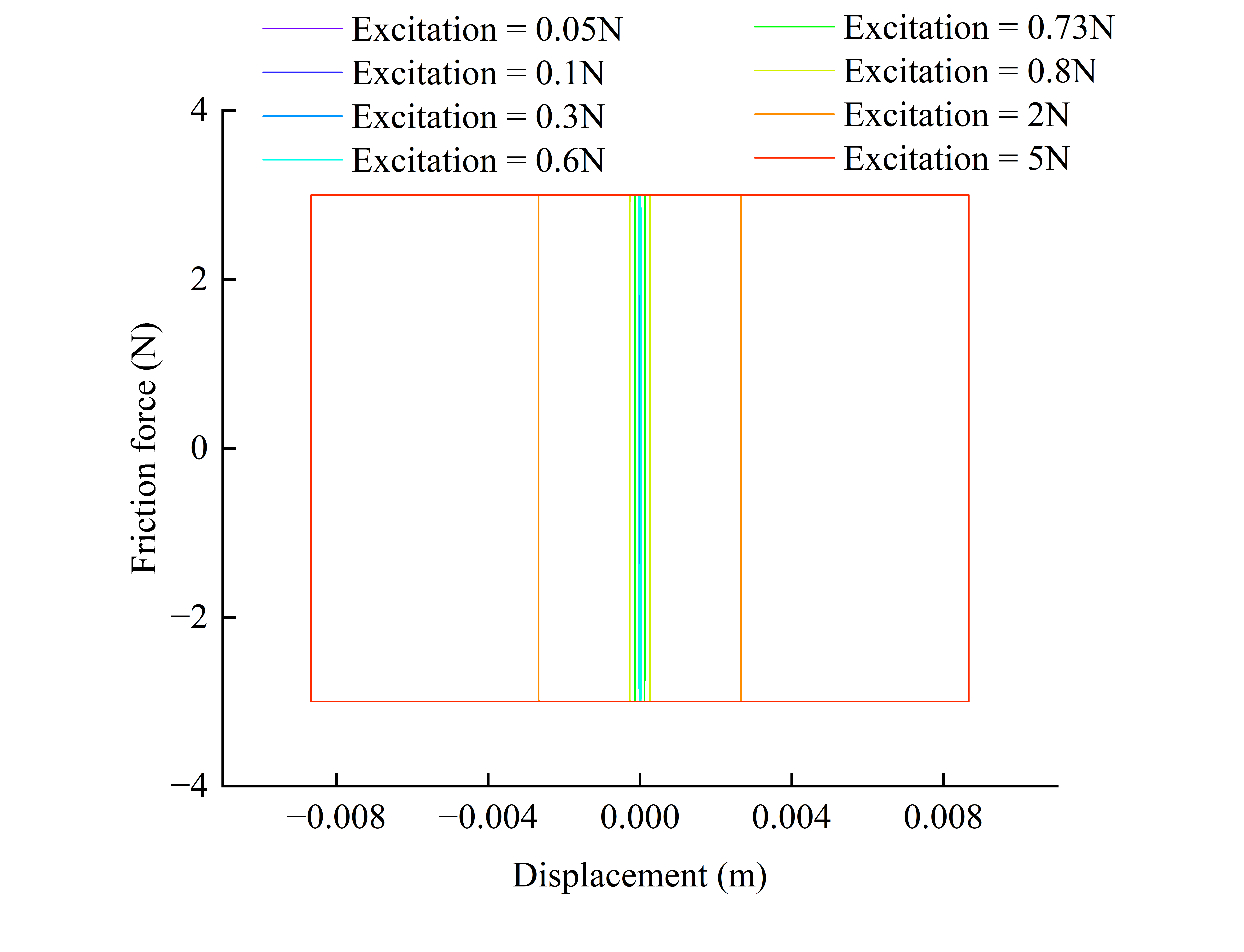}
        \caption{Smooth approximation of friction hysteresis at resonance points}        
        \label{fig:right_bot}
    \end{subfigure}  
    \caption{Numerical validation of the proposed backbone computation framework on a beam with localized tanh friction nonlinearity. (a) The computed resonance backbone curve precisely traces the locus of the resonance peaks of the FRCs under varying excitation levels. (b) Comparison against benchmark methods, where the FRCs are calculated via conventional HBM and the Shooting method (NLvib) to validate the accuracy of the proposed prediction. (c) Friction hysteresis loops at resonance points, demonstrating the smooth approximation of Coulomb friction behavior by the hyperbolic tangent function.}
    \label{fig:beam model with tanh nonlinearity FRCs}
\end{figure}

Retaining up to the 5th harmonic, we computed the forced response backbone curve corresponding to the first nonlinear mode, as shown in Figure~\ref{fig:beam model with tanh nonlinearity FRCs}. To validate the accuracy of the proposed method, the forced response curves were calculated using the conventional HBM and the shooting method (NLvib). It is evident that the resonance backbone curve precisely traces the locus of the resonance peaks under varying excitation levels.

To visually demonstrate this global consistency, Figure~\ref{fig:beam model with tanh nonlinearity FRCs}~(a) presents the 3D landscape where the computed backbone (black line) strictly coincides with the locus of the resonance peaks of the periodic response curves. Detailed quantitative verification is provided in Figure~\ref{fig:beam model with tanh nonlinearity FRCs}~(b), confirming the agreement between the proposed method and the benchmark solutions (HBM and Shooting). 

Furthermore, the corresponding force–displacement hysteresis loops at resonance are illustrated in Figure~\ref{fig:beam model with tanh nonlinearity FRCs}~(c). Notably, as the excitation amplitude increases, the loops expand and approach a rectangular shape characteristic of Coulomb friction. It is important to note that while the hyperbolic tangent function effectively approximates the sliding Coulomb friction, its smooth nature precludes the capture of a true "stick" state (zero relative velocity with non-zero holding force). However, since the primary objective of this case study is to validate the applicability of the proposed optimization framework to general non-polynomial nonlinearities rather than to refine the physical friction model, further discussion on the tribological limitations of the regularization is omitted.

\subsection{Blisk Element Model with tanh Friction Nonlinearity} \label{Blisk Element Model}
\begin{figure}
  \centering
  \begin{subfigure}[t]{0.48\textwidth}
  \includegraphics[width=\textwidth]{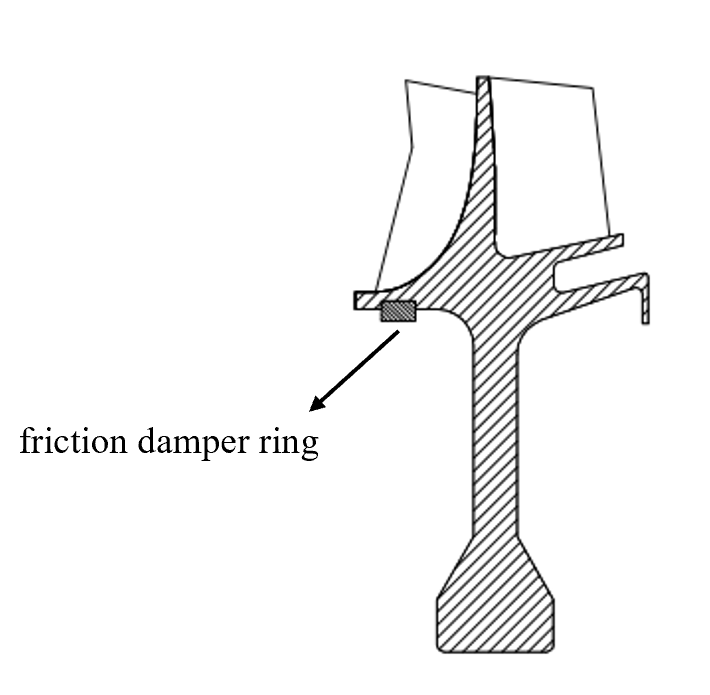}
  \caption{Configuration of the friction ring damper installation on the blisk.}  
  \end{subfigure}
  \hfill
  \begin{subfigure}[t]{0.48\textwidth}  
  \includegraphics[width=\textwidth]{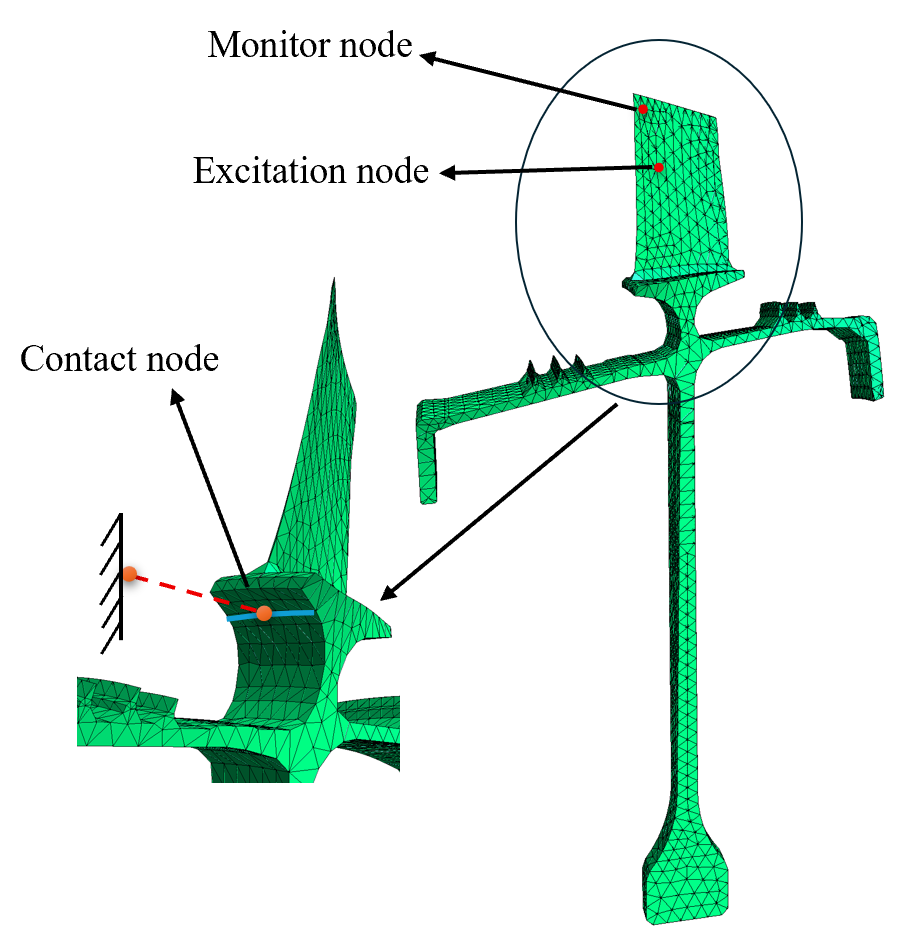}
  \caption{Finite element discretization of a single cyclic sector.}  
  \end{subfigure} 
  \caption{Schematic of the blisk–damper ring assembly and the finite element model of a cyclic sector.}
  \label{fig:blisk model with tanh nonlinearity}  
\end{figure}

To further demonstrate the scalability of the proposed resonance backbone computation method to finite element models of complex engineering structures, we consider the vibration response analysis of a compressor blisk (integral bladed disk) structure equipped with a friction damper ring, as illustrated in Figure~\ref{fig:blisk model with tanh nonlinearity}~(a).

Given the cyclic symmetry of the blisk and the localized nature of the contact interfaces, the complex dual Craig-Bampton method proposed in our previous work~\cite{wen2025complex} is employed to achieve efficient model order reduction. The DOFs at the excitation, response monitoring, and contact nodes are designated as master DOFs, while the first twenty free-interface normal modes are retained. The frictional interaction between the blisk and the damper ring is modeled using the previously validated tanh friction element. As illustrated in Figure~\ref{fig:blisk model with tanh nonlinearity}~(b), the damper ring is idealized as a grounded contact constraint.

The specific locations of the excitation and contact nodes are depicted in Figure~\ref{fig:blisk model with tanh nonlinearity}~(b). The geometric and material properties of the blisk, along with the parameters governing the contact element, are summarized in Table~\ref{tab:4}. 
\begin{table}[t]
\caption[Table]{Properties of the blisk FE model and the parameters of the contact element}\label{tab:4}
\centering{%
\begin{tabular}{llr}
\toprule
Parameter & Value \\
\midrule
Young' modulus &208 [Gpa] \\
Density & 7800 [kg/$m^3$] \\
Poisson's ratio ($v$) & 0.3 \\
Number of total elements & 2550 \\
Friction threshold($\mu N$) & $540$ [N] \\
constant($\epsilon$) & $1*10^{-3}$ \\
\bottomrule
\end{tabular}
}
\end{table}

\begin{figure}
  \centering
  \begin{subfigure}[t]{0.48\textwidth}
    \includegraphics[width=\textwidth]{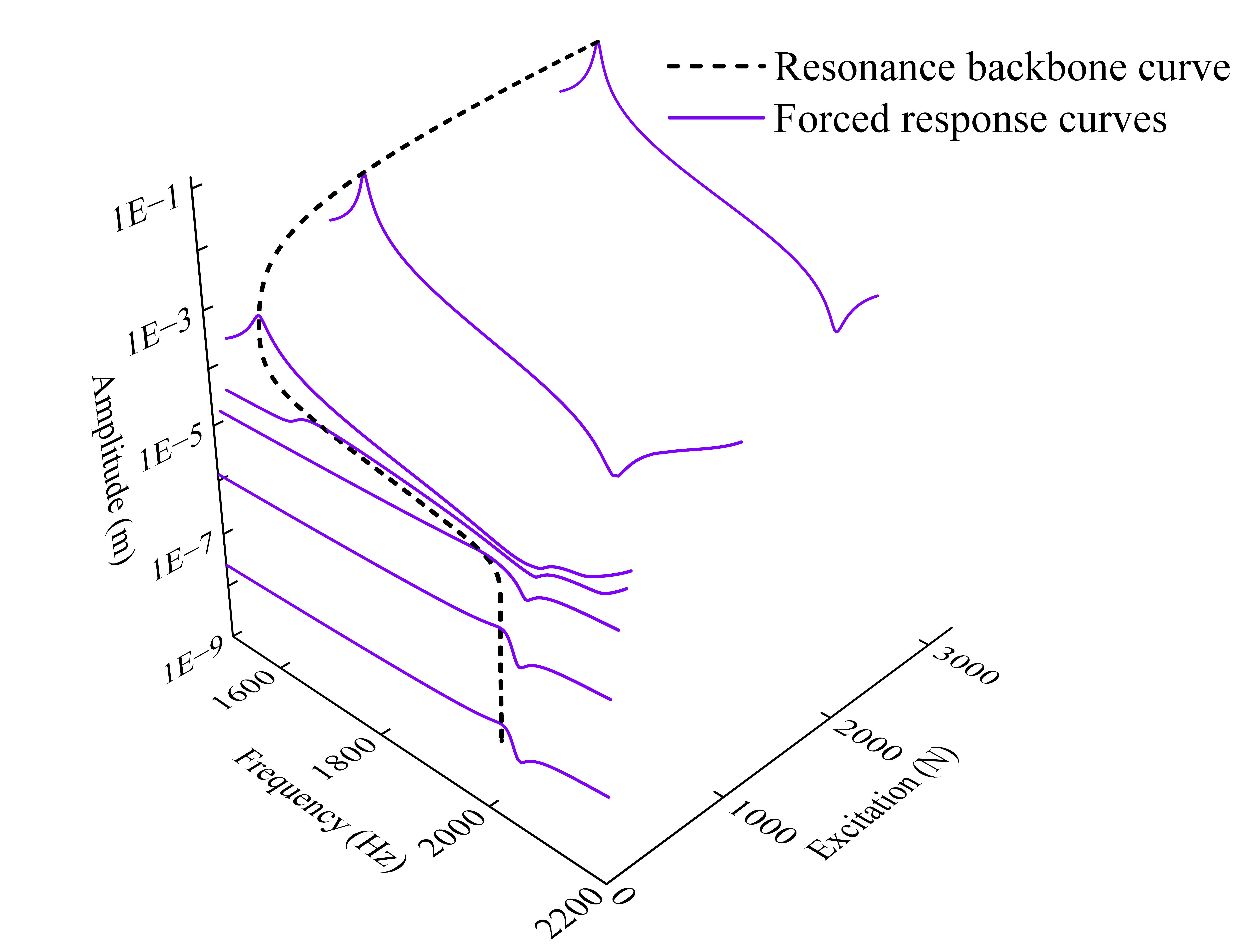}
    \caption{Backbone curve vs. HBM FRCs (0ND 1st mode)}  
  \end{subfigure}
  \hfill
  \begin{subfigure}[t]{0.48\textwidth} 
    \includegraphics[width=\textwidth]{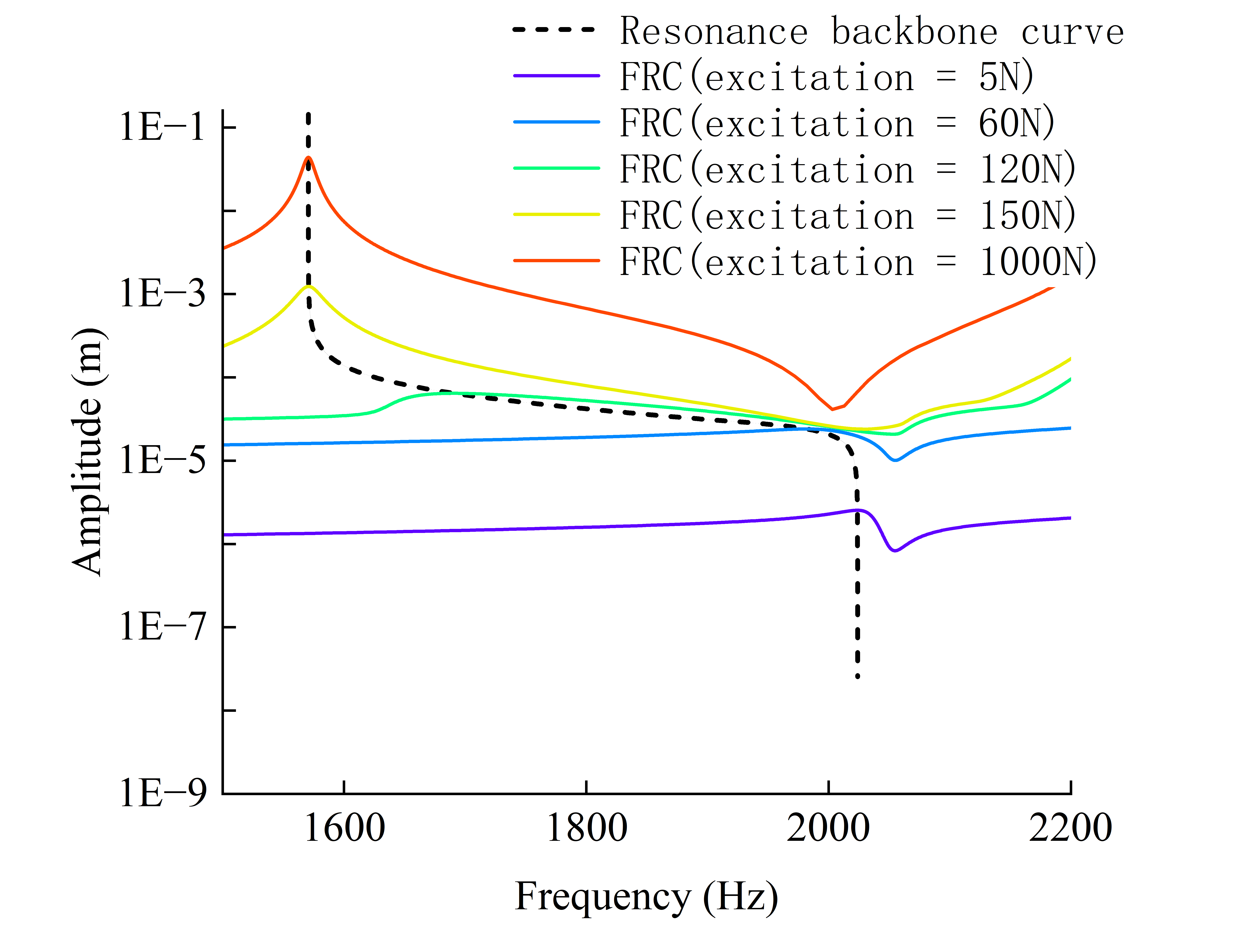}
    \caption{Evolution of forced response under varying excitation}  
  \end{subfigure} 
  \caption{ Forced response analysis of the compressor blisk with a friction damper ring. (a) Validation of the resonance backbone for the zero nodal diameter (0ND) 1st mode. The computed backbone is compared with FRCs obtained via HBM, demonstrating precise tracking of the resonance peak locus. (b) Evolution of the forced response under varying excitation levels, illustrating the amplitude-dependent damping and resonance frequency shifts induced by the friction ring.}
  \label{fig:blisk model with tanh nonlinearity 0ND FRCs}    
\end{figure}

Initially, the forced response backbone curve corresponding to the zero nodal diameter (0ND) fundamental mode was computed using a fundamental harmonic approximation. Figure~\ref{fig:blisk model with tanh nonlinearity 0ND FRCs}~(a) presents a comparison between this backbone curve and the FRCs obtained via the conventional HBM. It is evident that the computed backbone strictly traces the locus of the resonance peaks within the specified excitation range.

Furthermore, as illustrated in Figure~\ref{fig:blisk model with tanh nonlinearity 0ND FRCs}~(b), the results confirm that the computed backbone accurately captures both the amplitude-dependent damping and the hardening-induced frequency shifts exhibited by the system under varying excitation levels.

It is acknowledged that variations in normal contact pressure and the influence of tangential contact stiffness at the blisk–damper interface are not explicitly accounted for in this idealized friction model. However, as previously emphasized, the primary objective of this study is to validate the applicability of the proposed computational framework to complex nonlinear systems, rather than to refine the physical contact modeling. Consequently, detailed discussions on these specific tribological parameters are deemed beyond the scope of the present analysis.

\begin{figure}
  \centering
  \begin{subfigure}[t]{0.45\textwidth}
 
    \includegraphics[width=\textwidth]{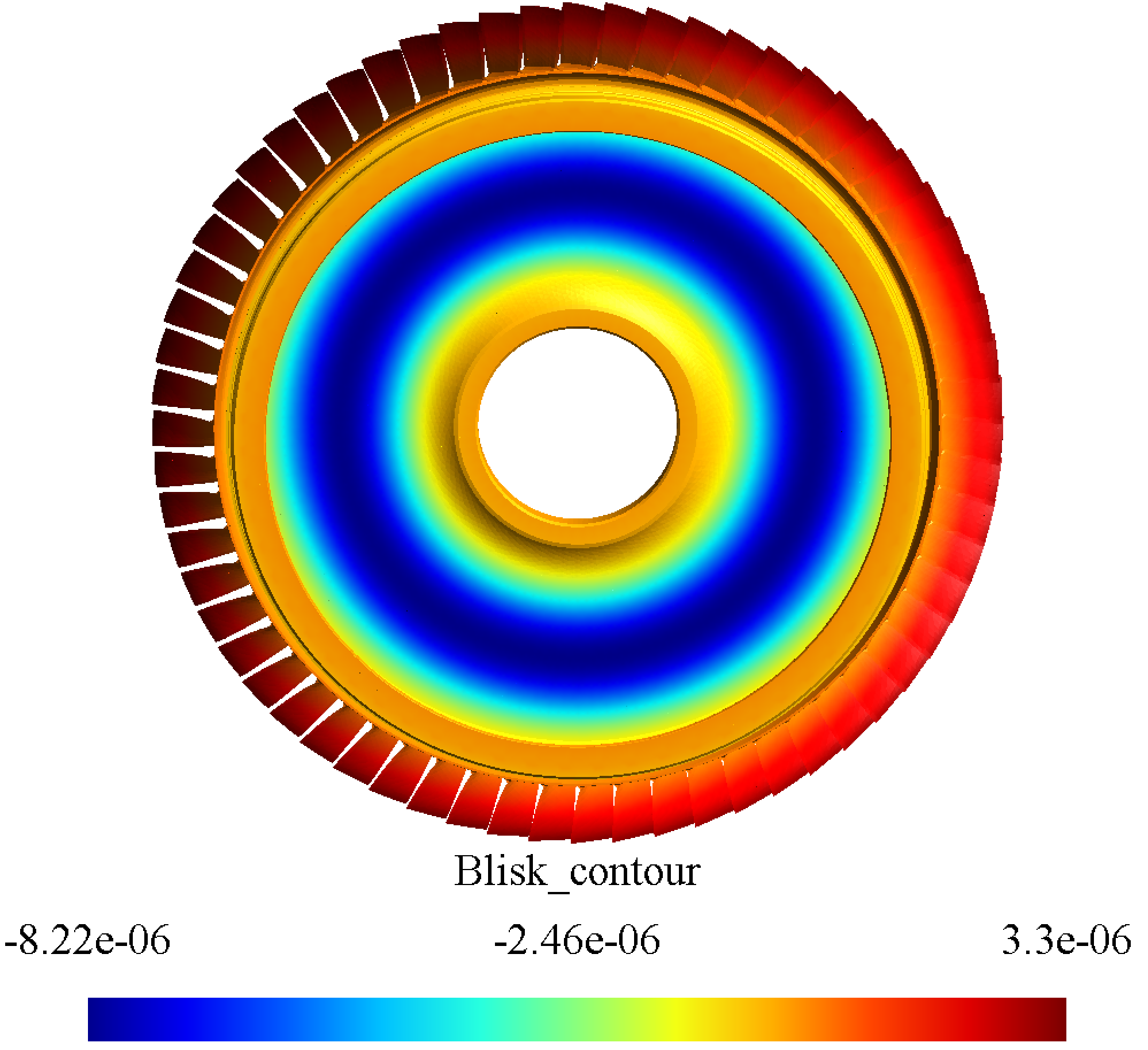}
  \caption{Resonant traveling displacement field at excitation level $F = 5$~N. } 
  \end{subfigure}
  \hfill
  \begin{subfigure}[t]{0.45\textwidth}
   
    \includegraphics[width=\textwidth]{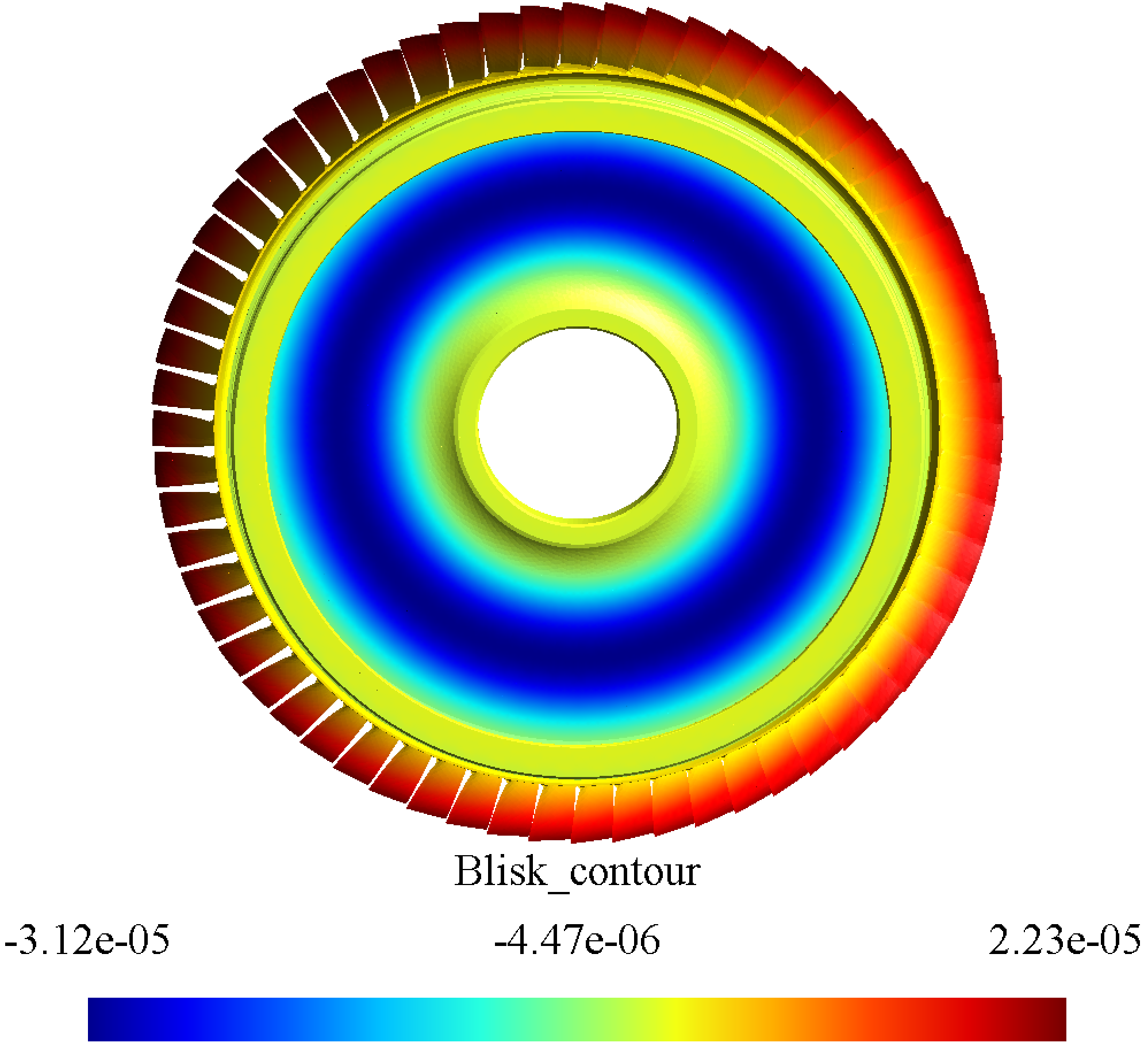}
  \caption{Resonant traveling displacement field at $F = 60$~N.} 
  \end{subfigure}
  
  \begin{subfigure}[t]{0.45\textwidth}
  
    \includegraphics[width=\textwidth]{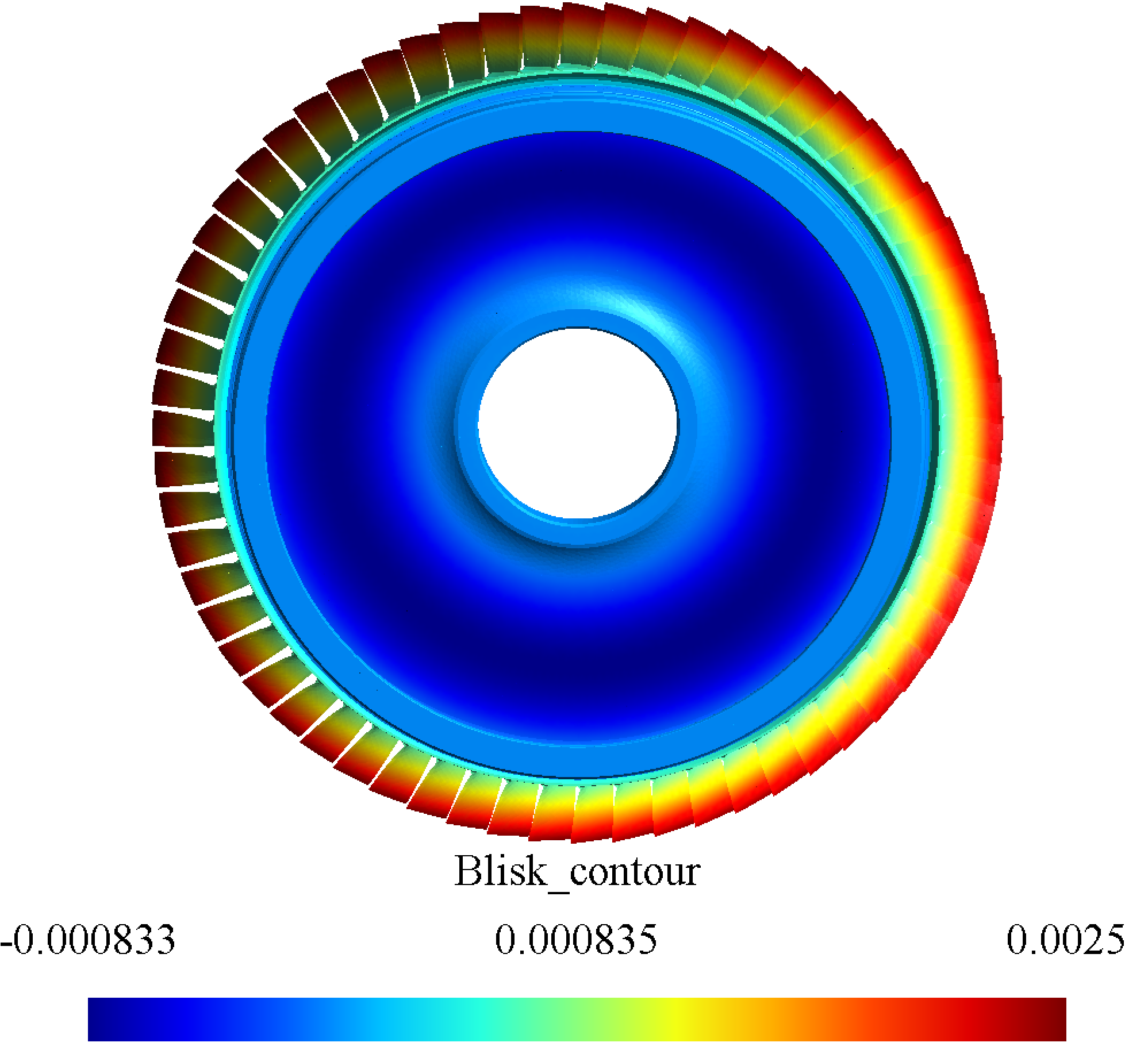}
  \caption{Resonant traveling displacement field at $F = 150$~N.}  
  \end{subfigure}
  \hfill
  \begin{subfigure}[t]{0.45\textwidth}
  
    \includegraphics[width=\textwidth]{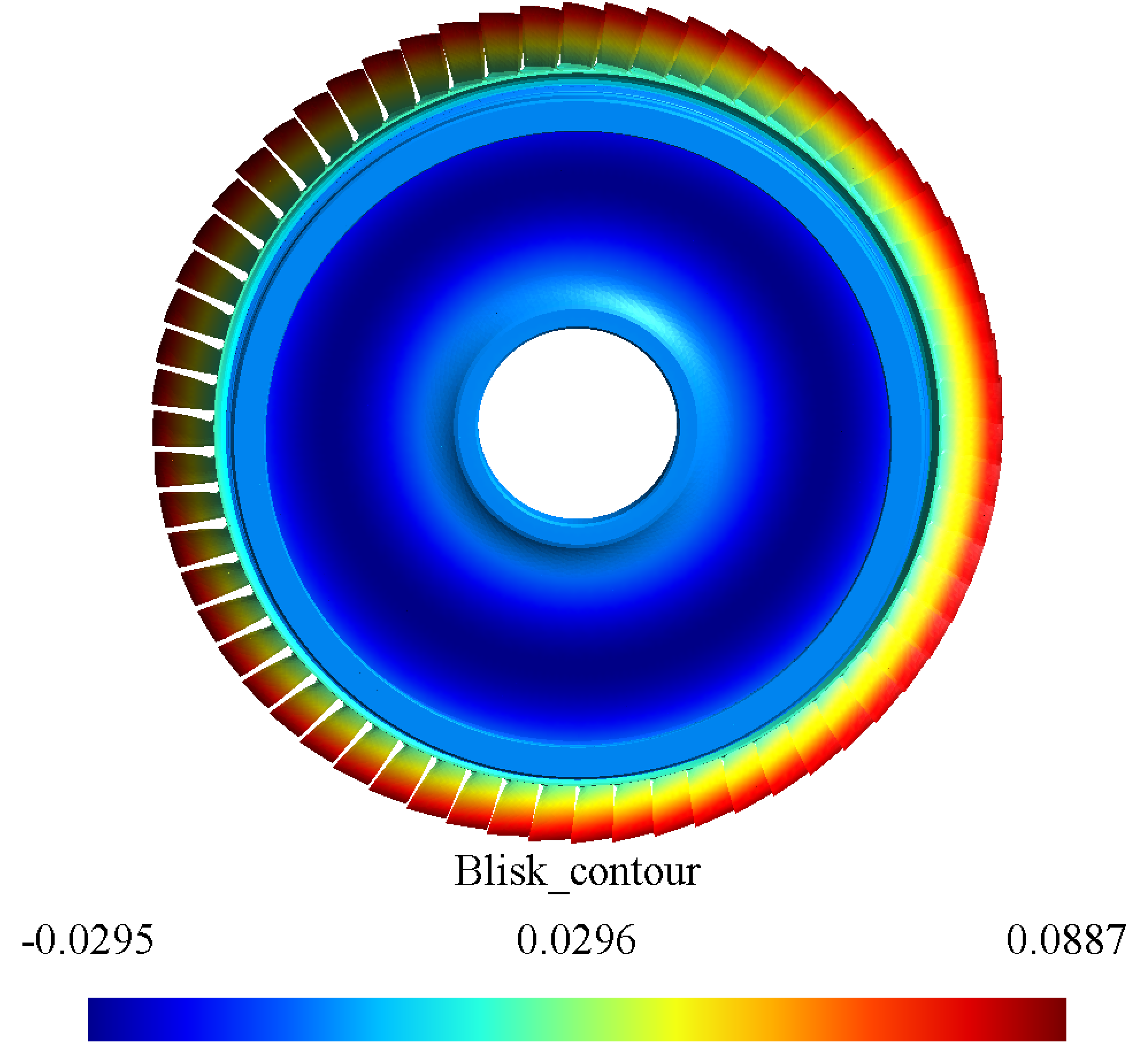}
  \caption{Resonant traveling displacement field at $F = 1000$~N.}  
  \end{subfigure}
  \caption{Reconstructed full-field axial traveling wave displacement contours of the compressor blisk at resonance, corresponding to the fundamental bending mode (0ND). The visualization is performed using the open-source finite element software Gmsh~\cite{geuzaine2009gmsh}.}
  \label{fig:blisk model with tanh nonlinearity 0ND mode shapes}   
\end{figure}

Concurrently with the backbone computation, the corresponding resonant mode shapes are extracted. Through the displacement recovery procedure applied to the reduced-order model, the full-field vibration displacement contours of the original blisk structure are reconstructed, as visualized in Figure~\ref{fig:blisk model with tanh nonlinearity 0ND mode shapes}. It is crucial to note that these contours depict the \textit{instantaneous} traveling wave displacement (a time snapshot). Although the vibration magnitude is identical across all cyclic sectors for a traveling wave, the visualized spatial wave-like pattern (sine/cosine distribution) inherently reflects the phase differences among sectors. The four subplots specifically correspond to the resonant states at the four distinct excitation levels marked in Figure~\ref{fig:blisk model with tanh nonlinearity 0ND FRCs}~(b).
From these reconstructed contours, visualized using the open-source finite element software Gmsh~\cite{geuzaine2009gmsh}, it is evident that as the excitation force increases, the structural response undergoes significant evolution, not only in magnitude and resonant frequency but also in its spatial deformation configuration. In stark contrast to linear systems, where the mode shapes remain invariant with respect to the energy level, the resonant deflection shapes in this nonlinear system exhibit distinct amplitude-dependent spatial variations. This phenomenon is fully consistent with the definition of NNMs, where the modal shape is intrinsically coupled to the vibration amplitude. Consequently, the proposed method demonstrates a distinct advantage: it does not merely track the resonance backbone at a discrete monitoring point but simultaneously captures the evolution of the complete nonlinear mode shape at every energy level, thereby providing a comprehensive representation of the system's global dynamic behavior.

\begin{figure}
  \centering
  \begin{subfigure}[b]{0.48\textwidth}
    \includegraphics[width=\textwidth]{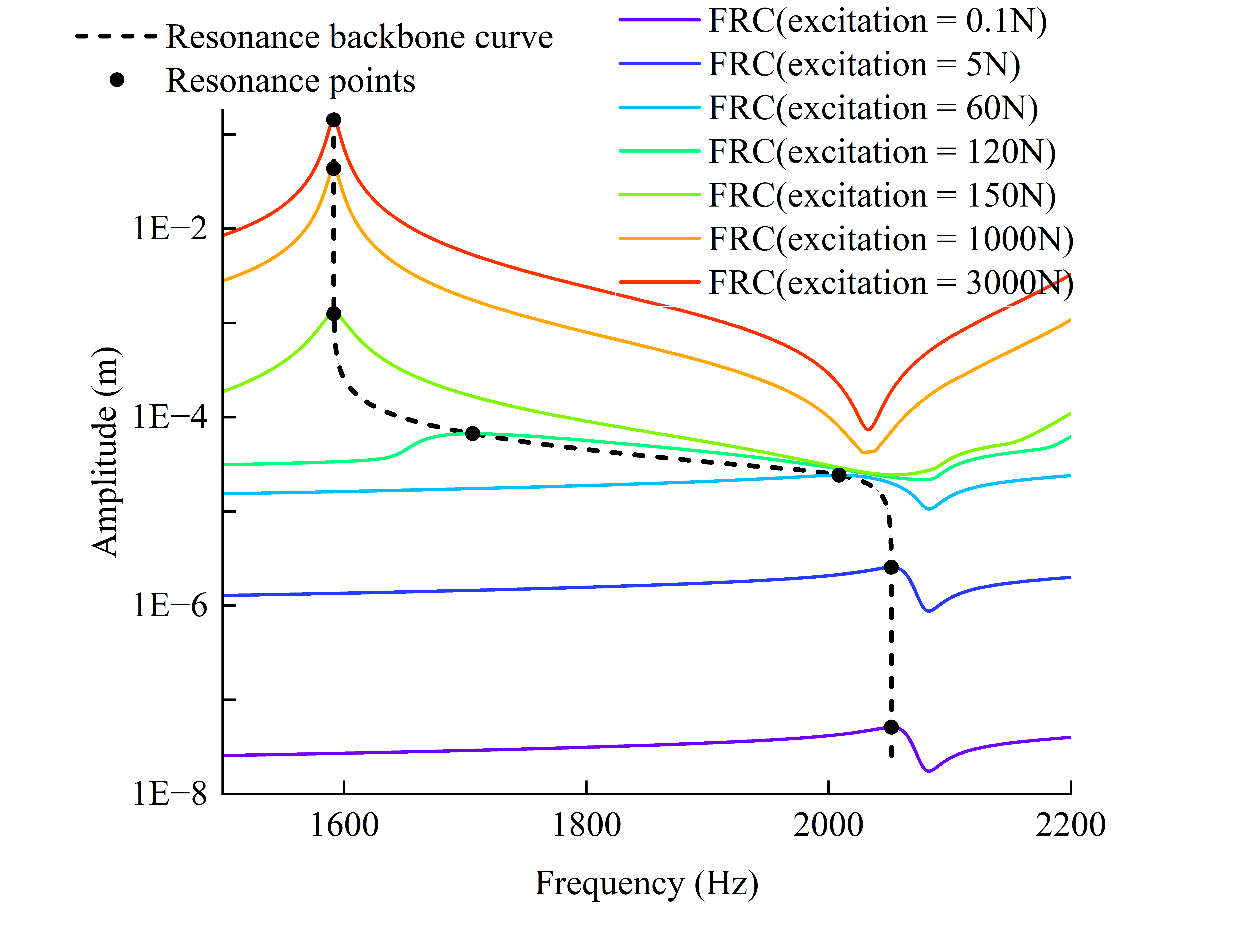}
     \caption{Validation of the resonance backbone curve for the 1ND mode.} 
  \end{subfigure}
  \hfill
  \begin{subfigure}[b]{0.48\textwidth}  
    \includegraphics[width=\textwidth]{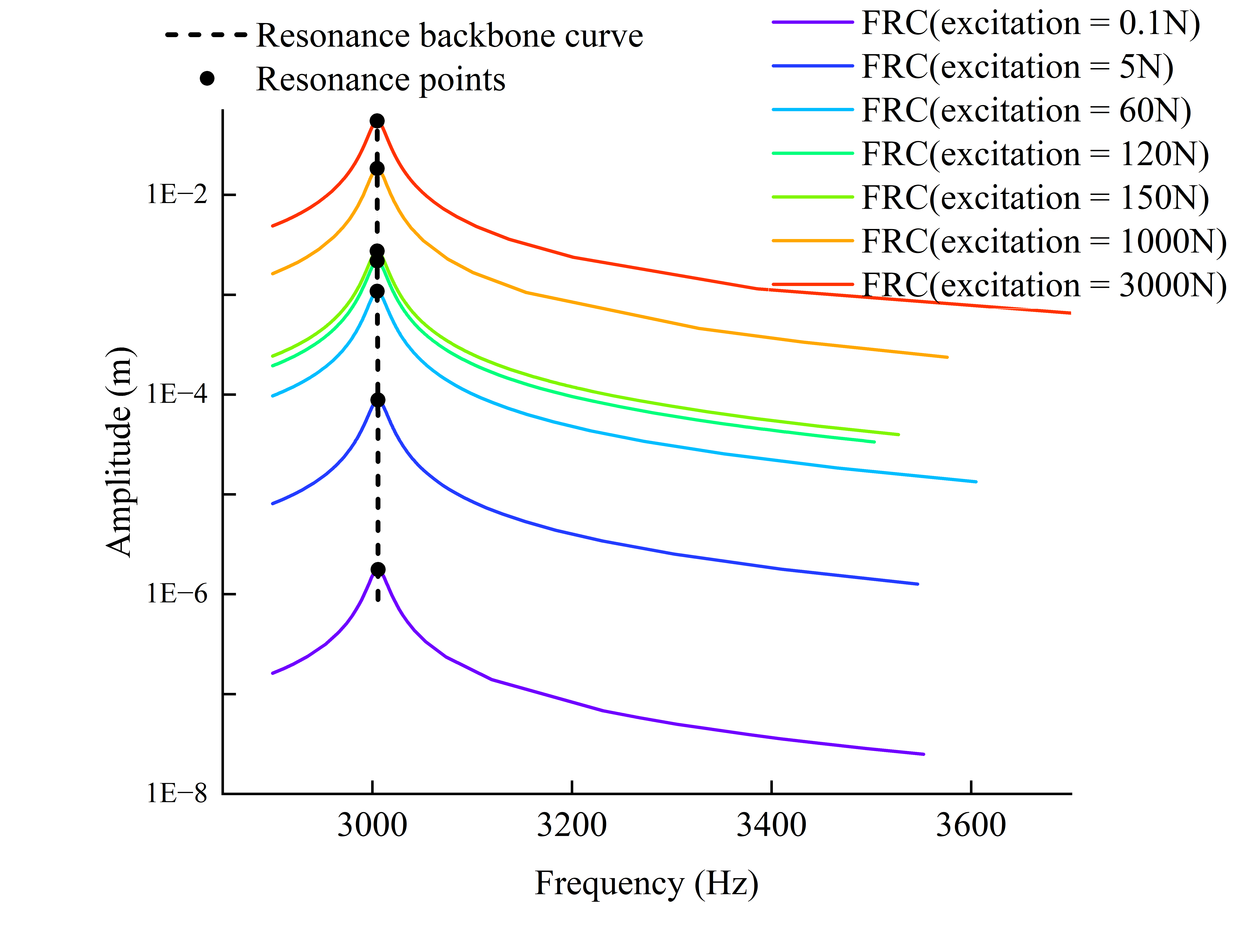}
    \caption{Validation of the resonance backbone curve for the 31ND mode.}  
  \end{subfigure} 
  \caption{Evolution of the resonance response for the first blade bending mode under varying excitation levels, demonstrating nonlinearity-induced frequency shifts.} 
  \label{fig:blisk model with tanh nonlinearity 1ND and 31ND FRCs}   
\end{figure}

To demonstrate the generality of the method, the forced response backbone curves corresponding to the first nonlinear modes of the 1-nodal diameter (1ND) and 31-nodal diameter (31ND) were computed. As shown in Figure~\ref{fig:blisk model with tanh nonlinearity 1ND and 31ND FRCs}, the obtained results are perfectly consistent with the resonance points on the FRCs calculated via HBM. This further validates the robustness and broad applicability of the proposed method.

\begin{figure}
  \centering
  \begin{subfigure}[t]{0.45\textwidth}

    \includegraphics[width=\textwidth]{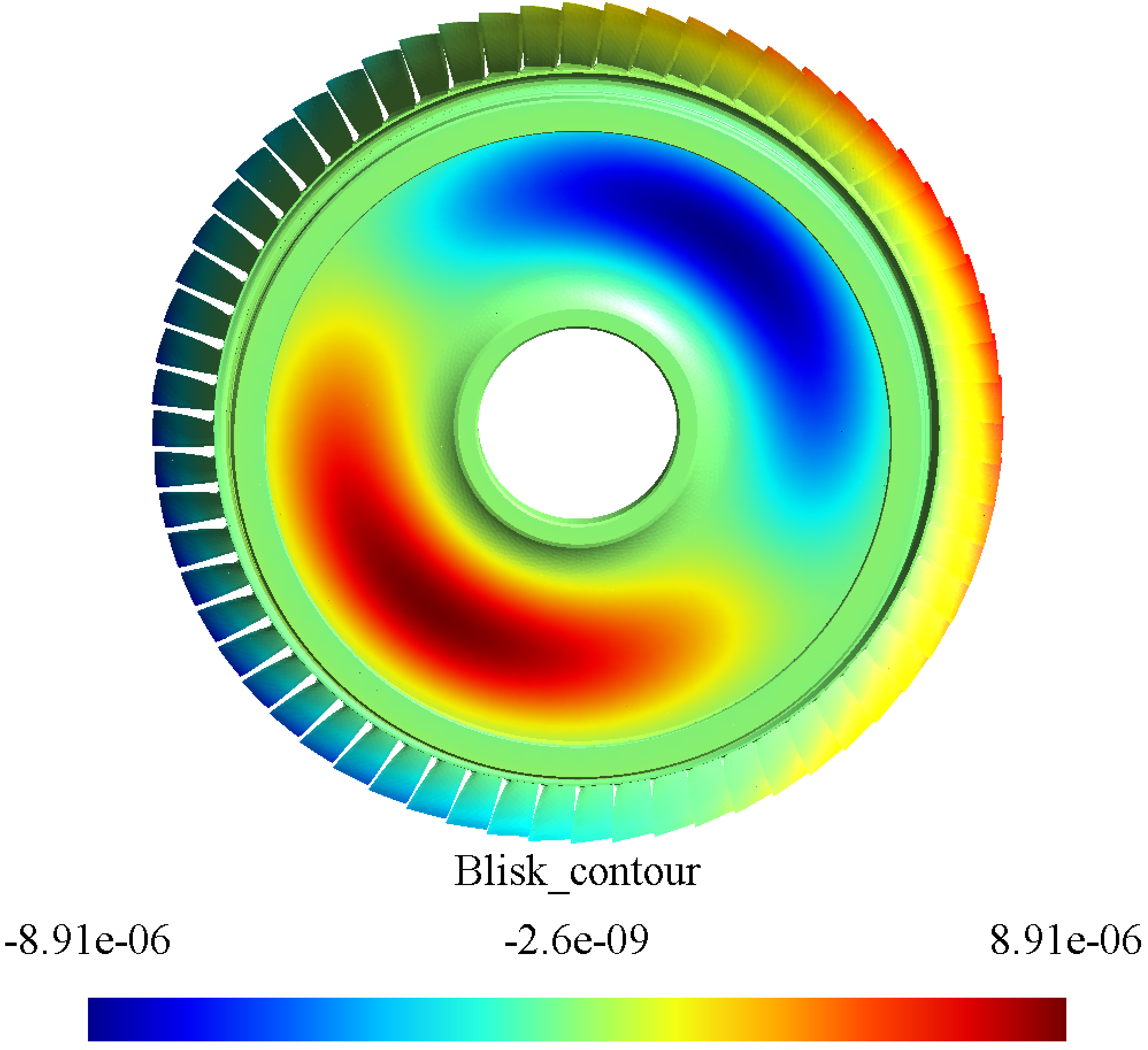}
    \caption{ Resonant traveling displacement field for the \textbf{1ND} mode at $F = 5$~N.}  
  \end{subfigure}
  \hfill
  \begin{subfigure}[t]{0.45\textwidth}
 
    \includegraphics[width=\textwidth]{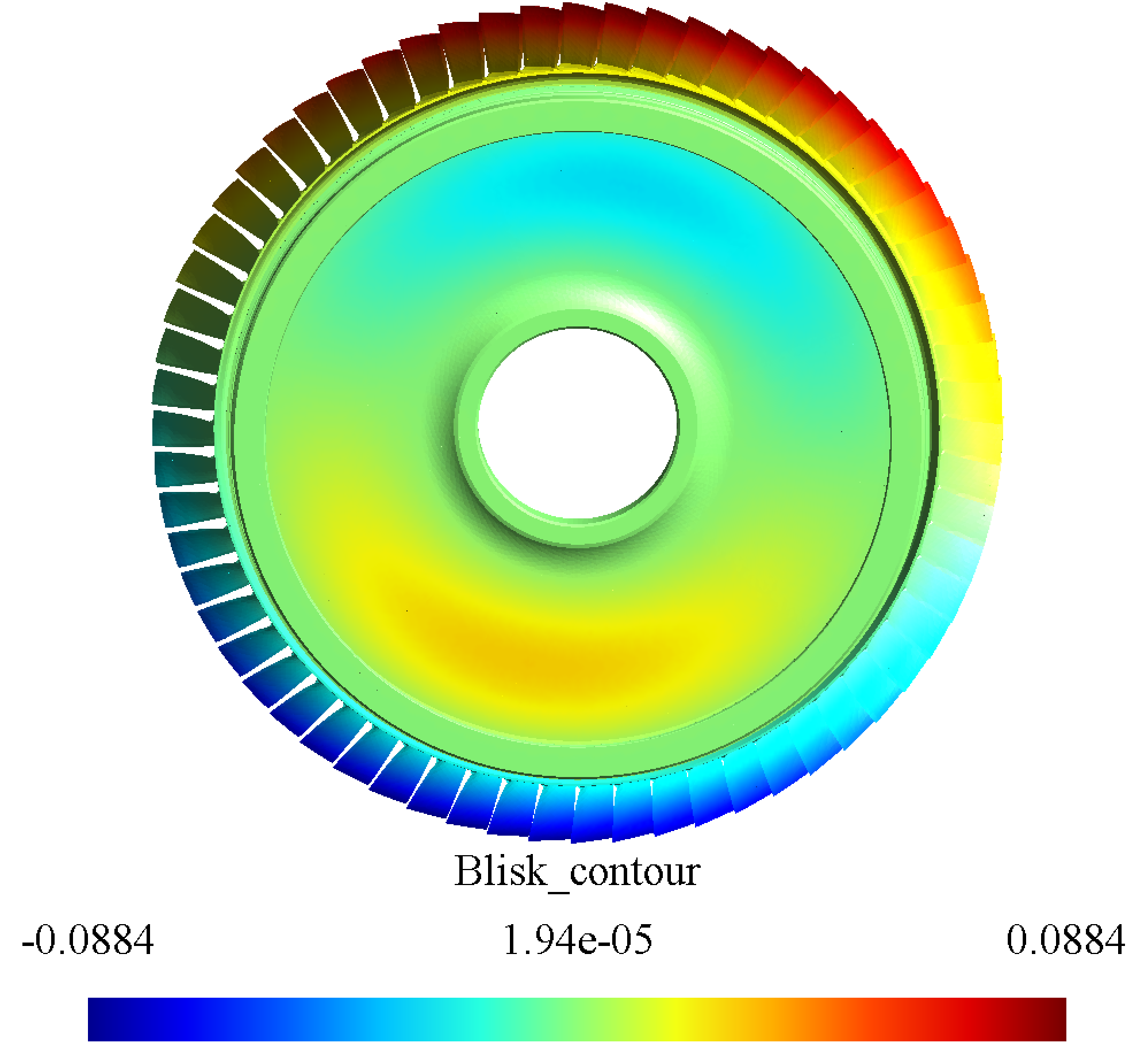}
  \caption{Resonant traveling displacement field for the \textbf{1ND} mode at $F = 1000$~N.}   
  \end{subfigure}
  
  \begin{subfigure}[t]{0.45\textwidth}
  
    \includegraphics[width=\textwidth]{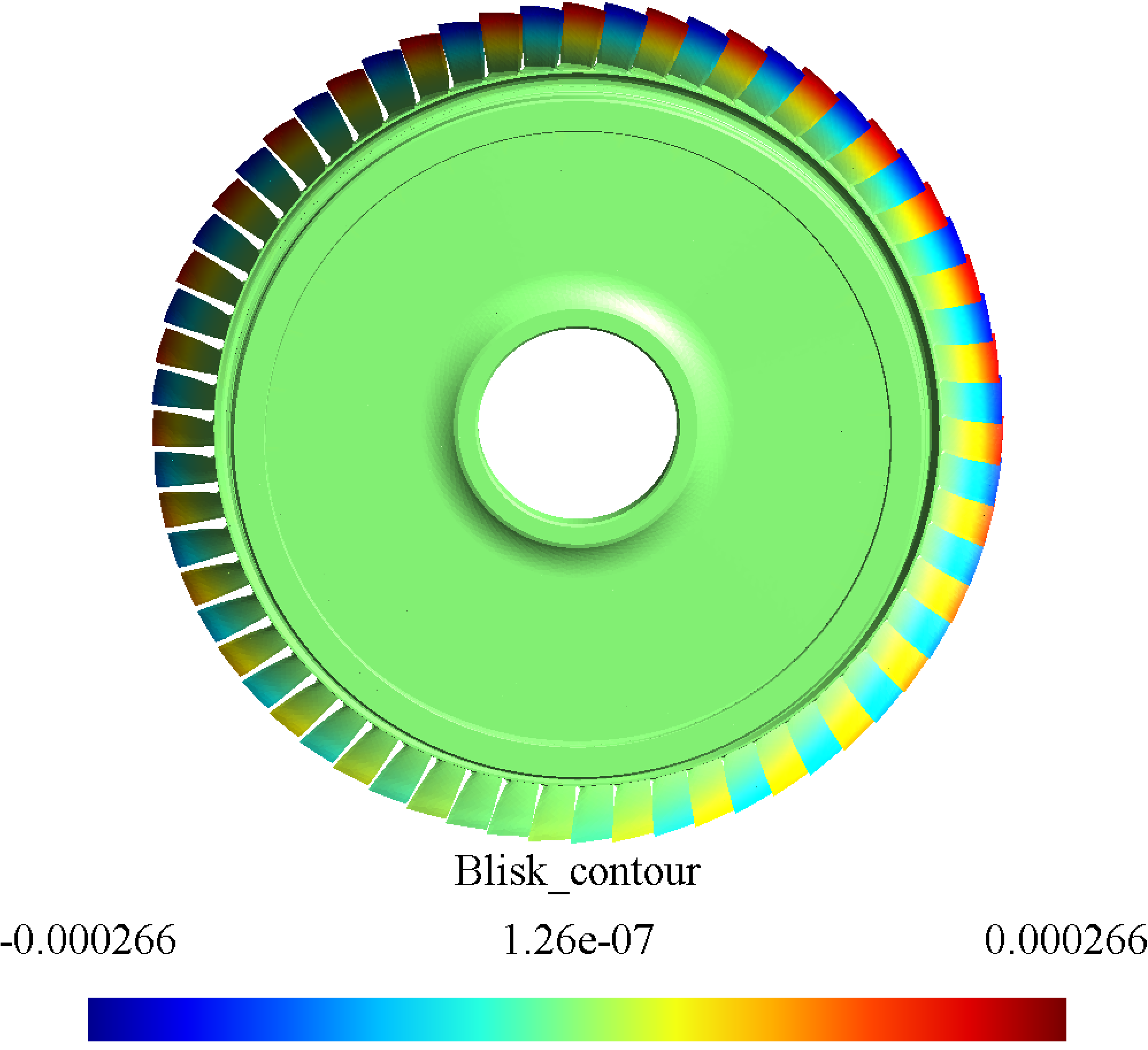}
  \caption{Resonant traveling displacement field for the \textbf{31ND} mode at $F = 5$~N.}  
  \end{subfigure}
  \hfill
  \begin{subfigure}[t]{0.45\textwidth}
  
    \includegraphics[width=\textwidth]{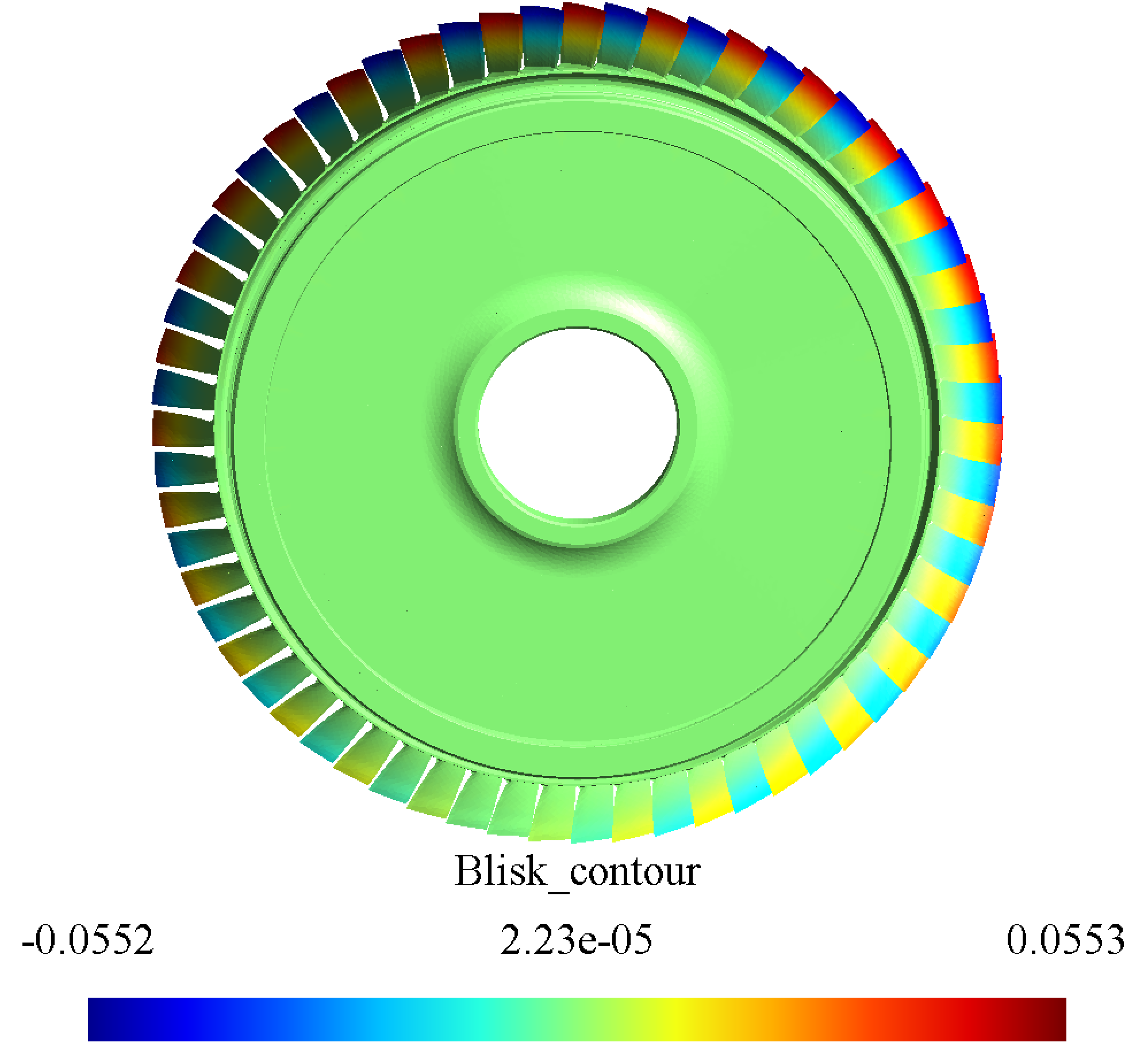}
   \caption{Resonant traveling displacement field for the \textbf{31ND} mode at $F = 1000$~N.} 
  \end{subfigure}
  \caption{Validation of the coupling-dependent nonlinearity hypothesis via resonant displacement reconstruction. The figure compares traveling wave axial displacement contours for the 1ND and 31ND modes at excitation levels of 5~N and 1000~N. The 31ND contours exhibit negligible topological variation, indicating that weak blade-disk coupling effectively minimizes relative interface motion, thereby isolating the mode from friction nonlinearity. In contrast, the 1ND contours undergo substantial spatial distortion due to strong coupling. This visual evidence implies that the significant frequency shifting observed in the 1ND mode stems from mode shape distortion, whereas the 31ND mode behaves in a quasi-linear manner. Visualization performed using Gmsh~\cite{geuzaine2009gmsh}.}
  \label{fig:blisk model with tanh nonlinearity 1ND and 31ND mode shapes}   
\end{figure}

From Figure~\ref{fig:blisk model with tanh nonlinearity 1ND and 31ND FRCs}, a distinct phenomenon is observed regarding the resonant response of the first blade bending mode. Unlike the 0ND and 1ND cases, the 31ND mode exhibits almost no frequency shift, and the resonance curve retains a linear characteristic without significant distortion. We hypothesize that this behavior arises because the first blade bending mode at this high nodal diameter involves minimal disk-blade coupling, leading to negligible relative motion at the friction interface. To verify this, the resonant traveling wave displacement contours for the 1ND and 31ND modes under excitation forces of 5 N and 1000 N are Visualized in Figure~\ref{fig:blisk model with tanh nonlinearity 1ND and 31ND mode shapes}.

It is clearly observed that for the 1ND case, the resonant mode shapes differ significantly between the two excitation levels, with particularly pronounced variations in the displacement pattern of the disk. In contrast, for the 31ND case, the resonant mode shapes remain virtually invariant. As illustrated in Figure~\ref{fig:blisk model with tanh nonlinearity 1ND and 31ND mode shapes}, this is attributed to the fact that the blade bending mode at 31ND involves minimal disk participation, with the disk response remaining negligible. This observation corroborates the aforementioned hypothesis.

In summary, the presented examples validate the proposed method for computing resonance backbone curves. Results demonstrate accurate tracking of resonance and anti-resonance curves. High efficiency was confirmed by benchmarking the analytical Hessian against numerical integration. Furthermore, the application to the blisk–damper system proves its applicability to complex structures with localized nonlinearities. Thus, the proposed method provides a robust and efficient tool for analyzing general nonlinear mechanical systems.

\section{Conclusion} \label{Conclusion}
This paper presents a high-efficiency numerical framework for computing the forced vibration resonance and anti-resonance backbone curves of mechanical systems. Based on the frequency-domain dynamic equations derived via the HBM, the proposed approach integrates Lagrange multipliers to rapidly trace the backbone curves and corresponding vibration modes of any non-delayed, second-order smooth ($C^2$-continuous) nonlinear system. By extending the classical AFT method, we derive the analytical Hessian tensors of nonlinear elements, which accelerates the computation of the Jacobian matrix for the Lagrange multiplier equations, thereby significantly enhancing Newton-iteration efficiency. Furthermore, for localized nonlinear elements requiring coordinate transformations, a tensor transformation method for the elemental Hessian is developed. This innovation substantially improves the method's applicability to localized nonlinear models, particularly within finite element environments. Compared to traditional forced vibration analysis, the proposed method directly captures the system's resonance backbone without tracking error, achieves superior computational efficiency, and is applicable to a broad class of nonlinear systems.

The validity and effectiveness of the proposed framework were demonstrated through three numerical examples. First, using a 2-DOF system with cubic stiffness, we verified the method's accuracy against the traditional HBM and confirmed its robustness against numerical instabilities in the resonance region. The capability to compute anti-resonance backbone curves was also successfully validated. A comparative analysis with numerical differentiation for Jacobian computation revealed that the analytical Hessian significantly improves solution efficiency. Second, in a beam finite element model with cubic nonlinearity, both resonance and anti-resonance backbones were computed, and the corresponding nonlinear modes were visualized. Subsequently, a tanh friction element was introduced to the beam model. The computed backbone curve precisely traced the resonance peaks obtained via HBM and shooting methods, demonstrating the method's versatility in handling general non-polynomial nonlinearities. Finally, the approach was applied to a compressor blisk finite element model equipped with a friction damper ring. The successful analysis of the blisk’s nonlinear behavior confirms that, when combined with model reduction, the proposed method is well-suited for industrial-scale finite element models.

Despite its distinct advantages, the proposed method has certain limitations. First, it is not directly applicable to systems lacking second-order smoothness (e.g., non-smooth contact mechanics without regularization). Second, since the formulation involves second-order derivatives of the governing equations, the computational complexity for nonlinear terms scales cubically with the number of nonlinear degrees of freedom. While nonlinear degree-of-freedom reduction effectively mitigates this issue for large-scale systems with localized nonlinearities, the computational burden remains significant for systems exhibiting global nonlinearity. Future work will focus on developing specialized algorithms leveraging GPU parallel computing to accelerate the large-scale matrix and tensor operations required for such cases.







\bibliographystyle{elsarticle-num}  
\bibliography{bibtex} 

\appendix
\section{Definitions of Frequency-Domain Matrices} \label{Appendix A}
This appendix details the explicit structures of the vectors and matrices presented in Equation (3). The vectors of Fourier coefficients for the displacement $\boldsymbol{Q}$, nonlinear force $\boldsymbol{F}_\mathrm{nl}$, and external excitation $\boldsymbol{F}_\mathrm{ex}$ are constructed by stacking the harmonic components as:
\begin{equation}
\begin{aligned}
\boldsymbol{Q} & =\left[\boldsymbol{Q}^0 ; \boldsymbol{Q}^{\mathrm{c}, 1} ; \boldsymbol{Q}^{\mathrm{s}, 1} ; \cdots ; \boldsymbol{Q}^{\mathrm{c}, N_h} ; \boldsymbol{Q}^{\mathrm{s}, N_h}\right], \\
\boldsymbol{F}_\mathrm{nl} & =\left[\boldsymbol{F}_\mathrm{nl}{ }^0 ; \boldsymbol{F}_\mathrm{nl}{ }^{\mathrm{c}, 1} ; \boldsymbol{F}_\mathrm{n l}{ }^{\mathrm{s}, 1} ; \cdots ; \boldsymbol{F}_\mathrm{nl}{ }^{\mathrm{c}, N_h} ; \boldsymbol{F}_\mathrm{nl}{ }^{\mathrm{s}, N_h}\right], \\
\boldsymbol{F}_\mathrm{ex} & =\left[\boldsymbol{F}_\mathrm{ex}{ }^0 ; \boldsymbol{F}_\mathrm{ex}{ }^{\mathrm{c}, 1} ; \boldsymbol{F}_\mathrm{ex}{ }^{\mathrm{s}, 1} ; \cdots ; \boldsymbol{F}_\mathrm{ex}{ }^{\mathrm{c}, N_h} ; \boldsymbol{F}_\mathrm{ex}{ }^{\mathrm{s}, N_h}\right].
\end{aligned}
\end{equation}

The dynamic stiffness matrix $\boldsymbol{D}(\omega)$ exhibits a block-diagonal structure, where $\boldsymbol{K}$, $\boldsymbol{M}$, and $\boldsymbol{C}$ denote the linear stiffness, mass, and damping matrices, respectively:
\begin{equation}
\left[\begin{array}{cccccc}
\boldsymbol{K} & 0 & 0 & \cdots & 0 & 0 \\
0 & \boldsymbol{K}-(\omega)^2 \boldsymbol{M} & \omega \boldsymbol{C} & \cdots & 0 & 0 \\
0 & -\omega \boldsymbol{C} & \boldsymbol{K}-(\omega)^2 \boldsymbol{M} & \cdots & 0 & 0 \\
\cdots & \cdots & \cdots & \cdots & \cdots & \cdots \\
0 & 0 & 0 & \cdots & \boldsymbol{K}-\left(N_\mathrm{h} \omega\right)^2 \boldsymbol{M} & N_\mathrm{h} \omega \boldsymbol{C} \\
0 & 0 & 0 & \cdots & -N_\mathrm{h} \omega \boldsymbol{C} & \boldsymbol{K}-\left(N_\mathrm{h} \omega\right)^2 \boldsymbol{M}
\end{array}\right].
\end{equation}

\section{Relationship between DFT Sequence and Fourier Series Coefficients} \label{Appendix B}
Apply the DFT to the sequence.
\begin{equation}
X[n]=\sum_{l=0}^{N-1} x[l] e^{\left(\frac{-i 2 \pi}{N} n l\right)} \quad n=0,1, \cdots, N-1.
\end{equation}
The Inverse Discrete Fourier Transform (IDFT) is defined as:
\begin{equation}
x[l]=\frac{1}{N} \sum_{n=0}^{N-1} X[n] e^{\left(\frac{i 2 \pi}{N} n l\right)} \quad l=0,1, \cdots, N-1.
\end{equation}
The Fourier series expansion of a periodic variable is given by:
\begin{equation}
x(t)=X^0+\sum_{k=1}^{\infty}\left[X^{\mathrm{c}, k} \cos (k \omega t)+X^{\mathrm{s}, k} \sin (k \omega t)\right] \quad X^0, X^{\mathrm{c}, k}, X^{\mathrm{s}, k} \in \mathbb{R}.
\end{equation}
Its complex form is expressed as:
\begin{equation}
x(t)=\sum_{k=-\infty}^{\infty} \tilde{X}^k \mathrm{e}^{i k \omega t} \quad \tilde{X}^k \in \mathbb{C}, \forall k \neq 0, \tilde{X}^k=\left(\tilde{X}^{-k}\right)^*.
\end{equation}
Substituting $t = \Delta T \cdot l$ and Equation (B.4) into Equation (B.1):
\begin{equation}
X[n]=\sum_{l=0}^{N-1} \sum_{k=-\infty}^{\infty} \tilde{X}^k e^{\left(\frac{i 2 \pi}{N} k l\right)} e^{\left(\frac{-i 2 \pi}{N} n l\right)} \quad n=0,1, \cdots, N-1.
\end{equation}
Combining the exponential terms and interchanging the order of summation yields:
\begin{equation}
X[n]=\sum_{k=-\infty}^{\infty} \tilde{X}^k\left(\sum_{l=0}^{N-1} e^{\left(\frac{i 2 \pi}{N}(k-n) l\right)}\right) \quad n=0,1, \cdots, N-1.
\end{equation}
Let $S$ be the sum within the parentheses:
\begin{equation}
\left.S=\left(\sum_{l=0}^{N-1} e^{\left(\frac{i 2 \pi}{N}(k-n) l\right.}\right)\right).
\end{equation}
$S$ is a geometric series, and its value is determined in two cases:
\begin{equation}
S=\left\{\begin{array}{lll}
N & k-n=r N & r \in \mathbb{Z} \\
0 & k-n \neq r N & r \in \mathbb{Z}
\end{array}\right..
\end{equation}
We substitute this expression into Equation (B.6):
\begin{equation}
X[n]=\sum_{r=-\infty}^{\infty} \tilde{X}^{n+r N} N \quad n=0,1, \cdots, N-1.
\end{equation}
When $N$ is greater than twice the highest retained harmonic (satisfying the sampling theorem), and the frequency components outside the analysis bandwidth are negligible, the above equation simplifies to:
\begin{equation}
X[n] \approx \tilde{X}^n N \quad n=0,1, \cdots, N-1.
\end{equation}
In the following, we neglect frequency components outside the bandwidth and use the equation:
\begin{equation}
\tilde{X}^n=X[n] / N \quad n=0,1, \cdots, N-1.
\end{equation}
This equation is of fundamental importance in signal analysis, as it enables us to process continuous signals using methods designed for discrete signals. The corresponding relationship with the coefficients of the trigonometric Fourier series is given by:
\begin{equation}
\begin{aligned}
& X[0] / N=X^0, \\
& X[n] / N=\left(X^{\mathrm{c}, k}-i X^{\mathrm{s}, k}\right) / 2.
\end{aligned}
\end{equation}

\section{Second-order Derivative Terms Involving Excitation Frequency} \label{Appendix C}
This appendix presents the explicit expressions for the second-order partial derivatives involving the excitation frequency, as discussed in Section \ref{sec:Extended AFT and Hessian}. By exploiting the symmetry of mixed derivatives and utilizing the discrete time-domain samples from Equation (46), these derivative terms are obtained as follows:
\begin{equation}
\frac{\partial^2 F_\mathrm{n l}^0}{\partial \omega \partial X^{\mathrm{c}, n}}=\frac{1}{N} \sum_{l=0}^{N-1}\left\{\begin{array}{l}
-n \frac{\partial f_\mathrm{n l}}{\partial \dot{x}}(l) \sin \left(\frac{2 \pi n l}{N}\right) \\
+\sum_{m=1}^{N_\mathrm{h}} n m \omega X^{\mathrm{c}, m} \frac{\partial^2 f_\mathrm{n l}}{\partial \dot{x}^2}(l) \sin \left(\frac{2 \pi m l}{N}\right) \sin \left(\frac{2 \pi n l}{N}\right) \\
-\sum_{m=1}^{N_\mathrm{h}}n m \omega X^{\mathrm{s}, m} \frac{\partial^2 f_\mathrm{n l}}{\partial \dot{x}^2}(l) \cos \left(\frac{2 \pi m l}{N}\right) \sin \left(\frac{2 \pi n l}{N}\right) \\
-\sum_{m=1}^{N_\mathrm{h}}m X^{\mathrm{c}, m} \frac{\partial^2 f_\mathrm{n l}}{\partial \dot{x} \partial x}(l) \sin \left(\frac{2 \pi m l}{N}\right) \cos \left(\frac{2 \pi n l}{N}\right) \\
+\sum_{m=1}^{N_\mathrm{h}}m X^{\mathrm{s}, m} \frac{\partial^2 f_\mathrm{n l}}{\partial \dot{x} \partial x}(l) \cos \left(\frac{2 \pi m l}{N}\right) \cos \left(\frac{2 \pi n l}{N}\right)
\end{array}\right\},
\end{equation}
\begin{equation}
\frac{\partial^2 F_\mathrm{n l}^0}{\partial \omega \partial X^{\mathrm{s}, n}}=\frac{1}{N} \sum_{l=0}^{N-1}\left\{\begin{array}{l}
-n \frac{\partial f_\mathrm{n l}}{\partial \dot{x}}(l) \sin \left(\frac{2 \pi n l}{N}\right) \\
+\sum_{m=1}^{N_\mathrm{h}}n m \omega X^{\mathrm{c}, m} \frac{\partial^2 f_\mathrm{n l}}{\partial \dot{x}^2}(l) \sin \left(\frac{2 \pi m l}{N}\right) \cos \left(\frac{2 \pi n l}{N}\right) \\
-\sum_{m=1}^{N_\mathrm{h}}n m \omega X^{\mathrm{s}, m} \frac{\partial^2 f_\mathrm{n l}}{\partial \dot{x}^2}(l) \cos \left(\frac{2 \pi m l}{N}\right) \cos \left(\frac{2 \pi n l}{N}\right) \\
-\sum_{m=1}^{N_\mathrm{h}}m X^{\mathrm{c}, m} \frac{\partial^2 f_\mathrm{n l}}{\partial \dot{x} \partial x}(l) \sin \left(\frac{2 \pi m l}{N}\right) \sin \left(\frac{2 \pi n l}{N}\right) \\
+\sum_{m=1}^{N_\mathrm{h}}m X^{\mathrm{s}, m} \frac{\partial^2 f_\mathrm{n l}}{\partial \dot{x} \partial x}(l) \cos \left(\frac{2 \pi m l}{N}\right) \sin \left(\frac{2 \pi n l}{N}\right)
\end{array}\right\},
\end{equation}
\begin{equation}
\frac{\partial^2 F_\mathrm{n l}^{\mathrm{c}, k}}{\partial \omega \partial X^{\mathrm{c}, n}}=\frac{2}{N} \sum_{l=0}^{N-1}\left\{\begin{array}{l}
-n \frac{\partial f_\mathrm{n l}}{\partial \dot{x}}(l) \cos \left(\frac{2 \pi k l}{N}\right) \sin \left(\frac{2 \pi n l}{N}\right) \\
+\sum_{m=1}^{N_\mathrm{h}}n m \omega X^{\mathrm{c}, m} \frac{\partial^2 f_\mathrm{n l}}{\partial \dot{x}^2}(l) \cos \left(\frac{2 \pi k l}{N}\right) \sin \left(\frac{2 \pi m l}{N}\right) \sin \left(\frac{2 \pi n l}{N}\right) \\
-\sum_{m=1}^{N_\mathrm{h}}n m \omega X^{\mathrm{s}, m} \frac{\partial^2 f_\mathrm{n l}}{\partial \dot{x}^2}(l) \cos \left(\frac{2 \pi k l}{N}\right) \cos \left(\frac{2 \pi m l}{N}\right) \sin \left(\frac{2 \pi n l}{N}\right) \\
-\sum_{m=1}^{N_\mathrm{h}}m X^{\mathrm{c}, m} \frac{\partial^2 f_\mathrm{n l}}{\partial \dot{x} \partial x}(l) \cos \left(\frac{2 \pi k l}{N}\right) \sin \left(\frac{2 \pi m l}{N}\right) \cos \left(\frac{2 \pi n l}{N}\right) \\
+\sum_{m=1}^{N_\mathrm{h}}m X^{\mathrm{s}, m} \frac{\partial^2 f_\mathrm{n l}}{\partial \dot{x} \partial x}(l) \cos \left(\frac{2 \pi k l}{N}\right) \cos \left(\frac{2 \pi m l}{N}\right) \cos \left(\frac{2 \pi n l}{N}\right)
\end{array}\right\},
\end{equation}
\begin{equation}
\frac{\partial^2 F_\mathrm{n l}^{\mathrm{c}, k}}{\partial \omega \partial X^{\mathrm{s}, n}}=\frac{2}{N} \sum_{l=0}^{N-1}\left\{\begin{array}{l}
-n \frac{\partial f_\mathrm{n l}}{\partial \dot{x}}(l) \cos \left(\frac{2 \pi k l}{N}\right) \sin \left(\frac{2 \pi n l}{N}\right) \\
+\sum_{m=1}^{N_\mathrm{h}}n m \omega X^{\mathrm{c}, m} \frac{\partial^2 f_\mathrm{n l}}{\partial \dot{x}^2}(l) \cos \left(\frac{2 \pi k l}{N}\right) \sin \left(\frac{2 \pi m l}{N}\right) \cos \left(\frac{2 \pi n l}{N}\right) \\
-\sum_{m=1}^{N_\mathrm{h}}n m \omega X^{\mathrm{s}, m} \frac{\partial^2 f_\mathrm{n l}}{\partial \dot{x}^2}(l) \cos \left(\frac{2 \pi k l}{N}\right) \cos \left(\frac{2 \pi m l}{N}\right) \cos \left(\frac{2 \pi n l}{N}\right) \\
-\sum_{m=1}^{N_\mathrm{h}}m X^{\mathrm{c}, m} \frac{\partial^2 f_\mathrm{n l}}{\partial \dot{x} \partial x}(l) \cos \left(\frac{2 \pi k l}{N}\right) \sin \left(\frac{2 \pi m l}{N}\right) \sin \left(\frac{2 \pi n l}{N}\right) \\
+\sum_{m=1}^{N_\mathrm{h}}m X^{\mathrm{s}, m} \frac{\partial^2 f_\mathrm{n l}}{\partial \dot{x} \partial x}(l) \cos \left(\frac{2 \pi k l}{N}\right) \cos \left(\frac{2 \pi m l}{N}\right) \sin \left(\frac{2 \pi n l}{N}\right)
\end{array}\right\},
\end{equation}
\begin{equation}
\frac{\partial^2 F_\mathrm{n l}^{\mathrm{s}, k}}{\partial \omega \partial X^{\mathrm{c}, n}}=\frac{2}{N} \sum_{l=0}^{N-1}\left\{\begin{array}{l}
-n \frac{\partial f_\mathrm{n l}}{\partial \dot{x}}(l) \sin \left(\frac{2 \pi k l}{N}\right) \sin \left(\frac{2 \pi n l}{N}\right) \\
+\sum_{m=1}^{N_\mathrm{h}}n m \omega X^{\mathrm{c}, m} \frac{\partial^2 f_\mathrm{n l}}{\partial \dot{x}^2}(l) \sin \left(\frac{2 \pi k l}{N}\right) \sin \left(\frac{2 \pi m l}{N}\right) \sin \left(\frac{2 \pi n l}{N}\right) \\
-\sum_{m=1}^{N_\mathrm{h}}n m \omega X^{\mathrm{s}, m} \frac{\partial^2 f_\mathrm{n l}}{\partial \dot{x}^2}(l) \sin \left(\frac{2 \pi k l}{N}\right) \cos \left(\frac{2 \pi m l}{N}\right) \sin \left(\frac{2 \pi n l}{N}\right) \\
-\sum_{m=1}^{N_\mathrm{h}}m X^{\mathrm{c}, m} \frac{\partial^2 f_\mathrm{n l}}{\partial \dot{x} \partial x}(l) \sin \left(\frac{2 \pi k l}{N}\right) \sin \left(\frac{2 \pi m l}{N}\right) \cos \left(\frac{2 \pi n l}{N}\right) \\
+\sum_{m=1}^{N_\mathrm{h}}m X^{\mathrm{s}, m} \frac{\partial^2 f_\mathrm{n l}}{\partial \dot{x} \partial x}(l) \sin \left(\frac{2 \pi k l}{N}\right) \cos \left(\frac{2 \pi m l}{N}\right) \cos \left(\frac{2 \pi n l}{N}\right).
\end{array}\right\},
\end{equation}
\begin{equation}
\frac{\partial^2 F_\mathrm{n l}^{\mathrm{s}, k}}{\partial \omega \partial X^{\mathrm{s}, n}}=\frac{2}{N} \sum_{l=0}^{N-1}\left\{\begin{array}{l}
-n \frac{\partial f_\mathrm{n l}}{\partial \dot{x}}(l) \sin \left(\frac{2 \pi k l}{N}\right) \sin \left(\frac{2 \pi n l}{N}\right) \\
+\sum_{m=1}^{N_\mathrm{h}}n m \omega X^{\mathrm{c}, m} \frac{\partial^2 f_\mathrm{n l}}{\partial \dot{x}^2}(l) \sin \left(\frac{2 \pi k l}{N}\right) \sin \left(\frac{2 \pi m l}{N}\right) \cos \left(\frac{2 \pi n l}{N}\right) \\
-\sum_{m=1}^{N_\mathrm{h}}n m \omega X^{\mathrm{s}, m} \frac{\partial^2 f_\mathrm{n l}}{\partial \dot{x}^2}(l) \sin \left(\frac{2 \pi k l}{N}\right) \cos \left(\frac{2 \pi m l}{N}\right) \cos \left(\frac{2 \pi n l}{N}\right) \\
-\sum_{m=1}^{N_\mathrm{h}} m X^{\mathrm{c}, m} \frac{\partial^2 f_\mathrm{n l}}{\partial \dot{x} \partial x}(l) \sin \left(\frac{2 \pi k l}{N}\right) \sin \left(\frac{2 \pi m l}{N}\right) \sin \left(\frac{2 \pi n l}{N}\right) \\
+\sum_{m=1}^{N_\mathrm{h}}m X^{\mathrm{s}, m} \frac{\partial^2 f_\mathrm{n l}}{\partial \dot{x} \partial x}(l) \sin \left(\frac{2 \pi k l}{N}\right) \cos \left(\frac{2 \pi m l}{N}\right) \sin \left(\frac{2 \pi n l}{N}\right).
\end{array}\right\}.
\end{equation}

The second-order partial derivative terms of the nonlinear force Fourier coefficients with respect to the fundamental excitation frequency can be written as follows:
\begin{equation}
\frac{\partial^2 F_\mathrm{n l}^0}{\partial \omega^2}=\frac{1}{N} \sum_{l=0}^{N-1} \sum_{n=1}^{N_\mathrm{h}} \sum_{m=1}^{N_\mathrm{h}}\left\{\begin{array}{l}
+n m X^{\mathrm{c}, n} X^{\mathrm{c}, m} \frac{\partial^2 f_{n l}}{\partial \dot{x}^2}(l) \sin \left(\frac{2 \pi n l}{N}\right) \sin \left(\frac{2 \pi m l}{N}\right) \\
-n m X^{\mathrm{c}, n} X^{\mathrm{s}, m} \frac{\partial^2 f_{n l}}{\partial \dot{x}^2}(l) \sin \left(\frac{2 \pi n l}{N}\right) \cos \left(\frac{2 \pi m l}{N}\right) \\
-n m X^{\mathrm{s}, n} X^{\mathrm{c}, m} \frac{\partial^2 f_{n l}}{\partial \dot{x}^2}(l) \cos \left(\frac{2 \pi n l}{N}\right) \sin \left(\frac{2 \pi m l}{N}\right) \\
+n m X^{\mathrm{s}, n} X^{\mathrm{s}, m} \frac{\partial^2 f_{n l}}{\partial \dot{x}^2}(l) \cos \left(\frac{2 \pi n l}{N}\right) \cos \left(\frac{2 \pi m l}{N}\right)
\end{array}\right\},
\end{equation}
\begin{equation}
\frac{\partial^2 F_\mathrm{n l}^{\mathrm{c}, k}}{\partial \omega^2}=\frac{2}{N} \sum_{l=0}^{N-1} \sum_{n=1}^{N_\mathrm{h}} \sum_{m=1}^{N_\mathrm{h}}\left\{\begin{array}{l}
+n m X^{\mathrm{c}, n} X^{\mathrm{c}, m} \frac{\partial^2 f_\mathrm{n l}}{\partial \dot{x}^2}(l) \cos \left(\frac{2 \pi k l}{N}\right) \sin \left(\frac{2 \pi n l}{N}\right) \sin \left(\frac{2 \pi m l}{N}\right) \\
-n m X^{\mathrm{c}, n} X^{\mathrm{s}, m} \frac{\partial^2 f_\mathrm{n l}}{\partial \dot{x}^2}(l) \cos \left(\frac{2 \pi k l}{N}\right) \sin \left(\frac{2 \pi n l}{N}\right) \cos \left(\frac{2 \pi m l}{N}\right) \\
-n m X^{\mathrm{s}, n} X^{\mathrm{c}, m} \frac{\partial^2 f_\mathrm{n l}}{\partial \dot{x}^2}(l) \cos \left(\frac{2 \pi k l}{N}\right) \cos \left(\frac{2 \pi n l}{N}\right) \sin \left(\frac{2 \pi m l}{N}\right) \\
+n m X^{\mathrm{s}, n} X^{\mathrm{s}, m} \frac{\partial^2 f_\mathrm{n l}}{\partial \dot{x}^2}(l) \cos \left(\frac{2 \pi k l}{N}\right) \cos \left(\frac{2 \pi n l}{N}\right) \cos \left(\frac{2 \pi m l}{N}\right)
\end{array}\right\},
\end{equation}
\begin{equation}
\frac{\partial^2 F_\mathrm{n l}^{\mathrm{c}, k}}{\partial \omega^2}=\frac{2}{N} \sum_{l=0}^{N-1} \sum_{n=1}^{N_\mathrm{h}} \sum_{m=1}^{N_\mathrm{h}}\left\{\begin{array}{l}
+n m X^{\mathrm{c}, n} X^{\mathrm{c}, m} \frac{\partial^2 f_\mathrm{n l}}{\partial \dot{x}^2}(l) \sin \left(\frac{2 \pi k l}{N}\right) \sin \left(\frac{2 \pi n l}{N}\right) \sin \left(\frac{2 \pi m l}{N}\right) \\
-n m X^{\mathrm{c}, n} X^{\mathrm{s}, m} \frac{\partial^2 f_\mathrm{n l}}{\partial \dot{x}^2}(l) \sin \left(\frac{2 \pi k l}{N}\right) \sin \left(\frac{2 \pi n l}{N}\right) \cos \left(\frac{2 \pi m l}{N}\right) \\
-n m X^{\mathrm{s}, n} X^{\mathrm{c}, m} \frac{\partial^2 f_\mathrm{n l}}{\partial \dot{x}^2}(l) \sin \left(\frac{2 \pi k l}{N}\right) \cos \left(\frac{2 \pi n l}{N}\right) \sin \left(\frac{2 \pi m l}{N}\right) \\
+n m X^{\mathrm{s}, n} X^{\mathrm{s}, m} \frac{\partial^2 f_\mathrm{n l}}{\partial \dot{x}^2}(l) \sin \left(\frac{2 \pi k l}{N}\right) \cos \left(\frac{2 \pi n l}{N}\right) \cos \left(\frac{2 \pi m l}{N}\right)
\end{array}\right\}.
\end{equation}

\end{document}